\documentclass[12pt]{article}
\usepackage{lmodern}
\usepackage[english]{babel}
\usepackage{pdflscape, pslatex} 
\usepackage{epstopdf}
\usepackage{float}
\usepackage{bm}
\usepackage[authoryear]{natbib}
\usepackage{times}
\usepackage{adjustbox} 
\usepackage{rotating}
\usepackage{setspace}
\usepackage{listings} 
\usepackage{array}
\usepackage{multirow, multicol} 
\usepackage{tabulary,ctable,tabularx}
\usepackage{booktabs}
\usepackage{graphicx}
\usepackage{appendix}
\usepackage{enumerate}
\usepackage{times}
\usepackage{subfig}
\usepackage{amsmath,amsfonts,amssymb,amsthm,titling}

\usepackage[a4paper,width=160mm,top=30mm,bottom=30mm]{geometry}

\usepackage{authblk}

\title{Constant versus Covariate Dependent Threshold in the Peaks-Over Threshold Method}

\author{ Richard Minkah\textsuperscript{}\thanks{\noindent\textsuperscript{}Corresponding author. E-mail address: \textit {rminkah@ug.edu.gh}}\\Department of Statistics and Actuarial Science, University of Ghana, Ghana\\ ~and\\ Tertius de Wet\\Department of Statistics and Actuarial Science, Stellenbosch University, South Africa}
\date{}

\begin{document}
	\def\spacingset#1{\renewcommand{\baselinestretch}%
		{#1}\small\normalsize} \spacingset{1}
\maketitle
\begin{abstract}
	The Peaks-Over Threshold is a fundamental method in the estimation of rare events such as small exceedance probabilities, extreme quantiles and return periods. The main problem with the Peaks-Over Threshold method relates to the selection of threshold above and below which the asymptotic results are valid for large and small observations respectively. In addition, the main assumption leading to the asymptotic results is that the observations are independent and identically distributed. However, in practice, many real life processes yield data that are non-stationary and/or related to some covariate variables. As a result, threshold selection gets complicated as it may depend on the covariates. Strong arguments have been made against the use of constant threshold as observation that is considered extreme at some covariate level may not qualify as an extreme observation at another covariate level. Some authors have attempted to obtain covariate dependent thresholds in different ways: the most appealing one relies on quantile regression. In this paper, we propose a covariate dependent threshold based on expectiles. We compare this threshold with the constant and the quantile regression in a simulation study for estimating the tail index of the Generalised Pareto distribution. As may be expected, no threshold is universally the best. However, certain general observations can be made for the exponential growth data considered. Firstly, we find that the expectile threshold outperforms the others when the response variable has smaller to medium values. Secondly, for larger values of the response variable, the constant threshold is generally the best method. The threshold selection methods are illustrated in the estimation of the tail index of an insurance claims data. 
	
\end{abstract}

\section{Introduction}\label{Int}

Let $Y_i,~ i = 1,\ldots,n$ be independent and identically distributed (i.i.d.) random variables with common continuous distribution function $F.$  We consider the behaviour of $Y$ given that it exceeds a high threshold, $u.$ The well-known \citet{Balkema1974} and \citet{Pickands1975} theorem proved that the generalised Pareto distribution (GPD) is the limiting distribution of the exceedances or excesses over $u,$ 

\begin{equation}\label{GPD}
H(y|\gamma, \sigma_u)=\left\{\begin{array}{ll}
1-\left(1+\frac{\gamma y}{\sigma_u}\right)^{-1/\gamma}, ~y\ge u,~&\mbox{if}~ \gamma\ne 0\\
1-\exp\left(-\frac{y}{\sigma_u}\right),~~~~~~y\ge u, & \mbox{if}~ \gamma=0,
\end{array}\right.
\end{equation} 
where $\gamma$ and $\sigma_u$ are the shape (extreme value index) and scale parameters of the distribution function $H.$ The goal of extreme value analysis is mainly to obtain estimates of high quantiles, exceedance probabilities and return periods. However, each of these parameters depend on the extreme value index, $\gamma,$ which measures the tail heaviness of the underlying distribution. In particular, the distribution belongs to the Pareto domain of attraction for $\gamma>0,$ Gumbel domain of attraction for $\gamma=0,$ and  the Weibull domain of attraction for $\gamma<0$ with a right endpoint. 

The GPD is often used to model the tails of underlying distribution of data and the main difficulty in this approach is the selection of the threshold, $u,$ above which the GPD assumption is appropriate. It is usually a delicate balance between bias and variance. A high threshold may result in less excesses leading to large variance and a low threshold may violate the GPD assumption leading to large bias. This remains a central issue in the Peaks-Over Threshold (POT) method in statistics of extremes \citep[see e.g.][]{Tancredi2006,Thompson2009,Scarrott2012}. In the iid case, \citet{Scarrott2012} provide a plethora of methods for threshold selection. However, threshold selection in the presence of covariate information has not received much attention in the literature. This paper reviews the few available methods and provide a new method for selecting threshold in the presence of covariate information. 

Although the main assumption leading to the asymptotic results in \cite{Balkema1974} and \citet{Pickands1975} theorem, is that the observations are independent and identically distributed, in practice, many real life processes yield data that may be related to some covariate variables. This occurs in environmental, financial and insurance datasets, among others. As a result, the parameters of the GPD and by extension other extreme events such as quantiles and small exceedance probabilities may be related to some covariate variables. In that case, the usual asymptotic assumptions may not hold. \citet{Davison1990} and \citet{Beirlant2004} argue that making use of regression techniques aimed at the tails of the distribution function enables the inclusion of covariate information for better point estimates and improved inference. The main idea of this approach is to model one or more of the parameters of the underlying distribution function, $F,$ as a function of the covariates \citep{Davison1990}. However, the issue of threshold selection gets complicated as the threshold may depend on the covariates. A a result, in \citet{Beirlant2005}, a constant and a covariate dependent (based on the \citet{Koenker1978}), thresholds were used in some of the examples. The latter was to incorporate the effect of the covariates in setting the threshold and hence the exceedances. Strong arguments have been made against the former as the observation(s) that is(are) considered extreme at some covariate level may not qualify for an extreme case at another covariate level \citep{Northrop2011}. 

\begin{figure}
	\centering
	\subfloat[]{%
		\includegraphics[height=6cm,width=.48\textwidth]{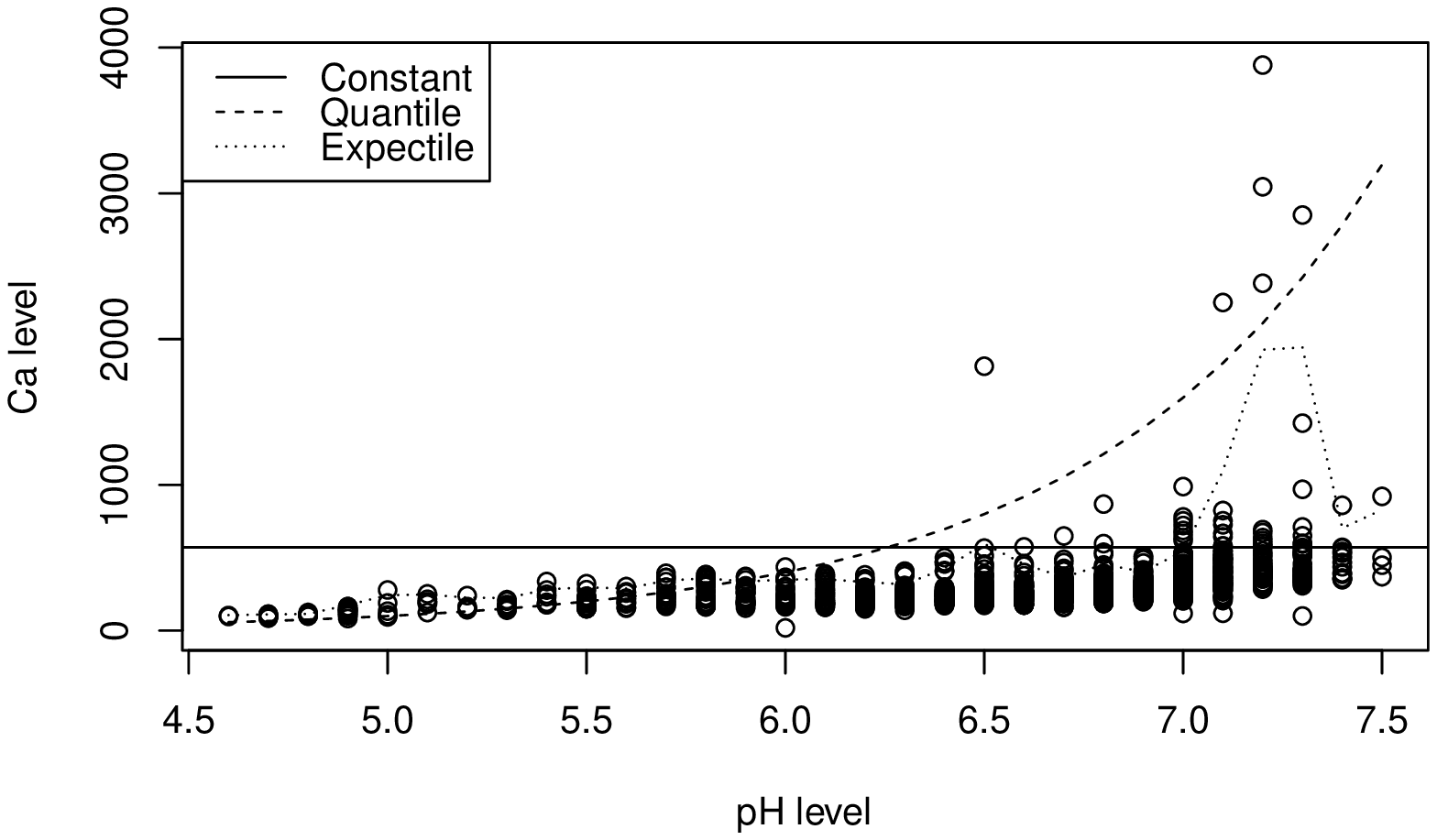}}\hfill
	\subfloat[]{%
		\includegraphics[height=6cm,width=.48\textwidth]{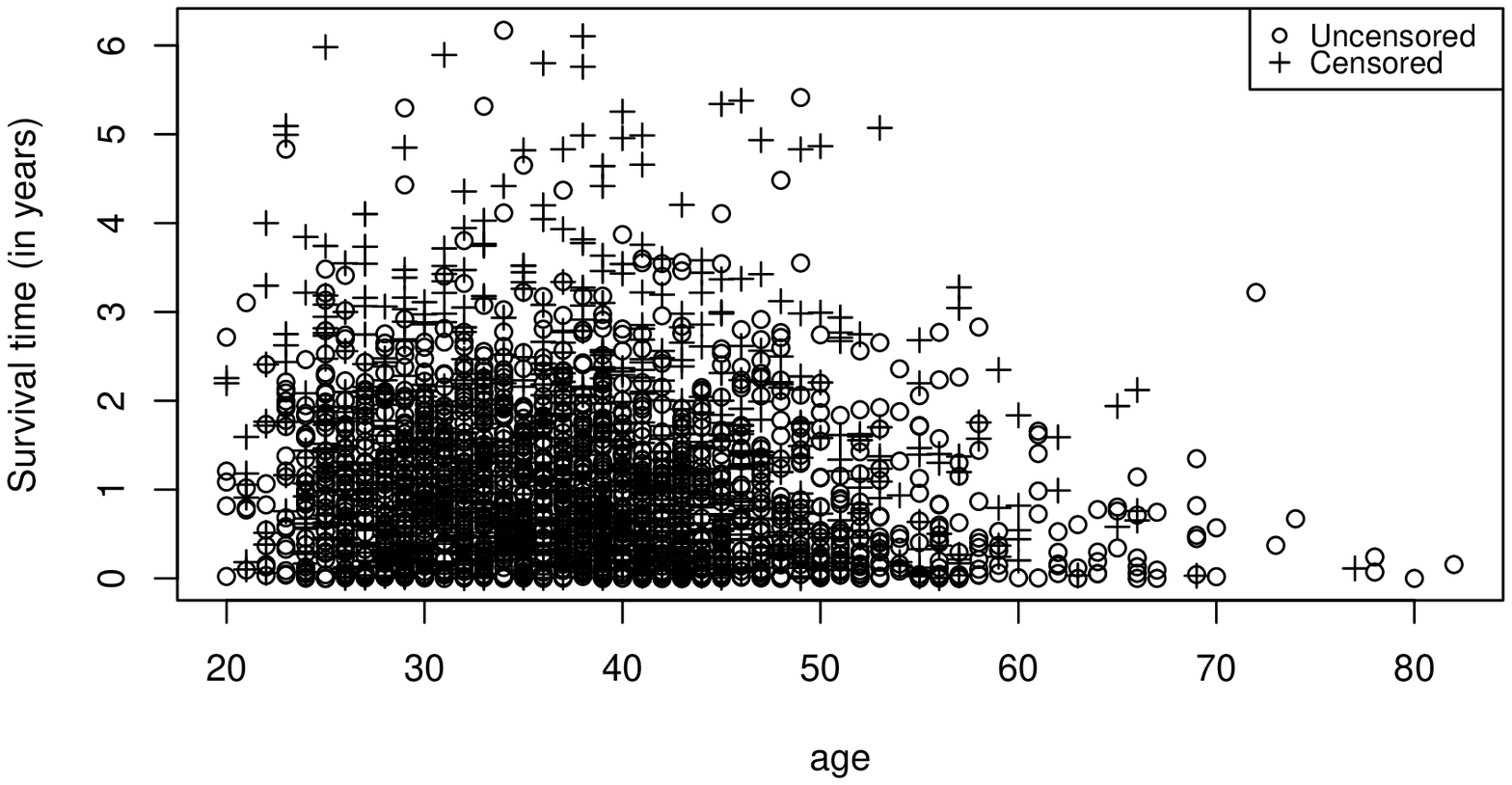}}\\
	
	\caption{Left: Condroz data with superimposed threshold types;
	Right: Australian HIV/AIDS Survival Data}
	\label{example}
\end{figure}

So for example, in the condroz data \citep{Beirlant2005}, as shown in Figure \ref{example}, the Ca level for pH level of 5 is quite different from \emph{pH} level of 7. Thus, setting a constant threshold classify only \emph{Ca} levels with \emph{pH} levels from 6 and above as exceedances for the estimation of the parameters of the GP distribution. As a result, the effect of small \emph{pH} levels i.e. $pH<6,$ will not be included in the estimation of the parameters and hence does not reflect in the estimation of extreme events such as high quantiles and exceedance probabilities.

Also, in the Australian AIDS survival data considered in \citep{Ndao2014}, the survival of patients aged 35 years is quite different from that of age 75 years. Therefore setting a constant threshold may classify most survival times at diagnosis of people aged 75 years and above as non-extreme observations as their survival times are shorter compared to ages around 35 years.

In these cases (and many others), it becomes reasonable to set a threshold that ensures exceedances spread across the covariate space or the exceedance probabilities within a covariate window remain (near) constant. In light of this, some authors have attempted to do so in different ways. \citet{Smith1989} splits data into different seasons and set different thresholds and \citet{Coles2001} expresses the threshold as a smooth function of time. However, in the latter case, the selection of a cyclical threshold for which the rate of exceedance is constant was done by trial-and-error. A more reasonable approach is presented in \citet{Beirlant2005} and \citet{Northrop2011}, is the quantile regression approach of \citet{Koenker1978} is used to set a threshold spanning across the covariate space. Despite these attempts, questions remain as to the performance of this covariate dependent threshold compared to the constant threshold. In this paper, we compare the performance of these two threshold selection methods in addition to a new proposal  based on expectiles \citep{Newey1987}. 

The use of expectiles in the extreme value literature is limited, to the best of our knowledge, only the recent paper by \citet{Daouia2018} where the authors use extreme expectiles to estimate extreme tail risk. Our approach is different from  \citet{Daouia2018} as we use expectiles to obtain exceedances for the estimation of the parameters of the GPD. The expected advantages of the expectiles over quantiles may depend on sensitivity to very large or small observations. I addition, whereas quantiles depend on the frequency of tail realisations of the respondent variable, expectiles depend on both tail realisations and their frequency \citep[see e.g.][]{Kuan2009,Newey1987}. Therefore, thresholds based on expectiles are expected to provide exceedances that span across the covariate space than quantiles. Finally, we compare the performance of these covariate dependent thresholds with the constant threshold in a simulation study.

The rest of the paper is organised as follows. In section \ref{P1}, we present the GP distribution when covariate information is present. Then in Section \ref{P2}, we compare the threshold methods using a simulation study. A practical application on an insurance dataset is presented in Section \ref{P3}. Lastly, we conclude the paper is Section  \ref{P4}.

\section{Framework and Threshold Selection Methods} \label{P1}

Consider a random variable  $Y$ with unknown distribution function $F,$ where $F$ belongs to the Fr\'{e}chet domain of attraction (i.e. $\gamma>0$). If we let $u$ be a sufficiently high threshold, then from the Pickands-Balkema-de Haan theorem, the conditional distribution of $V=Y-u$ given $Y>u$ can be approximated by the Generalised Pareto (GP) distribution. Thus, the excesses $V_j,~ j=1,...,k$ over the threshold, $u,$ is modelled by the GP distribution in (\ref{GPD}).

In the presence of associated covariate vectors $\bm{x}_1,...,\bm{x}_n$, \cite{Davison1990} proposed that the excesses be modelled with a GP distribution taking $\gamma$ and/or $\sigma$ as a function of the covariate vectors and a vector of regression coefficients. Let $\bm{\beta}_1$ and $\bm{\beta}_2$ be the vector of regression coefficients of $\gamma$ and $\sigma$ functions respectively. Thus, the representations of the $\gamma$ and the $\sigma$ functions are:

\[\gamma(\bm{x})=\kappa_1\left(\bm{x},\bm{\beta}_1\right)\] and \[\sigma(\bm{x})=\kappa_2\left(\bm{x},\bm{\beta}_2\right)\] where $\kappa_1$ and $\kappa_2$ are completely specified functions. 

In practice, the parameter covariate models for $\gamma(\bm{x})$ and $\sigma(\bm{x})$ are structured from generalised linear models \citep{Beirlant2005,Wang2009}. Some of the popular choices of link functions in this context are log-link and the identity link for the shape and scale parameters respectively \cite[see e.g.][]{Davison1990,Smith1989,Coles2001,Wang2009}. In the regression context, the distribution function (\ref{GPD}) is defined conditional on $\bm{x}.$ We denote the conditional distribution function by $H(v;\bm{x})$. 

The maximum likelihood estimation method can be used to estimate the regression parameter $\bm{\beta} = (\bm{\beta}_1',\bm{\beta}_2').$ In that case, the log-likelihood function is given by

\begin{equation}\label{GPDlik}
\log L(\bm{\beta})=\sum_{i=1}^{n}\log{h\left(v_i;\gamma(\bm{x}),\sigma(\bm{x})\right)}.
\end{equation}
where $h$ is density function,

\begin{equation}\label{GPDden}
h\left(v_i;\gamma(\bm{x}),\sigma(\bm{x})\right)=\left\{ 
\begin{array}{cc}
\frac{1}{\sigma(\bm{x})}\left( 1+\frac{\gamma(\bm{x})}{\sigma(\bm{x})}v\right)^{-(1/\gamma+1)}, & 1+\frac{\gamma(\bm{x})}{\sigma(\bm{x})}v>0,~\gamma(\bm{x})\neq 0, \\

\frac{1}{\sigma(\bm{x})}\exp{(-\frac{v}{\sigma(\bm{x})})}, & v>0,~ \gamma(\bm{x})=0,\\
\end{array}
\right.,
\end{equation}
of the conditional GP distribution. Maximisation of the log-likelihood function (\ref{GPDlik}) with respect to $\bm{\beta}$ results in the maximum likelihood estimator, $\hat{\bm{\beta}}.$ With $\hat{\bm{\beta}}$ and the completely specified functions, $\kappa_1$ and $\kappa_2,$ the maximum likelihood estimators of $\gamma(\bm{x})$ and $\sigma(\bm{x})$ can be obtained.

In Section \ref{Int}, we illustrated the difficulties in the threshold selection: a delicate balance between bias and variance. In the regression case, the problem is compounded as the threshold may depend on the covariate vector.



\cite{Beirlant2005} provides a procedure on the basis of the quantile regression approach of \cite{Koenker1978} for obtaining a covariate dependent threshold. Let $Q^q(p;\bm{x})$ be the $p$-th $(0<p<1)$ conditional quantile function associated with the distribution function $H(v;\bm{x})$.
Also, assume that the regression $Q^q(p;\bm{x})$ can be modelled by a completely specified function $u(\bm{x};\bm{\theta}_p).$ The $p$-th quantile regression estimator, $\hat{\bm{\theta}}^q_p,$ minimizes  the objective function

\begin{equation}\label{Koenker1}
Q^q(p;\bm{\theta}_p)=\sum_{i=1}^{n}r^q_p \left(V_i-u(\bm{x}_i;\bm{\theta}_p)\right)
\end{equation}
over $\bm{\theta}_p.$ Here, $r^q_p$ is a convex function of the form
\begin{equation}\label{RQ_fun}
r^q_p(\tau)= |\tau|\times|p-\mathbf{1}\left(\tau<0\right)|
\end{equation}
with $\mathbf{1}\left(.\right)$ the indicator function. 

\citet{Beirlant2005} proposed setting a covariate dependent threshold, $u^q_{\bm{x}},$ as a particular regression quantile, $u^q_{\bm{x}}=u(\bm{x};\bm{\theta}_p).$ The estimated conditional quantile function is then used to compute the exceedances for the estimation of the parameters of the GP distribution.

Following a similar analogy, we propose a covariate dependent threshold based on an asymmetric least squares function  \citep{Newey1987}, instead of (\ref{RQ_fun}), of the form

\begin{equation}\label{EX_fun}
r^e_p(\tau)= \tau^2\times|p-\mathbf{1}\left(\tau<0\right)|.
\end{equation}
The $p$th expectile (a class of asymmetric least squares) estimators,  $\hat{\bm{\theta}}^e_p,$ are obtained by minimising 

\begin{equation}\label{Expectile}
Q_e(p;\bm{\theta}_p)=\sum_{i=1}^{n}r^e_p \left(V_i-u(\bm{x}_i;\bm{\theta}_p)\right).
\end{equation}
over $\bm{\theta}_p$ with $r^e_p$ given in (\ref{EX_fun}). 
From this asymmetric least squares estimator, the proposed covariate dependent threshold is set at a particular expectile, $u^e_{\bm{x}}=u(\bm{x};\bm{\theta}_p).$ The exceedances (or excesses) over the expectile threshold are used for fitting the GP distribution. This threshold takes into account the advantages of the asymmetric least least squares outlined in Section \ref{Int}.

\section{Simulation Study}\label{P2}

In this section, we compare the performance of the threshold selection procedures discussed in estimating the conditional tail index through a simulation study.

\subsection{Simulation Design}
In the present simulation study, we generate $R=1000$ samples of size $n$ ($n =1000,~2000,\\~5000$) of independent pairs of observations $(y_i, x_i), i=1, ..., n$, where $x_i$ is uniformly distributed on [0,1]. The conditional distribution function of $Y$ given $x$ is chosen from three different distributions, namely Burr, Pareto and the Fr\'{e}chet  distributions.  

The Burr distribution, introduced by \citet{Burr1942}, is a flexible family of distributions with a range of shapes.  Its limiting form includes the Pareto, log logistic and gamma distributions. It is widely used in most simulation studies in the EVT literature due to its flexibility. Some of the practical applications of the Burr distribution includes modelling household incomes, lifetime, travel time, among others. In addition, the Fr\'{e}chet distribution is a heavy tailed distribution used widely in the EVT literature. Its application includes modelling maximal temperature and interfacial damage in microelectronic packages in Hydrology and engineering respectively \citep{Harlow2002}. 

For each distribution, our interest is to estimate the conditional tail index function, 

\begin{equation}\label{gam.x}
\gamma(x) = \exp(-0.05-2x).
\end{equation}
The motivation for the function of the form, (\ref{gam.x}), is that it generates  data sets of the form as shown in Figure  \ref{example}(a). Specifically, we assess the estimators of $\gamma(x)$ at three $x$-values, 0.32, 0.57 and 0.99, with corresponding parameter values of $\gamma(x)$ respectively 0.50, 0.30 and 0.13. These  values of $\gamma(x)$ are considered as large, medium and small conditional tail indices respectively.  

At each value of $x,$ we estimate the parameter  $\gamma(x)$ using the constant threshold and the two covariate dependent thresholds based on expectile and quantile regressions. Whereas the constant threshold is obtained by using one of the order statistics, the quantile regression and the expectile based covariates are implemented in R using the packages \emph{quantreg} and \emph{expectreg} respectively. In order to properly evaluate the threshold selection methods, we compute the Median Absolute Deviation (MAD) and Median Bias (hereafter referred to as bias) at different values of the number of exceedances, $k.$ 

Therefore, for the $R$ samples generated, we obtain $\hat{\bm{\gamma}_k}(x)=\left(\hat{\gamma}_{k,1}(x),\ldots, \hat{\gamma}_{k,R}(x)\right)'$ at each $k$ and the MAD and bias are obtained as
 \begin{equation}\label{MAD}
 MAD_k(\hat{\gamma}(x))= \underset{1\le r\le R}{\mbox{median}}\left|\hat{\gamma}_{k,r}(x)-\gamma(x) \right|
 \end{equation}

and

\begin{equation}\label{Bias}
\mbox{Bias}\left(\hat{\gamma}_k(x)\right)=\underset{1\le r\le R}{\mbox{median}}\left(\gamma_{k,r}(x)\right)-\gamma(x)
\end{equation} 
 
respectively.

\subsection{Simulation Results and Discussion}

\begin{figure}[htp!]
	\centering
	
	\subfloat[]{%
		\includegraphics[height=6cm,width=.33\textwidth]{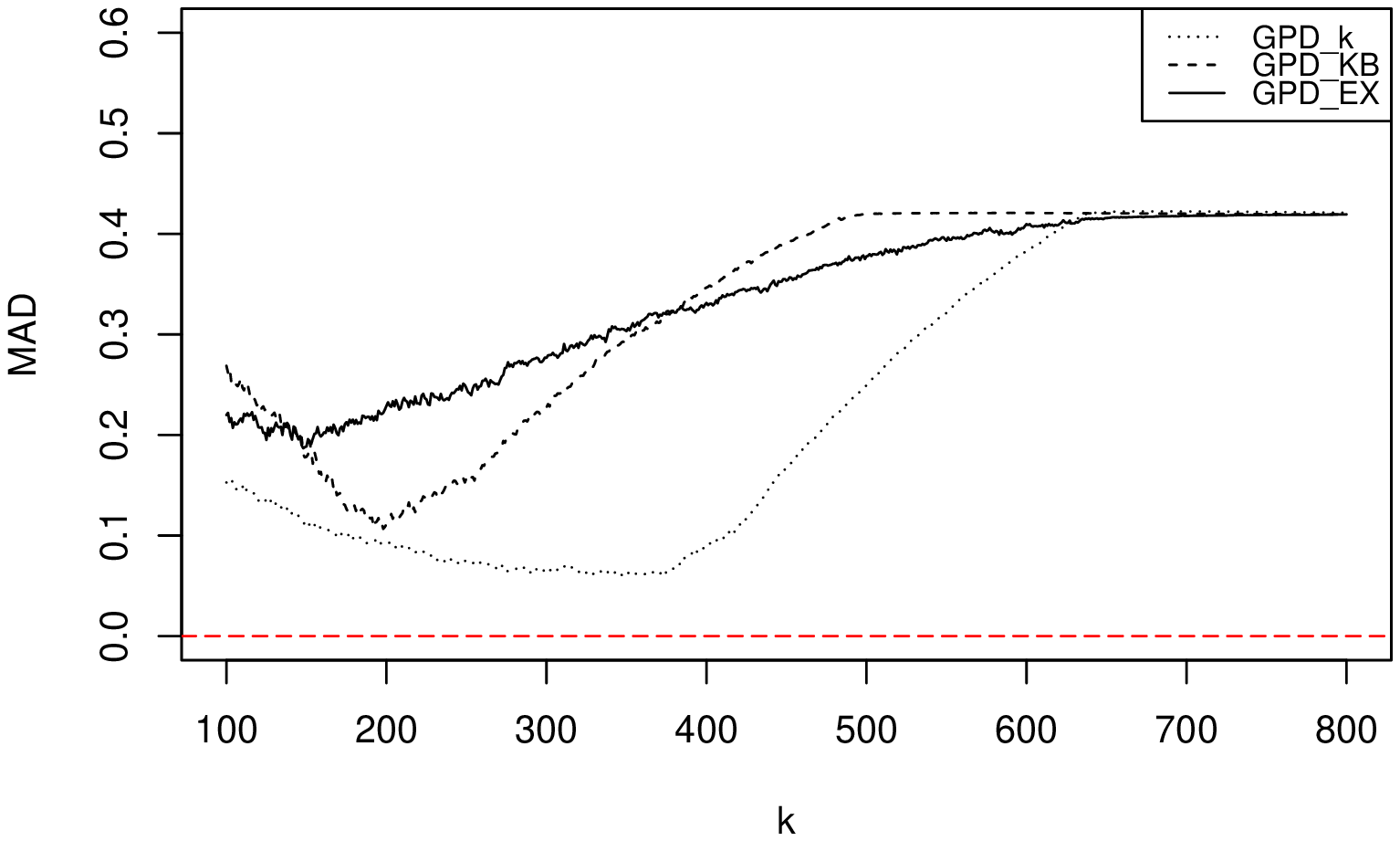}}\hfill
	\subfloat[]{%
		\includegraphics[height=6cm,width=.33\textwidth]{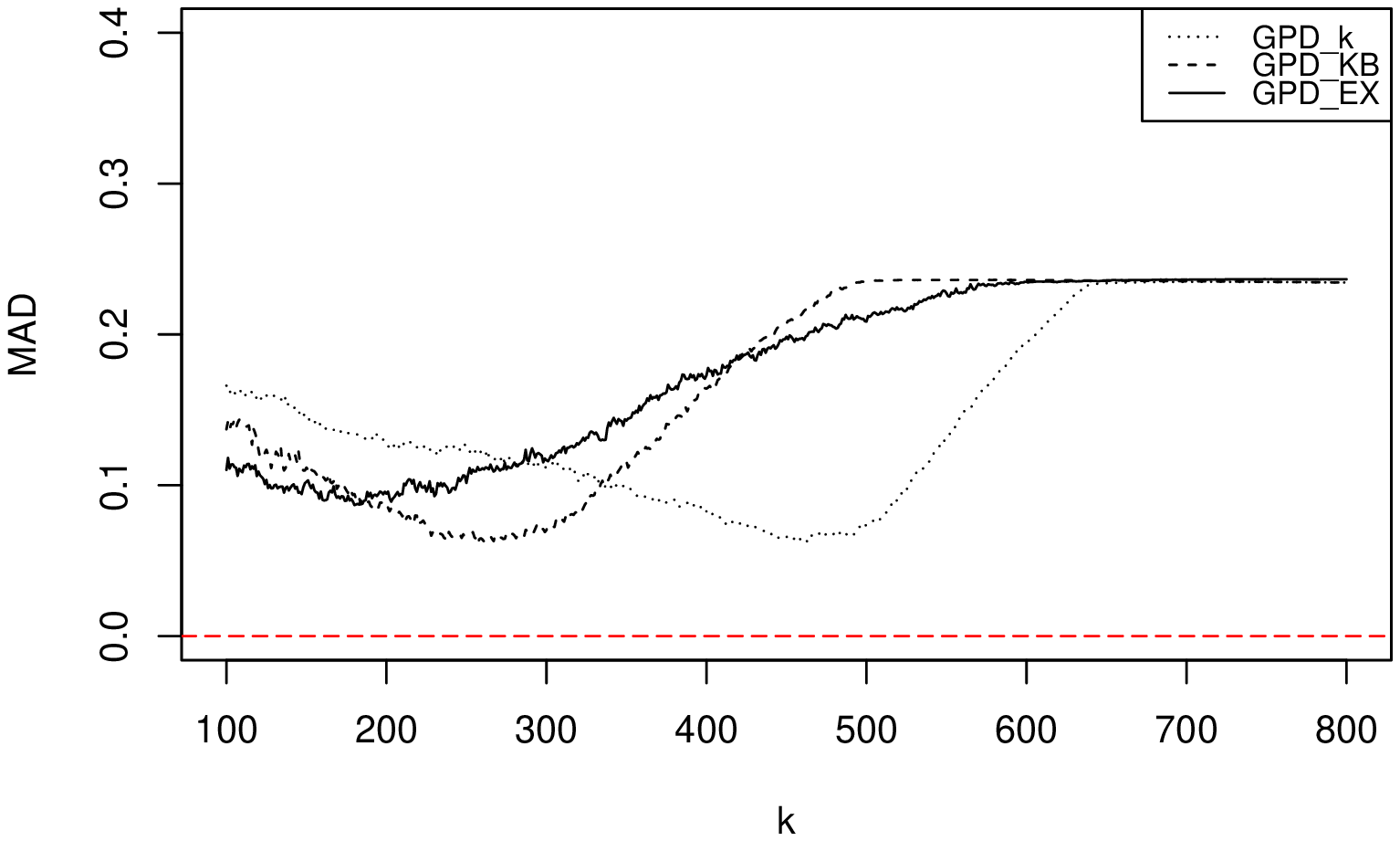}}\hfill
	\subfloat[ ]{%
		\includegraphics[height=6cm,width=.33\textwidth]{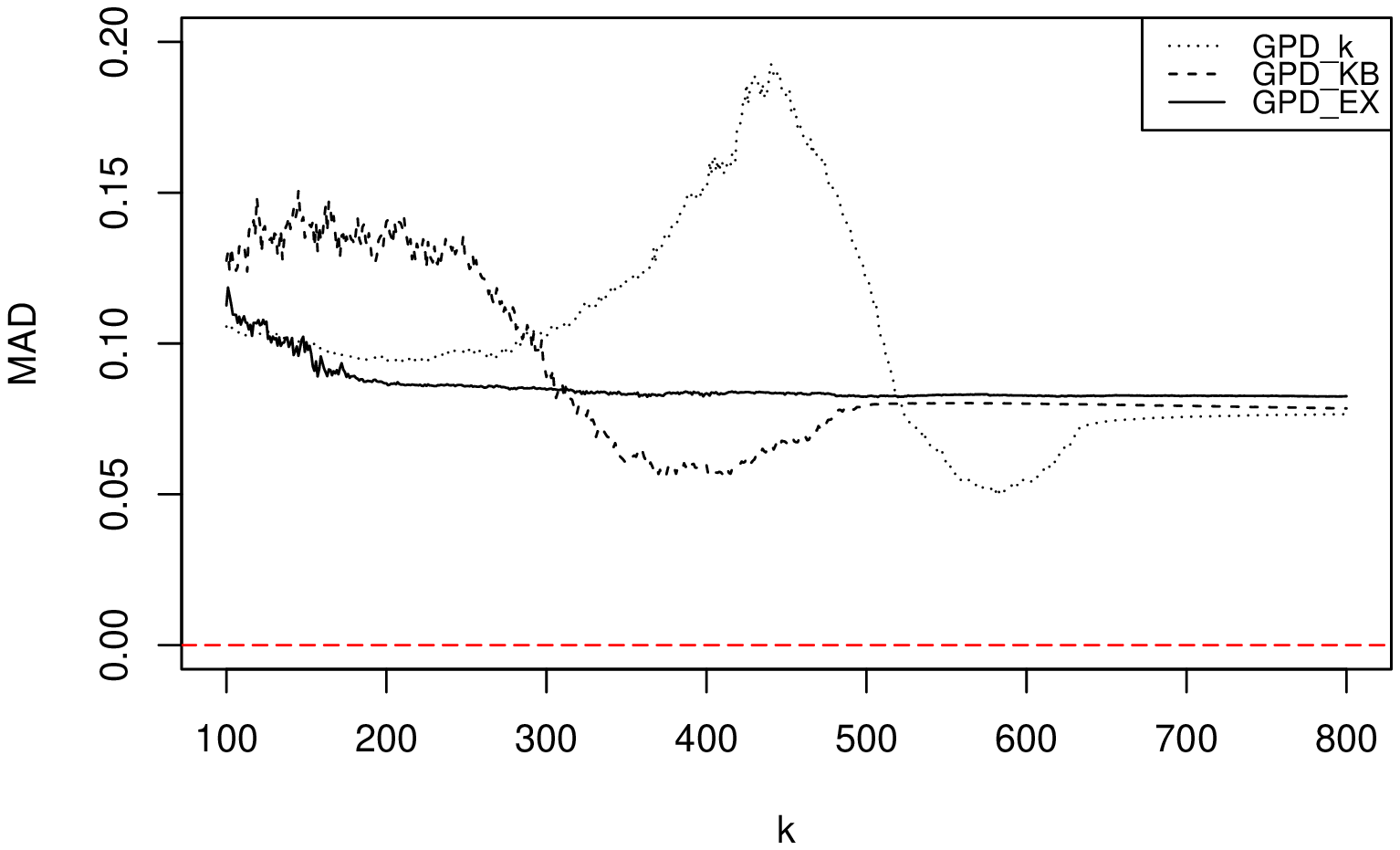}}\\
	\subfloat[]{%
		\includegraphics[height=6cm,width=.33\textwidth]{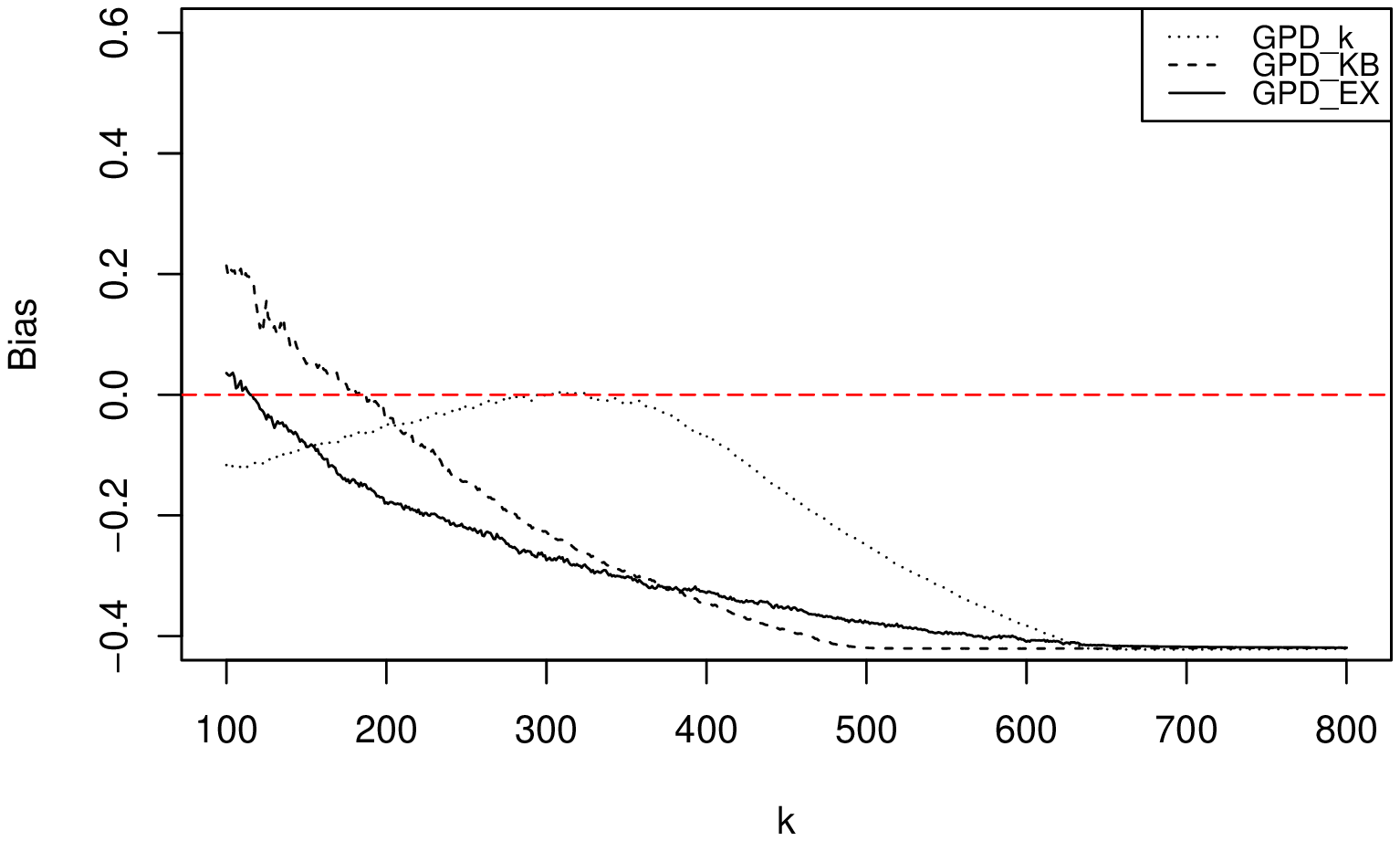}}\hfill
	\subfloat[]{%
		\includegraphics[height=6cm,width=.33\textwidth]{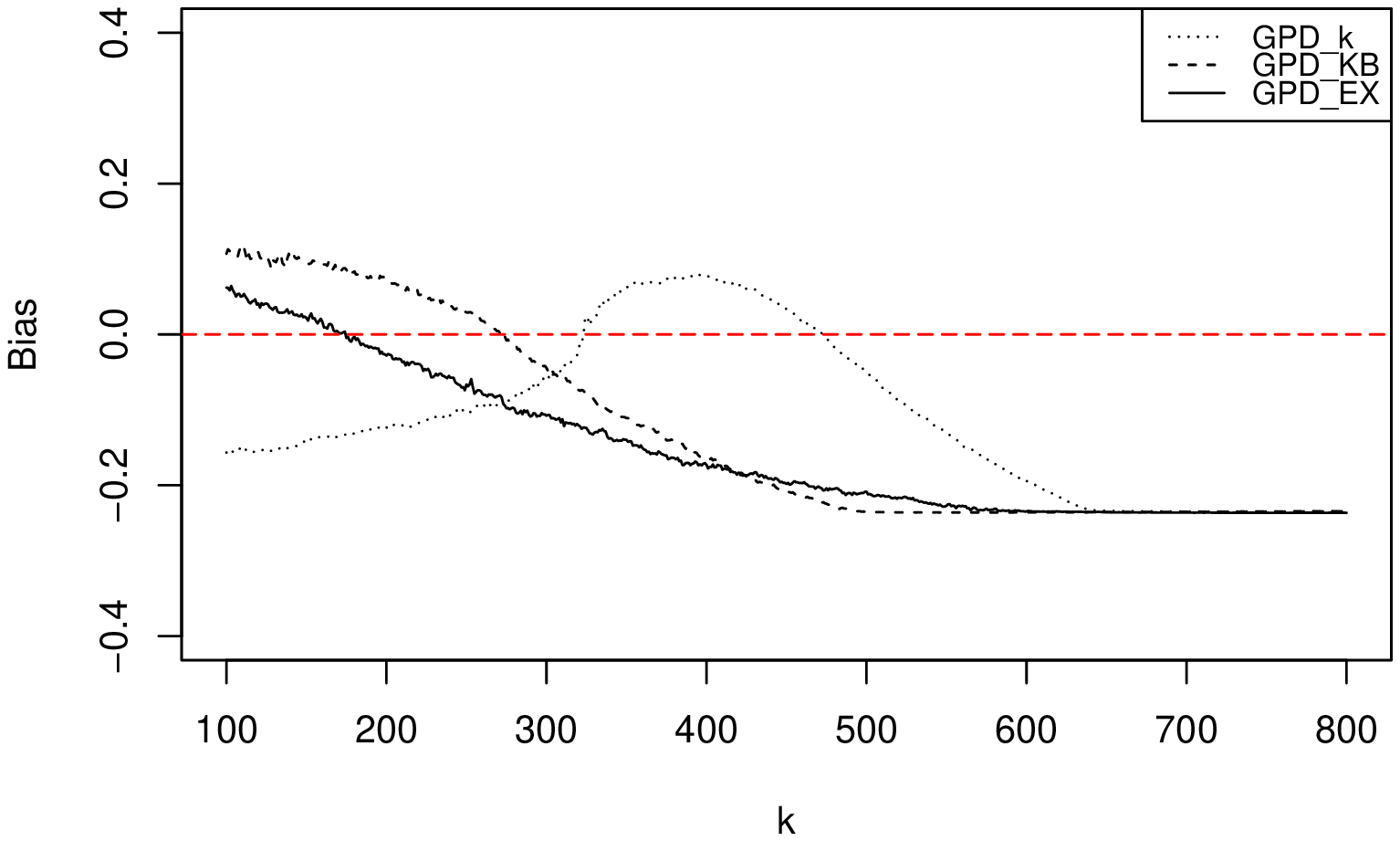}}\hfill
	\subfloat[ ]{%
		\includegraphics[height=6cm,width=.33\textwidth]{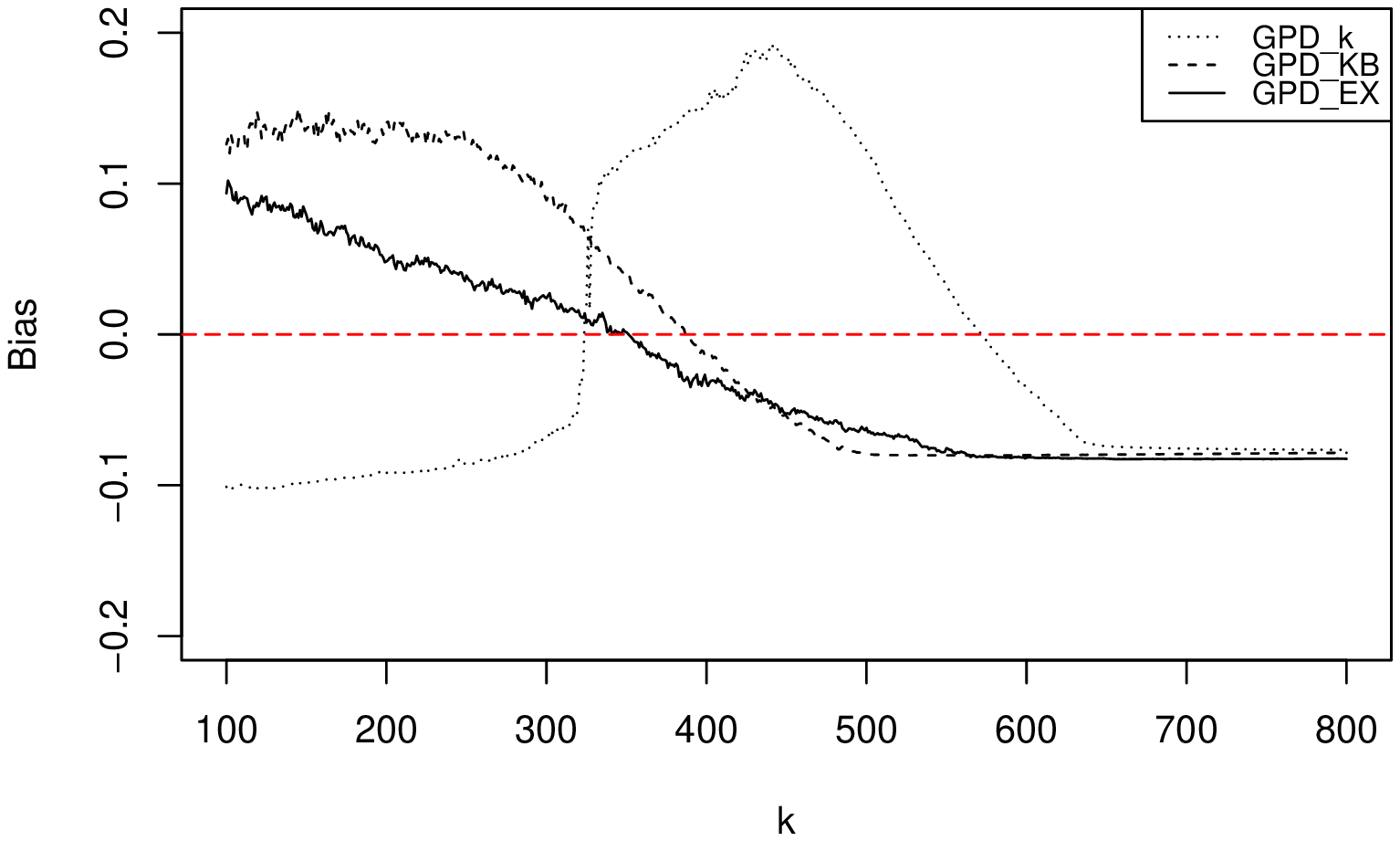}}\\
	\caption{Pareto Distribution with $n=1000:$ Left Column: $\gamma(x)=0.50;$ Middle Column: $\gamma(x)=0.30;$ and Right Column: $\gamma(x)=0.13$}
	\label{Bur1}
\end{figure}

\begin{figure}[htp!]
	\centering
	
	\subfloat[]{%
		\includegraphics[height=6cm,width=.33\textwidth]{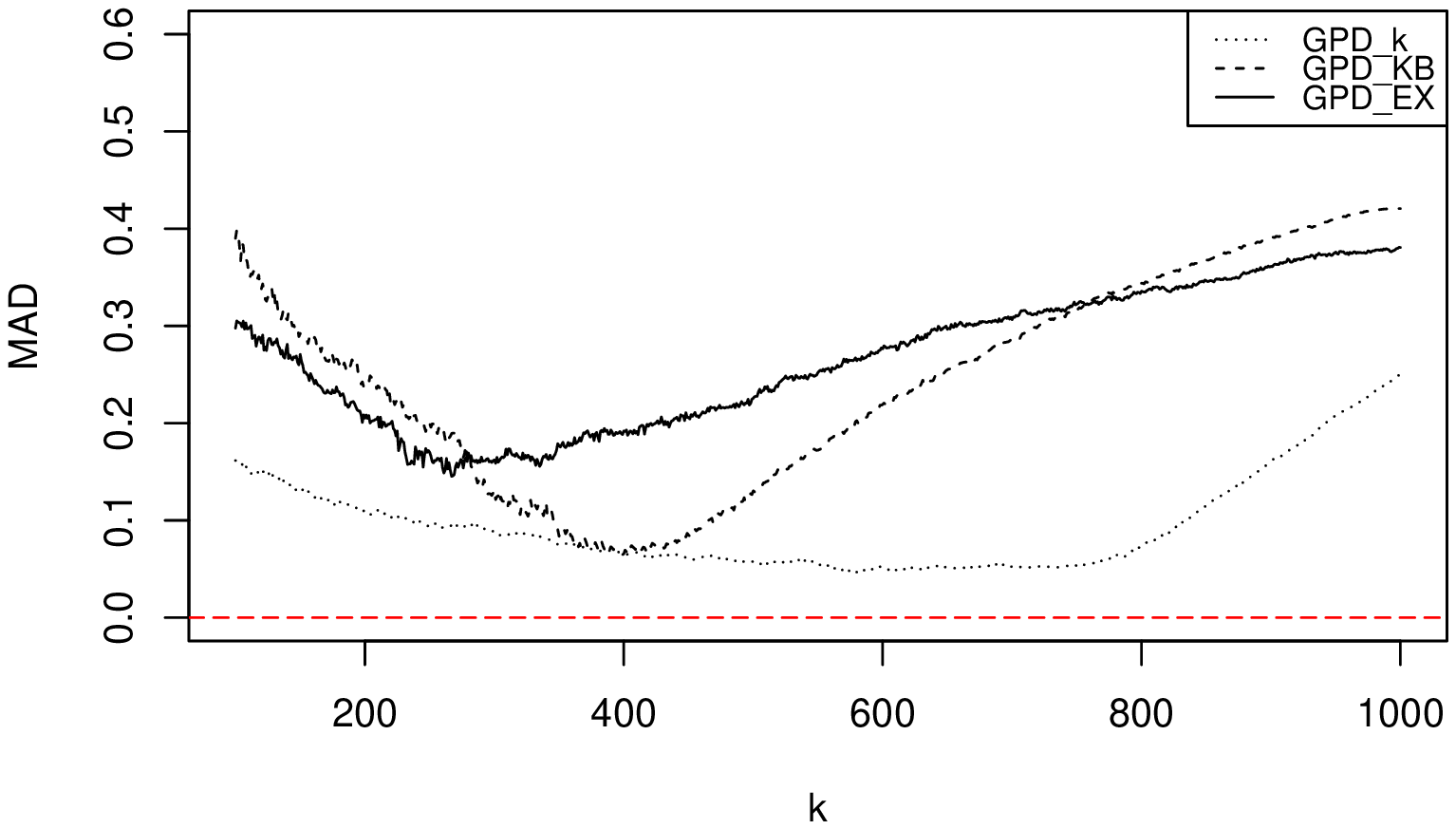}}\hfill
	\subfloat[]{%
		\includegraphics[height=6cm,width=.33\textwidth]{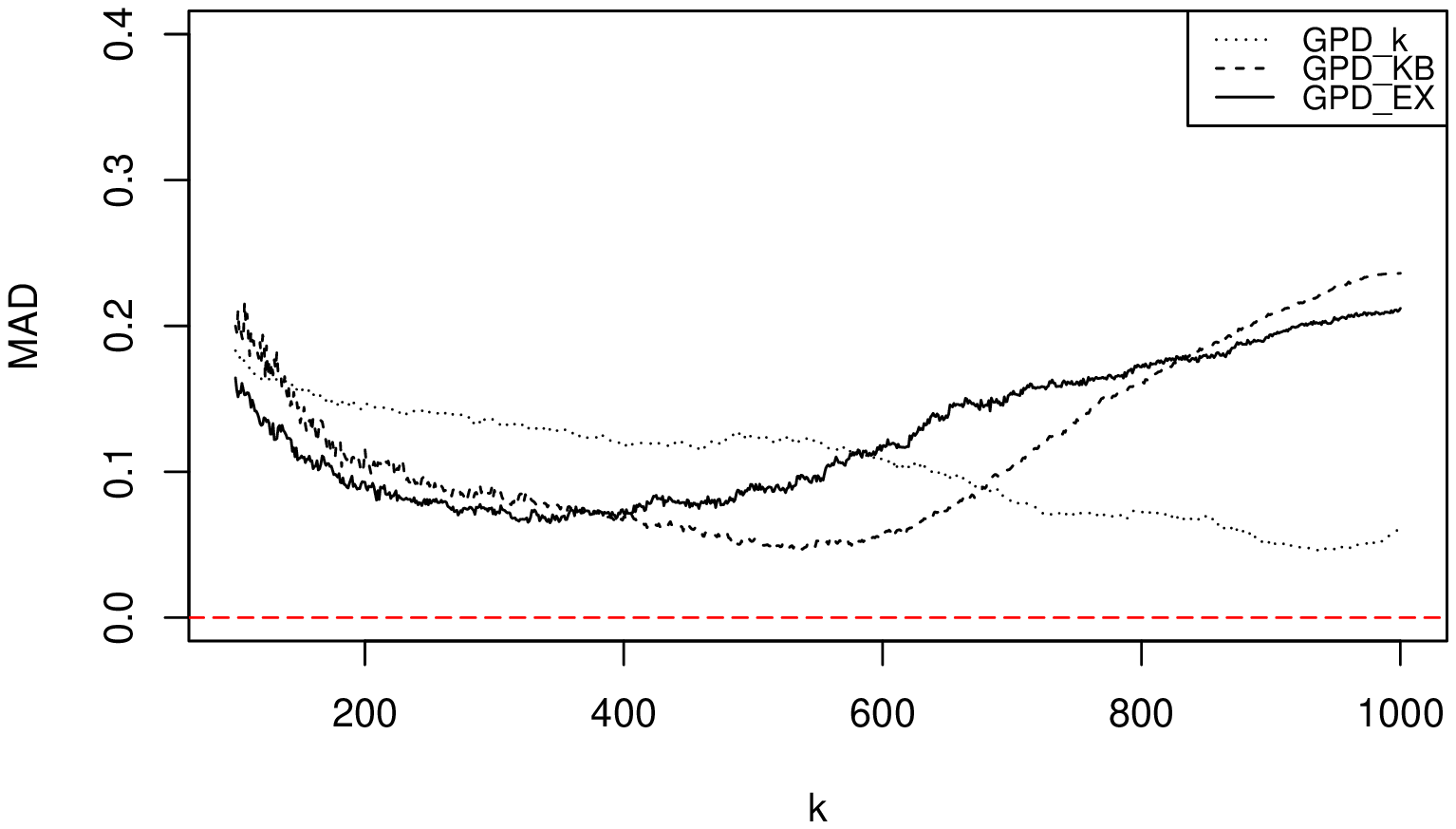}}\hfill
	\subfloat[ ]{%
		\includegraphics[height=6cm,width=.33\textwidth]{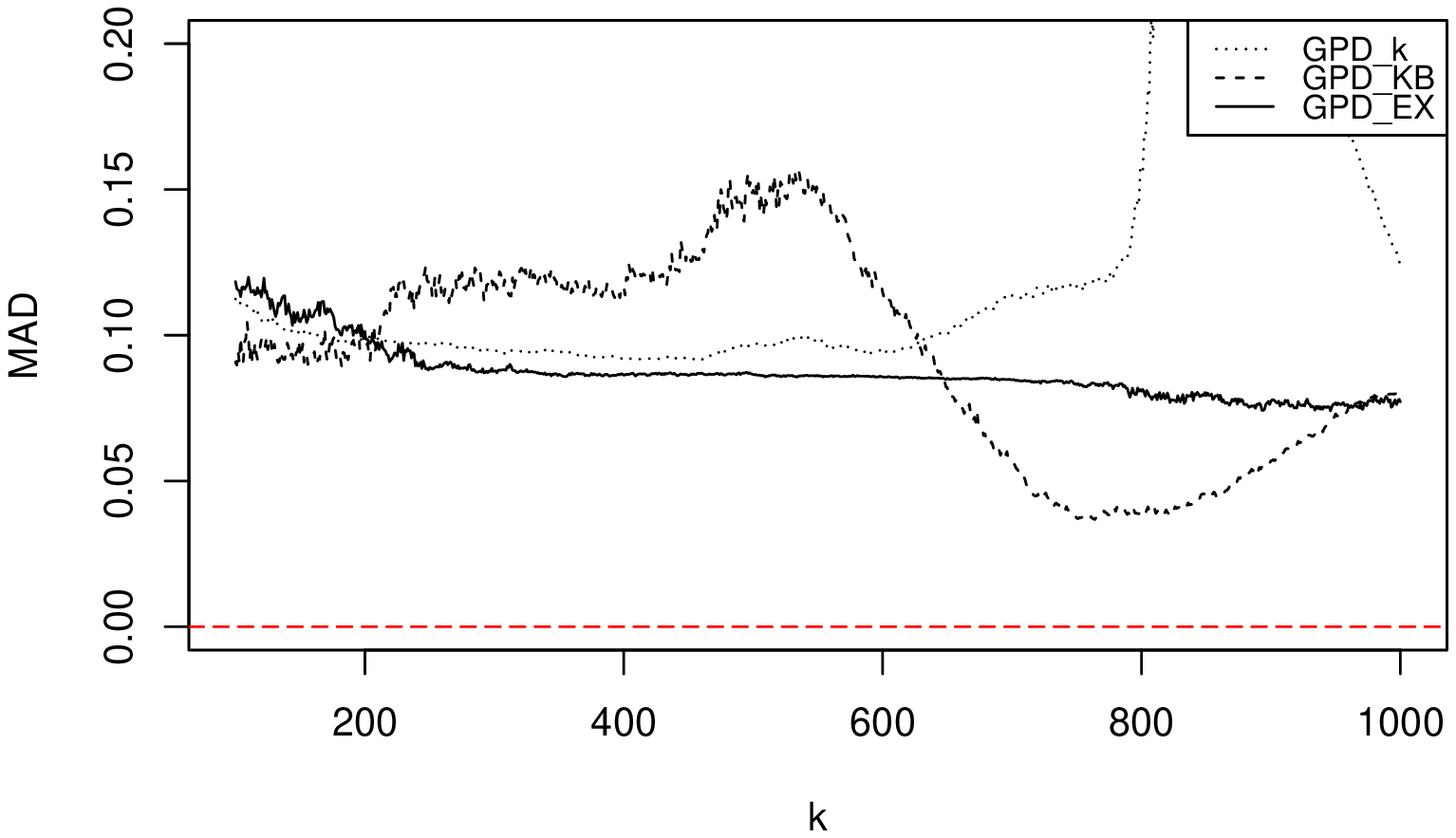}}\\
	\subfloat[]{%
		\includegraphics[height=6cm,width=.33\textwidth]{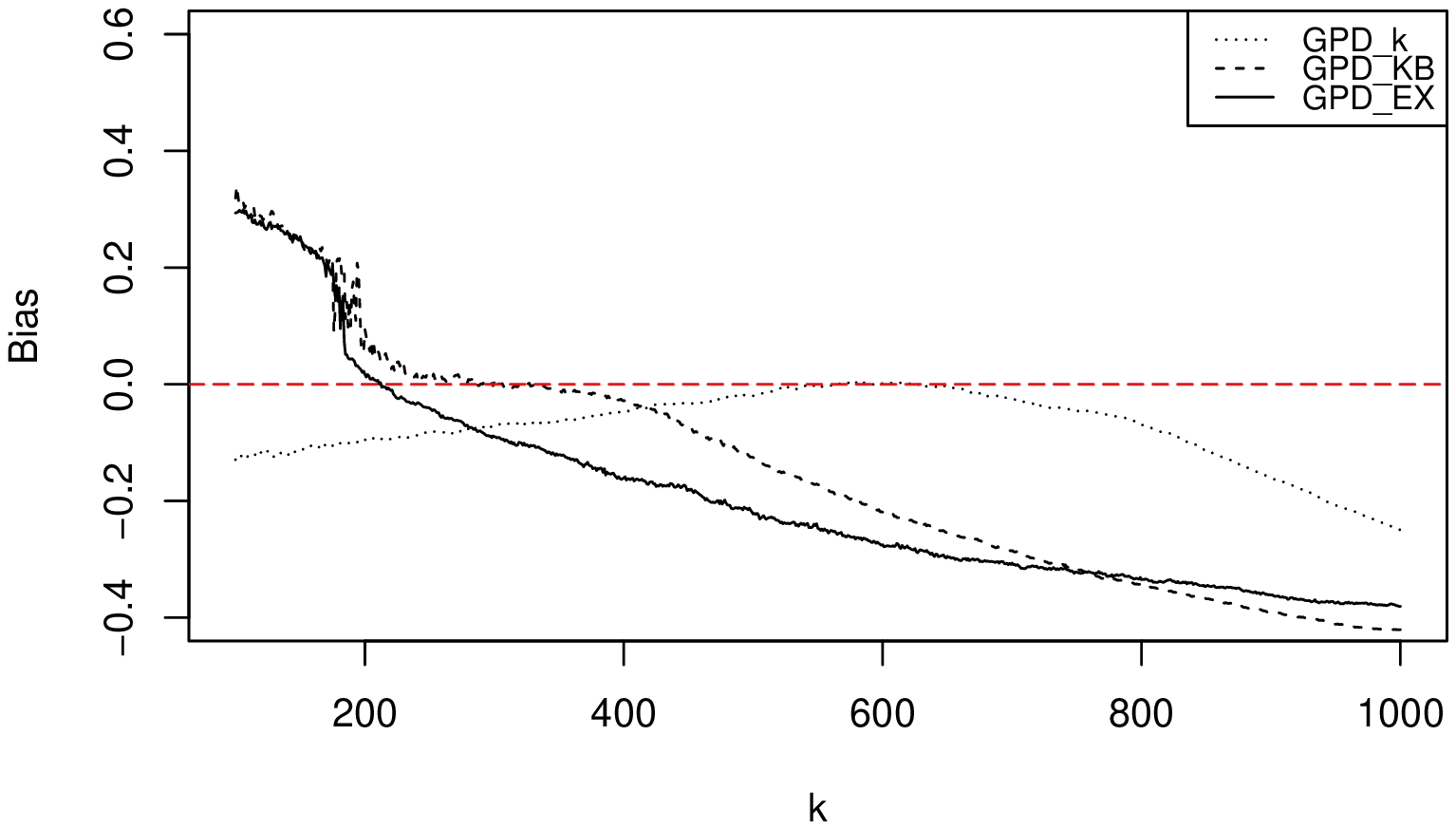}}\hfill
	\subfloat[]{%
		\includegraphics[height=6cm,width=.33\textwidth]{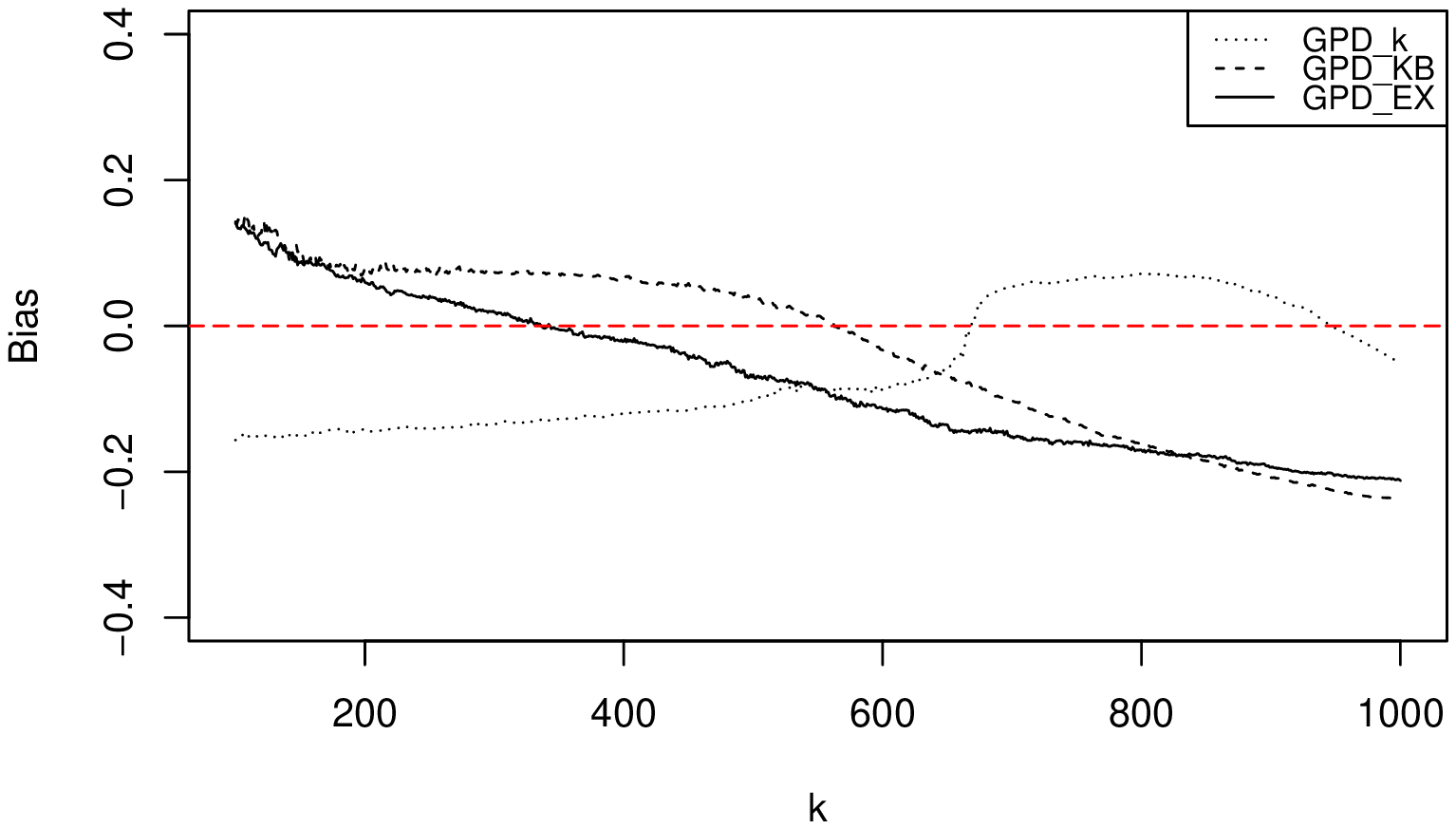}}\hfill
	\subfloat[ ]{%
		\includegraphics[height=6cm,width=.33\textwidth]{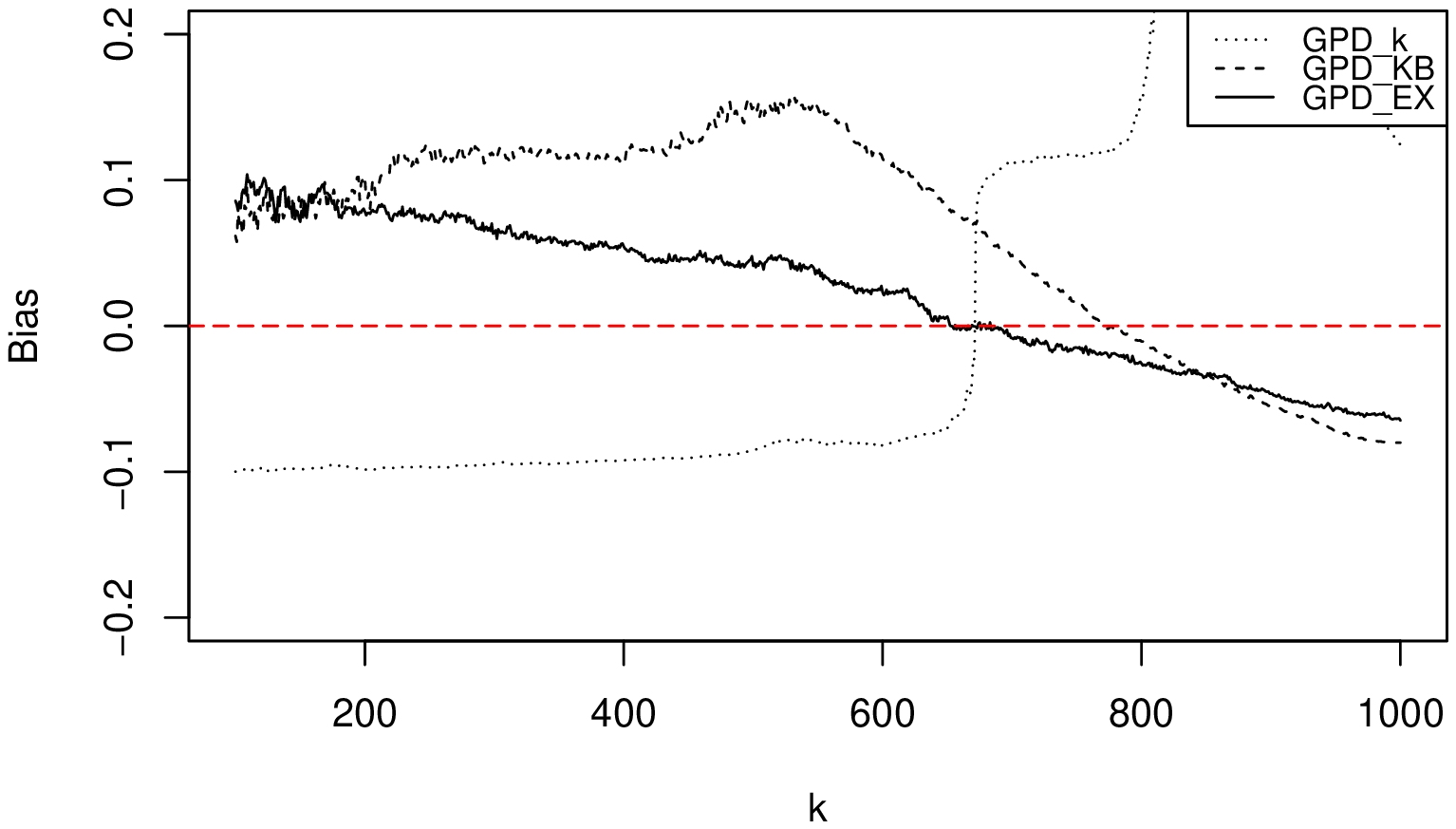}}\\
	\caption{Pareto Distribution with $n=2000:$ Left Column: $\gamma(x)=0.50;$ Middle Column: $\gamma(x)=0.30;$ and Right Column: $\gamma(x)=0.13$}
	\label{Bur2}
\end{figure}

\begin{figure}[htp!]
	\centering
	
	\subfloat[]{%
		\includegraphics[height=6cm,width=.33\textwidth]{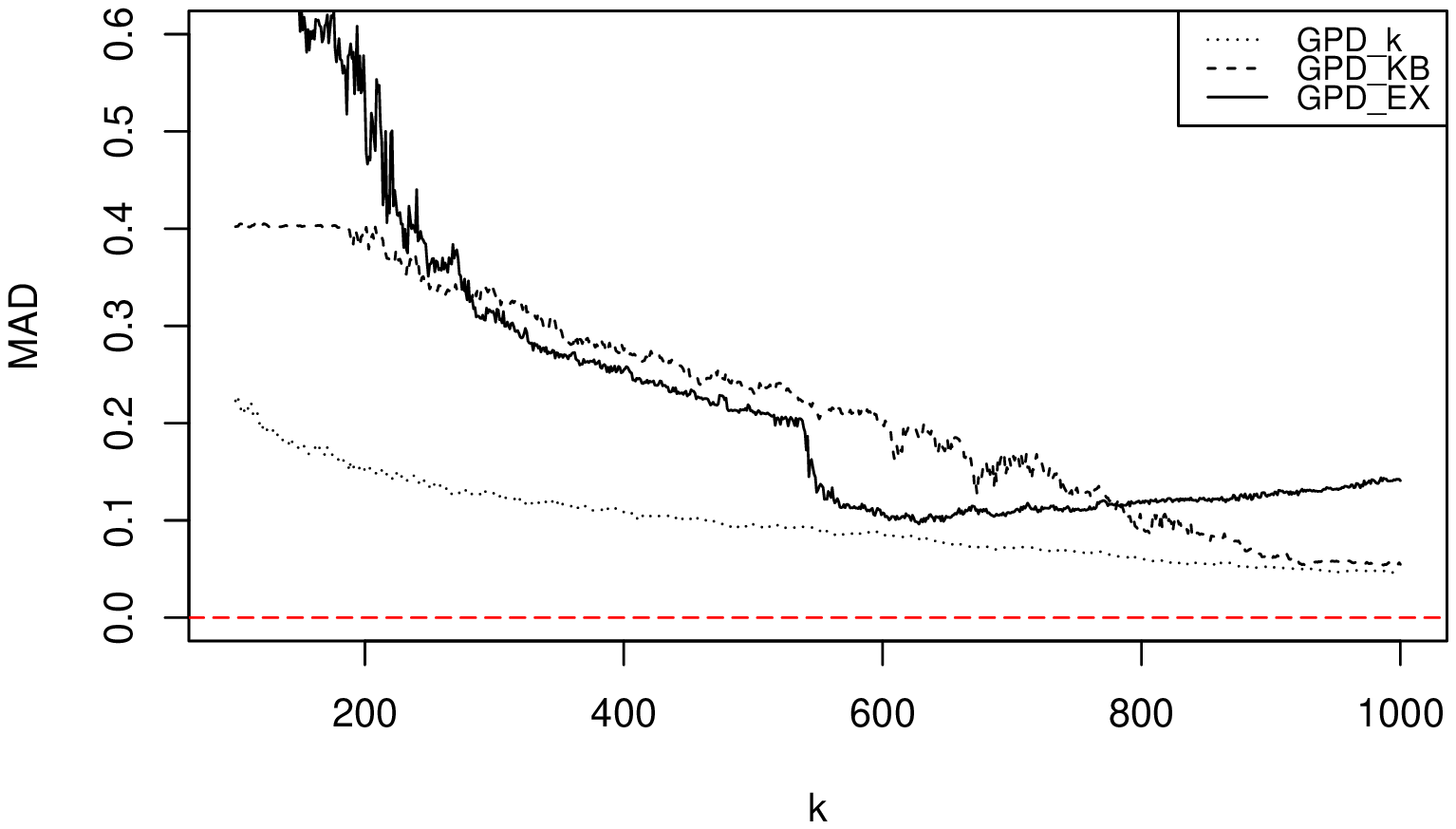}}\hfill
	\subfloat[]{%
		\includegraphics[height=6cm,width=.33\textwidth]{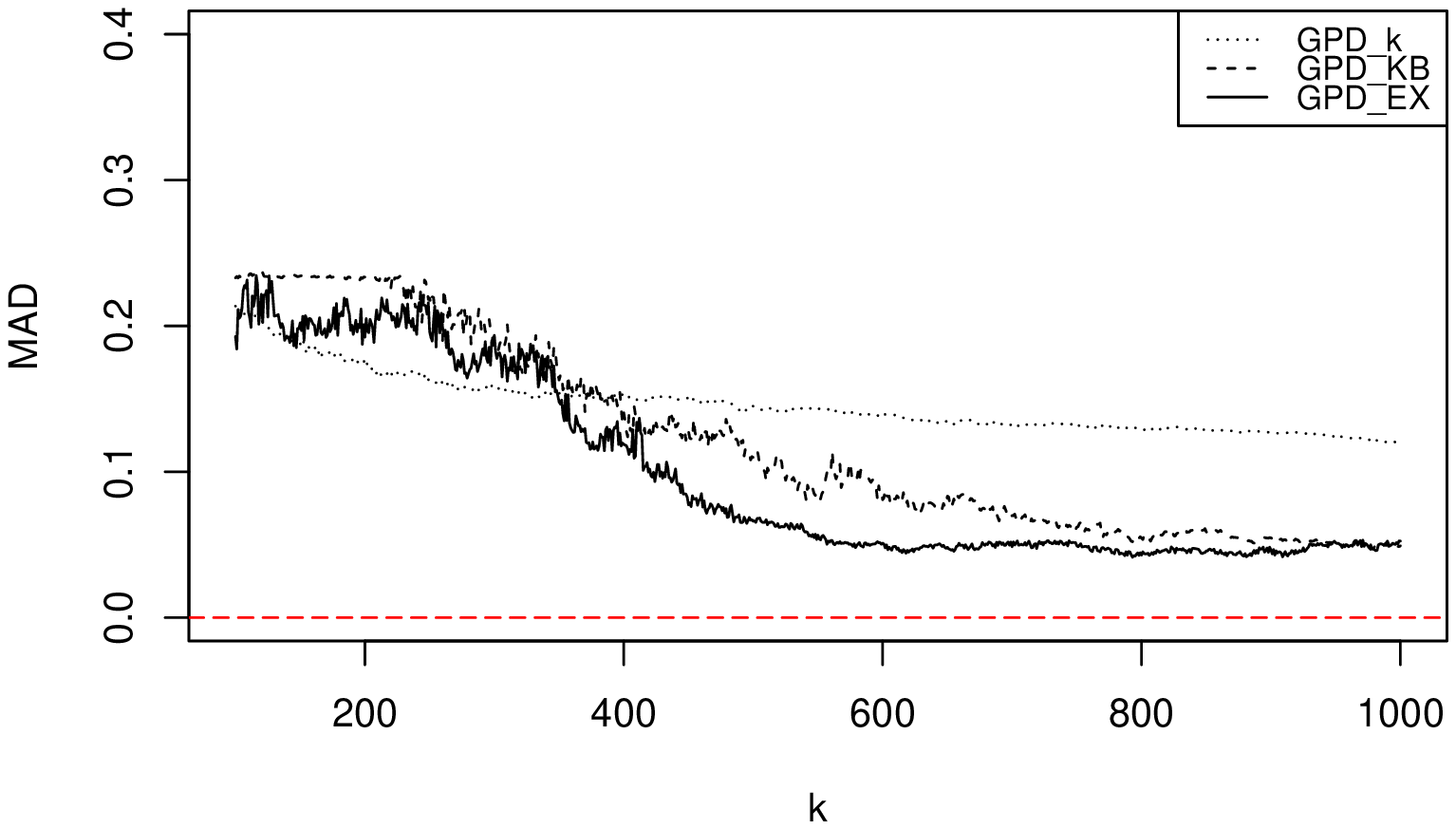}}\hfill
	\subfloat[ ]{%
		\includegraphics[height=6cm,width=.33\textwidth]{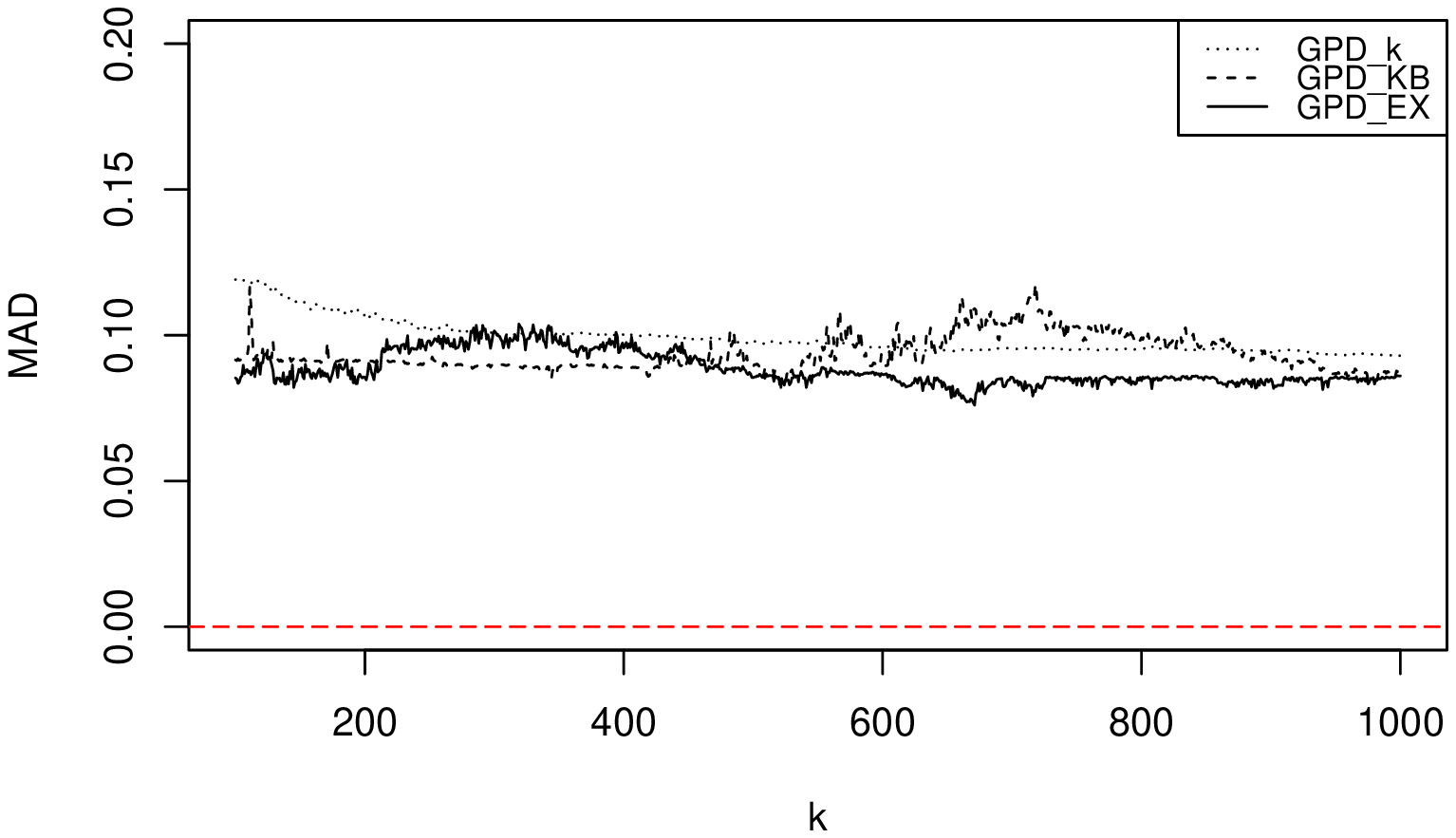}}\\
	\subfloat[]{%
		\includegraphics[height=6cm,width=.33\textwidth]{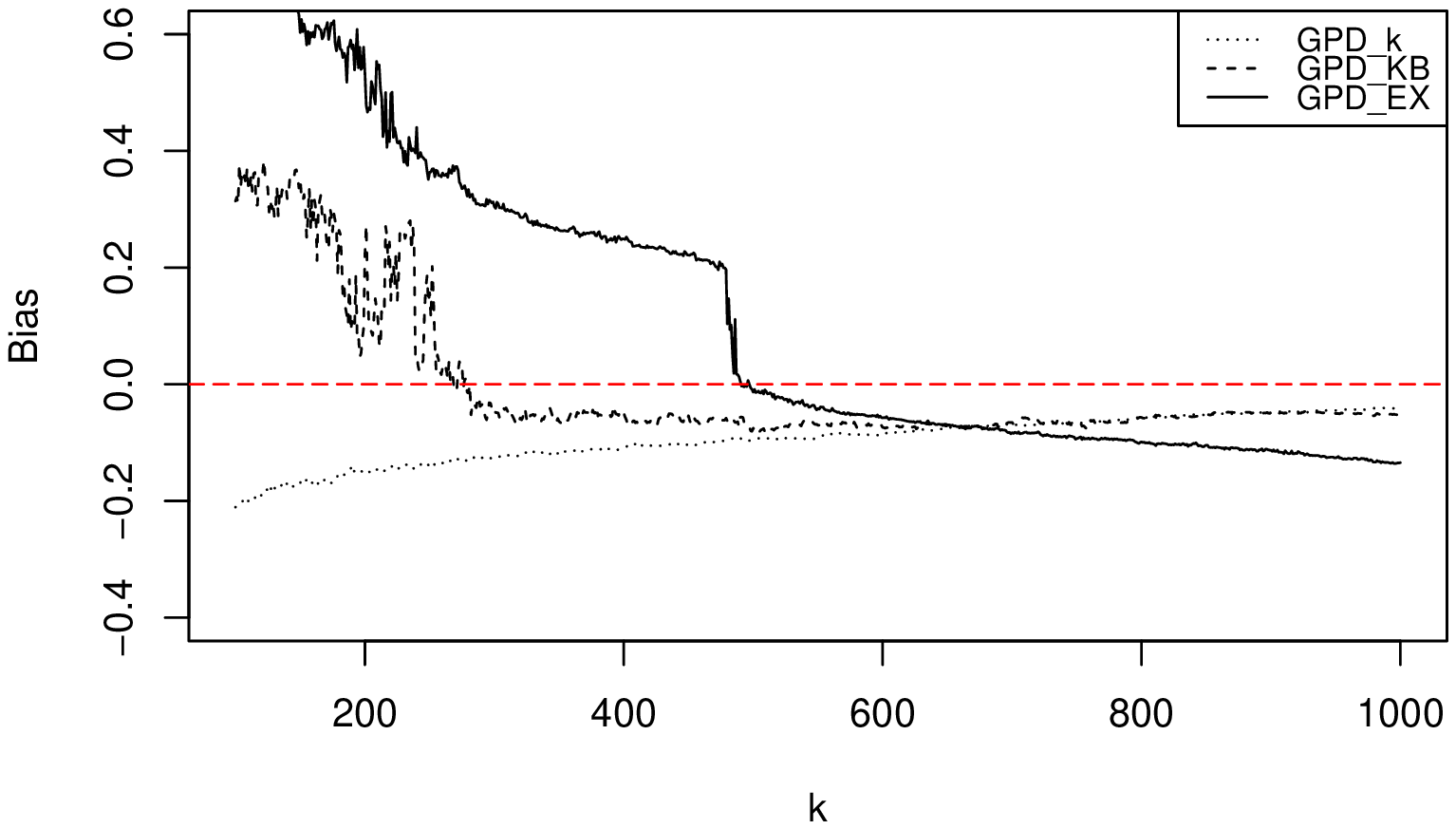}}\hfill
	\subfloat[]{%
		\includegraphics[height=6cm,width=.33\textwidth]{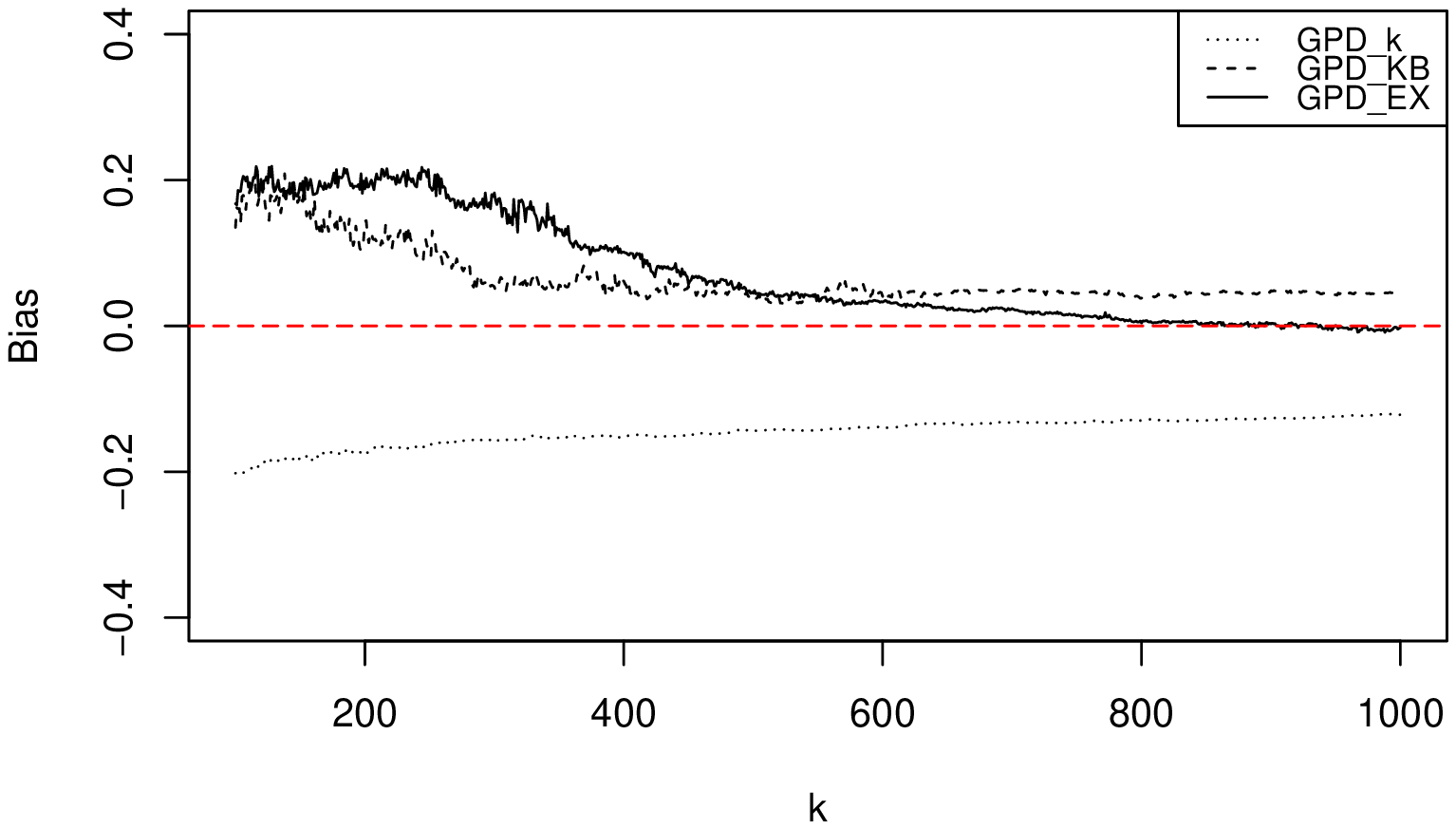}}\hfill
	\subfloat[ ]{%
		\includegraphics[height=6cm,width=.33\textwidth]{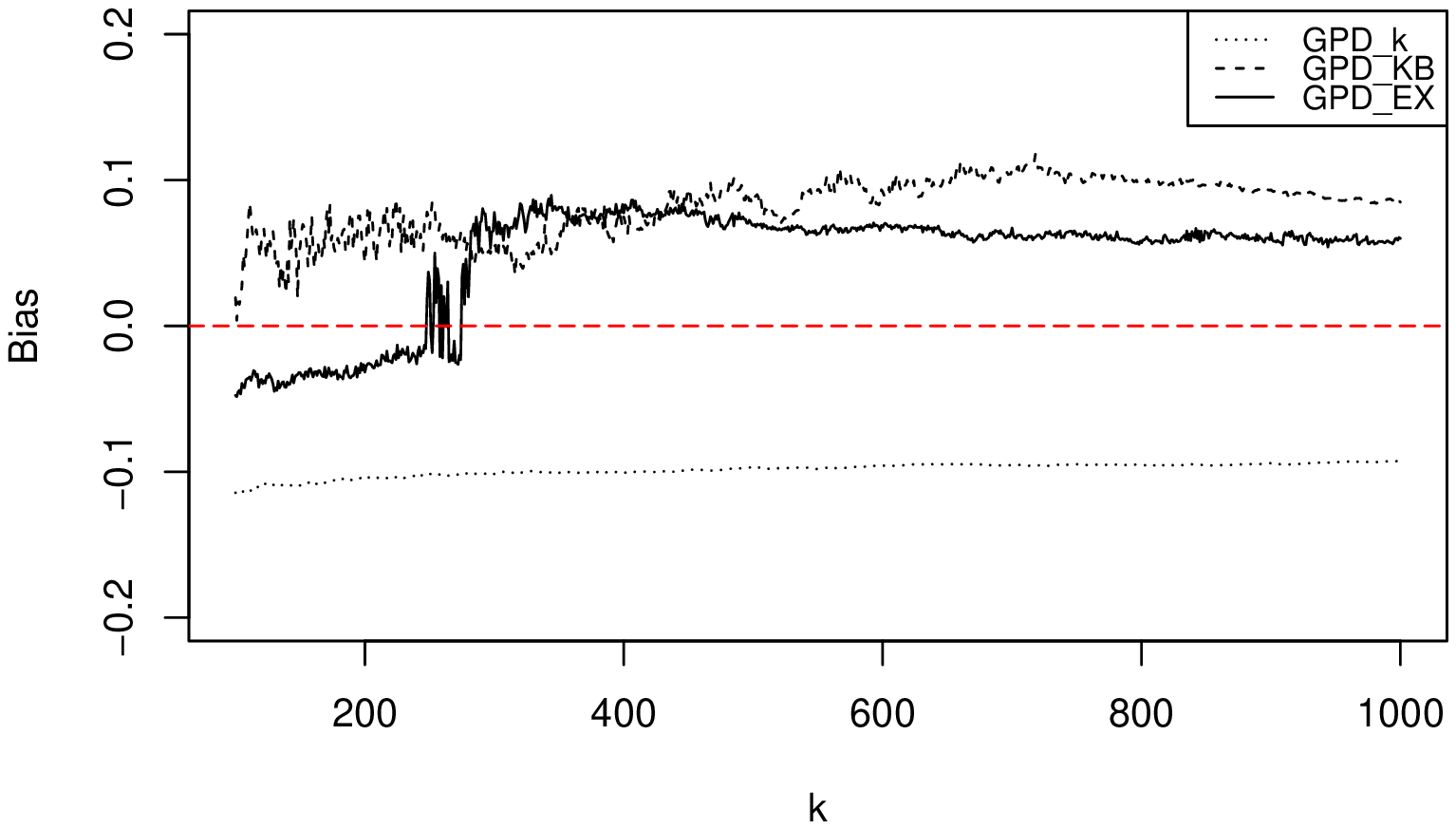}}\\
	\caption{Pareto Distribution with $n=5000:$  Left Column: $\gamma(x)=0.50;$ Middle Column: $\gamma(x)=0.30;$ and Right Column: $\gamma(x)=0.13$}
	\label{Bur3}
\end{figure}

The results of the simulation for the Burr distribution are presented in Figures \ref{Bur1} to \ref{Bur3}. For ease of presentation, we present the results for the Fr\'{e}chet and the Pareto distributions to the Appendix.

Considering the performance of the various threshold selection methods for the estimation of $\gamma(x)=0.50,$ we find that the constant threshold provides better MAD values. In terms of bias, unlike the covariate dependent thresholds, the constant threshold exhibits negative bias implying that it generally produces smaller values compared with the actual $\gamma(x).$ In the case of the covariate dependent thresholds, the result is mixed. The expectile threshold provides better MAD values for small values of $k.$ However, for large values of $k$ (i.e. inclusion of more moderate observations), the quantile regression based threshold has better MAD values. The observations made from the MAD values apply to the bias values as well. 

Also, for $\gamma(x)=0.30,$ overall the constant threshold again provides the smallest values of MAD. This is followed by the expectile threshold for smaller values of $k$ within 15\% of the sample size. Beyond this point, the quantile based threshold has the smallest MAD values. 

In addition, for small values of $\gamma(x),$ the expectile based threshold generally outperforms the constant and quantile regression thresholds: it has smaller MAD values and is stable along the path of $k.$ In terms of bias, the expectile regression threshold generally has values closer to zero than the rest of the thresholds.  

Furthermore, it can be seen from  Figures \ref{Par}-\ref{Par3}, that higher variability is recorded for large values of $\gamma(x)$ (i.e. at smaller values of $x$) and the variability decreases as $\gamma(x)$ decreases from 0.50 to 0.13. Also, in terms of bias, it follows a similar path with smaller bias values and stable results across $k$ values been recorded at $\gamma(x)=0.13.$ In addition, as $k$ increases beyond $50\%$ of the sample size, the values of the performance measures, bias and MAD, each approaches a common value. This occurs because as more observations are included in the exceedances, the difference between the competing thresholds become minimal and hence similar values of bias and MAD for each threshold. This observation is apparent in the  region $k>600$ for the figures with sample sizes $n=1000,$ across the three distributions.

In general, the performance of the constant threshold is best in the cases where there are exceedances over it at the covariate level.  For example, in our case, this occurs at the covariate values $x=0.32$ and $x=0.57.$ However, in the case of $x=0.99,$ most of the observations do not exceed the constant threshold, and hence, does not perform well. We find that, the expectile threshold performs better than the quantile regression based threshold in most instances except for larger values of $k$ where the result is mixed. Therefore, the expectile threshold provides a better estimator when $\gamma(x)$ is small.

We conclude that, no threshold selection procedure is universally the best. However, certain considerations can be made depending on the position of the covariate and the number of order statistics the researcher needs to be able to compute estimates of the parameters of the GP distribution. The expectile has been shown to perform well when the value of $\gamma(x)$ is small (i.e in the region where the response variable has smaller values). On the other hand, when $\gamma(x)$ is large, the constant threshold was found to give better estimates than the rest of the threshold selection methods. This is due to the fact that exceedances are recorded at this level in contrast to the previous case. Lastly, in the moderate part the result is mixed: the expectile threshold generally has better MAD and bias values for smaller values of $k.$ As $k$ increases the quantile and constant thresholds provide better bias and MAD values. However, for large $n,$ the expectile outperforms these thresholds.

The foregoing conclusions suggest combining the expectile and constant threshold to obtain a combined constant-expectile threshold that takes advantage of the performances of these thresholds. However, the performance of such a threshold  lies between the two thresholds: performing well in one situation and not well in the other. Therefore, the combined threshold does not provide much improvement upon the individual thresholds, and hence, we omitted the results. In addition, the results for the Pareto and Fr\'{e}chet distributions lead to similar conclusions.

\section{Practical Application}	\label{P3}
In this section, we illustrate the application of the covariate dependent threshold in estimating the conditional tail index of a motor insurance claim dataset from a company in Ghana. Due to competition and privacy issues, we have withheld the name of the company and the data have been modified. Specifically, we model the time until payment of claims as a function of the amount (in 1000 Ghana Cedis (GHC)). 

Under Ghana laws, it is mandatory for vehicle owners to purchase motor insurance to cover their legal liability to others who may suffer bodily injury (or death) or damage to vehicle and other properties due to the usage of the vehicle. This has made it virtually the most common insurance type; it is usually purchased for cars, trucks, motorcycles and all vehicles that ply the roads. Thus in effect, the risk of loss is transferred from the owner of the vehicle to an insurance company in exchange for a premium. The premiums are based on the type of insurance (comprehensive or a third-party) that the vehicle owner prefers.  

Claim administration in Ghana remains a problem for most policy-holders. Delays in claim administration arise from the nature of the procedures involved in settlement. A basic procedure for assessing claims includes:  Report to police; Notification of insurance company; Verification; Proof of claims; and Settlement. These processes can cause a significant delay to occur from the day an insurance company is notified of a claim to the day of payment. All companies through their adverts in the print and mass media tout fast delivery in the event of claims by policy-holders and thus, serves as the most attractive public relations strategy for insurance companies. However, there have been complaints to the insurance regulator, National Insurance Commission, on the length of time it takes to obtain a claim when the amount involved is large. To this end, we study the tails of the distribution of the time until claim payments for a Ghanaian insurance company.

\begin{table}[hptb]
	\centering
	\caption{Summary statistics of claims data}
\begin{tabular}{cc}
	\toprule 
	
	Statistic & Time (in days)\\\hline
Minimum	&  2.00\\
1st Quarter	& 4.00 \\ 
Median	&  16.00\\ 
Mean	& 69.88  \\
3rd Quarter&  64.00 \\
Maximum & 971.00 \\
Skewness & 3.69\\	
Standard deviation& 142.04 \\\bottomrule	 	
\end{tabular} 
\label{Summary}	
\end{table}

\begin{figure}
	\centering
	
		\includegraphics[height=8cm,width=0.7\textwidth]{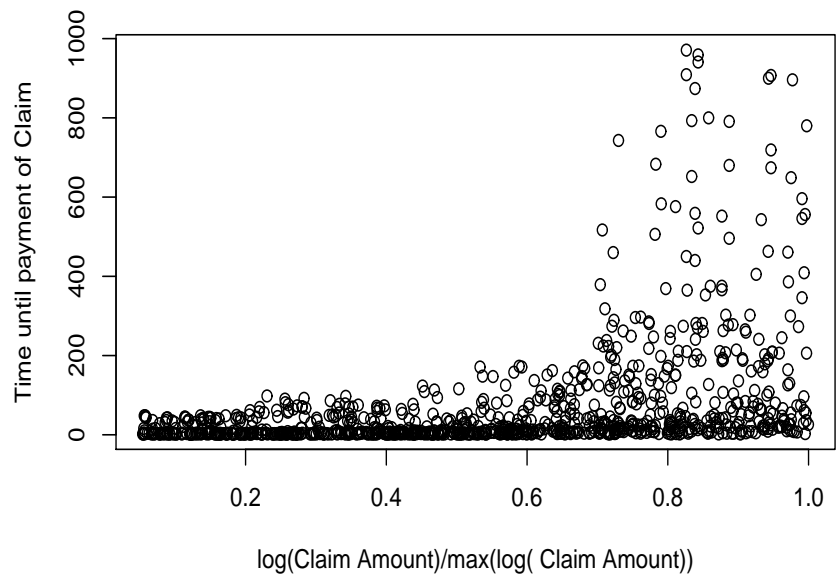}
	\caption{Scatter plot of the insurance data and estimates of $\gamma(x)$}
	\label{Prelimnary}

\end{figure}

Table \ref{Summary} shows summary statistics of the time until claims payment. The earliest payment time was two days and the longest was 971 days arising as a result of litigation. In addition, Figure \ref{Prelimnary} shows some graphical outputs for the number of days until a claim is settled as a function of the amount of claim. The exponential like growth of the time until claims payment in relation to the amount involved suggests modelling the conditional tail distribution as GP with the tail index and/or scale parameters taken as exponential functions of the claim amounts.

\begin{figure}[htp!]
	\centering
	
	\subfloat[]{%
		\includegraphics[height=6cm,width=.49\textwidth]{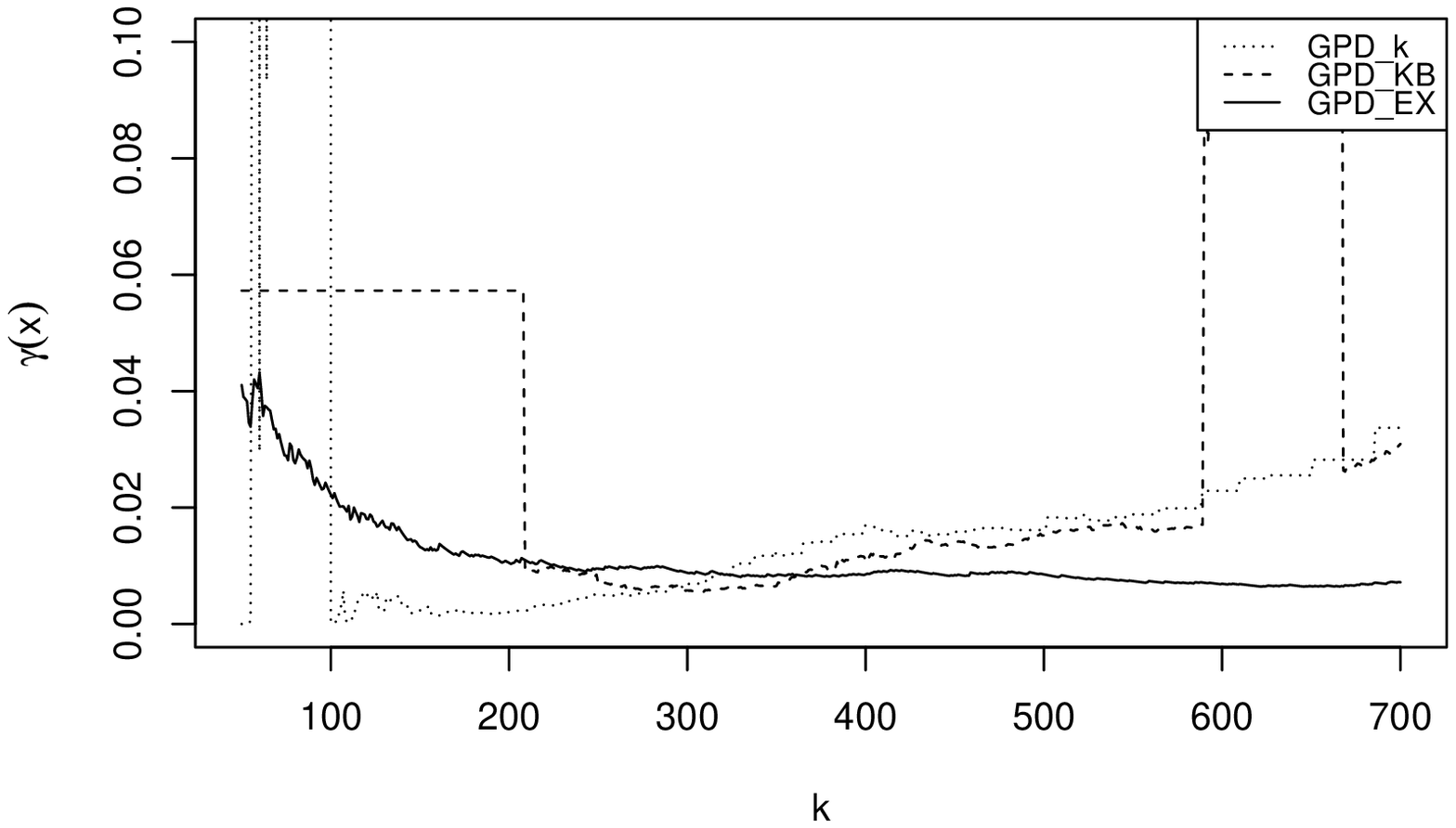}}\hfill
	\subfloat[]{%
		\includegraphics[height=6cm,width=.49\textwidth]{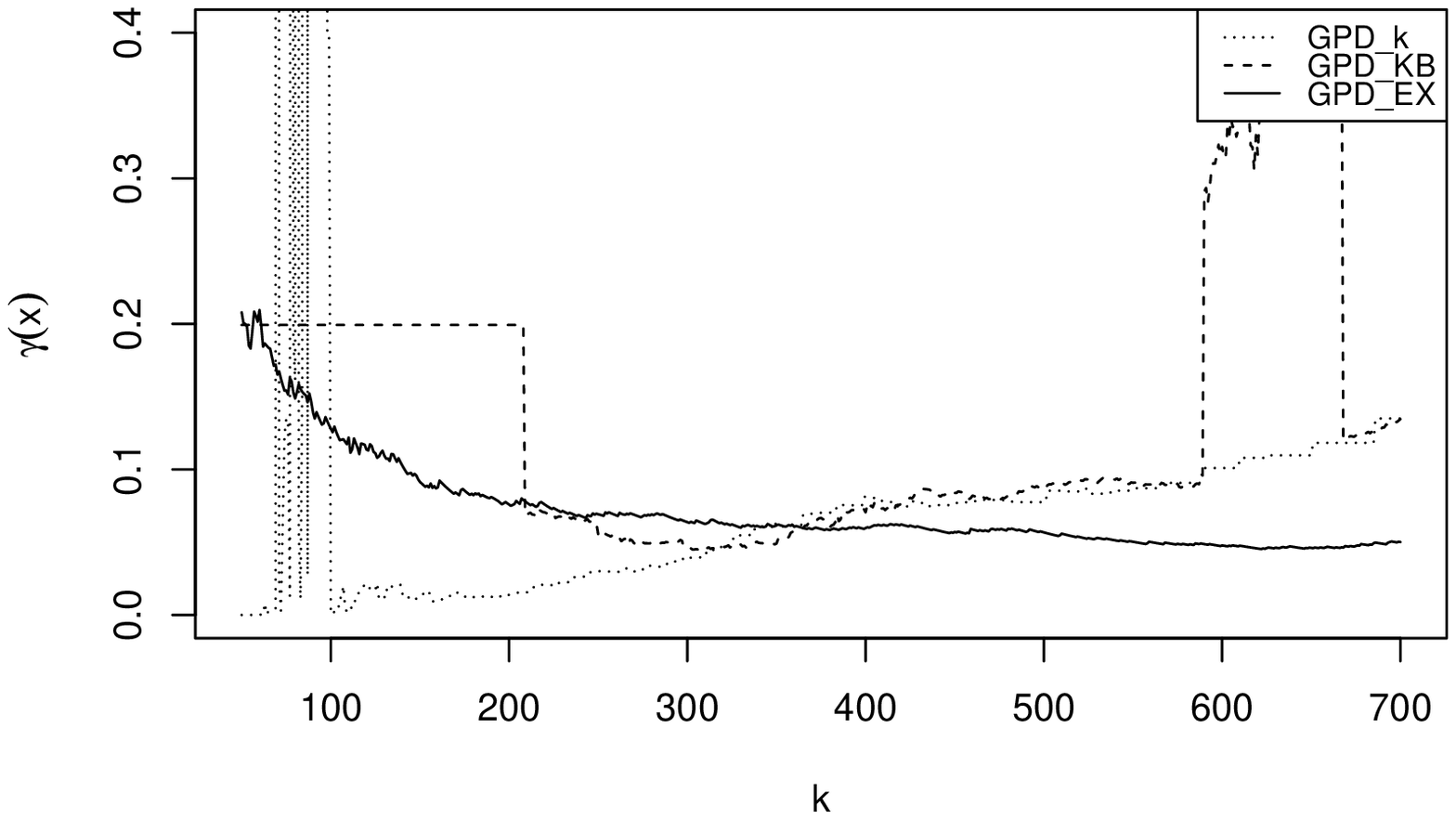}}\\
	\subfloat[ ]{%
		\includegraphics[height=6cm,width=.49\textwidth]{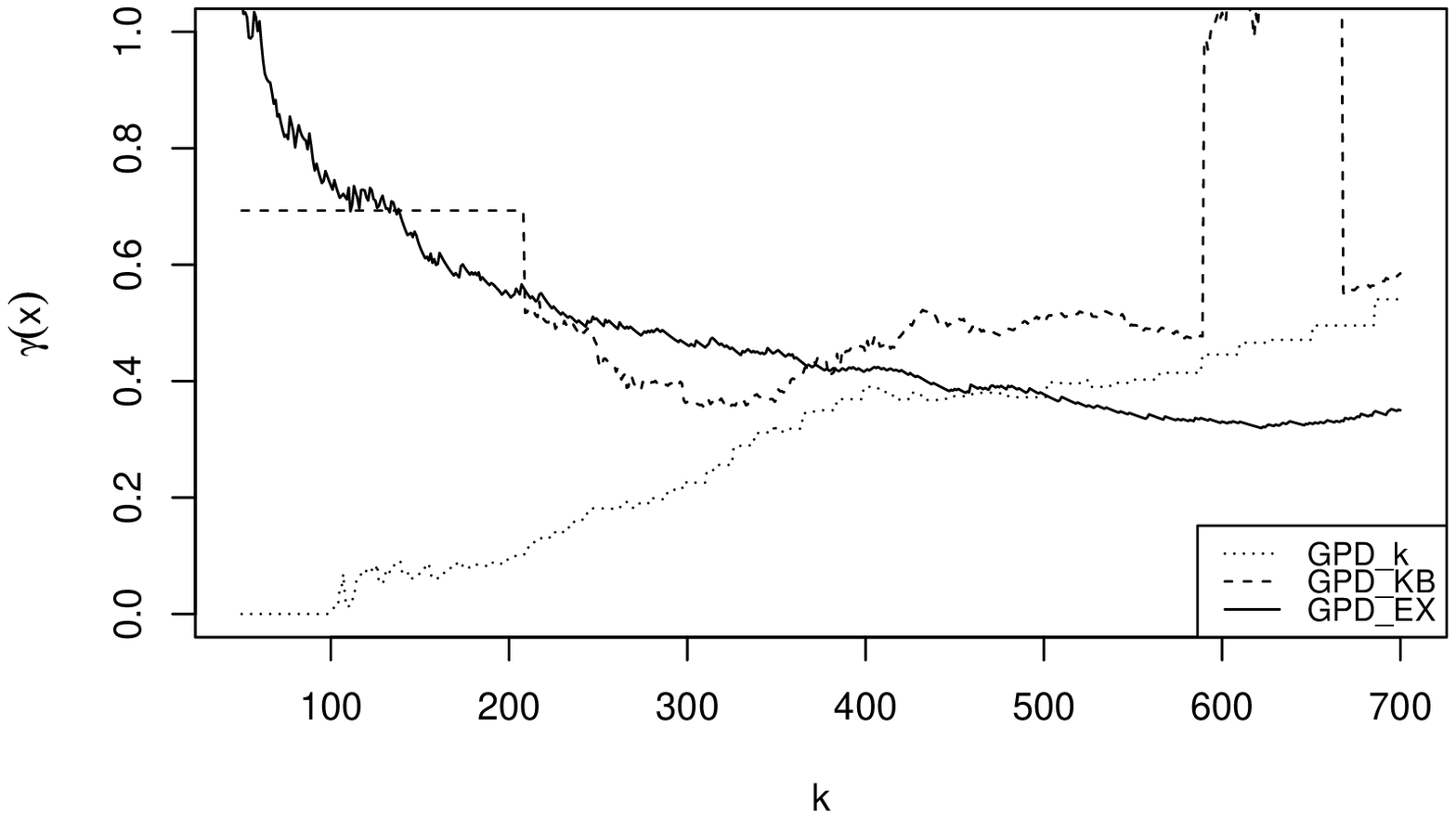}}\hfill
	\subfloat[]{%
		\includegraphics[height=6cm,width=.49\textwidth]{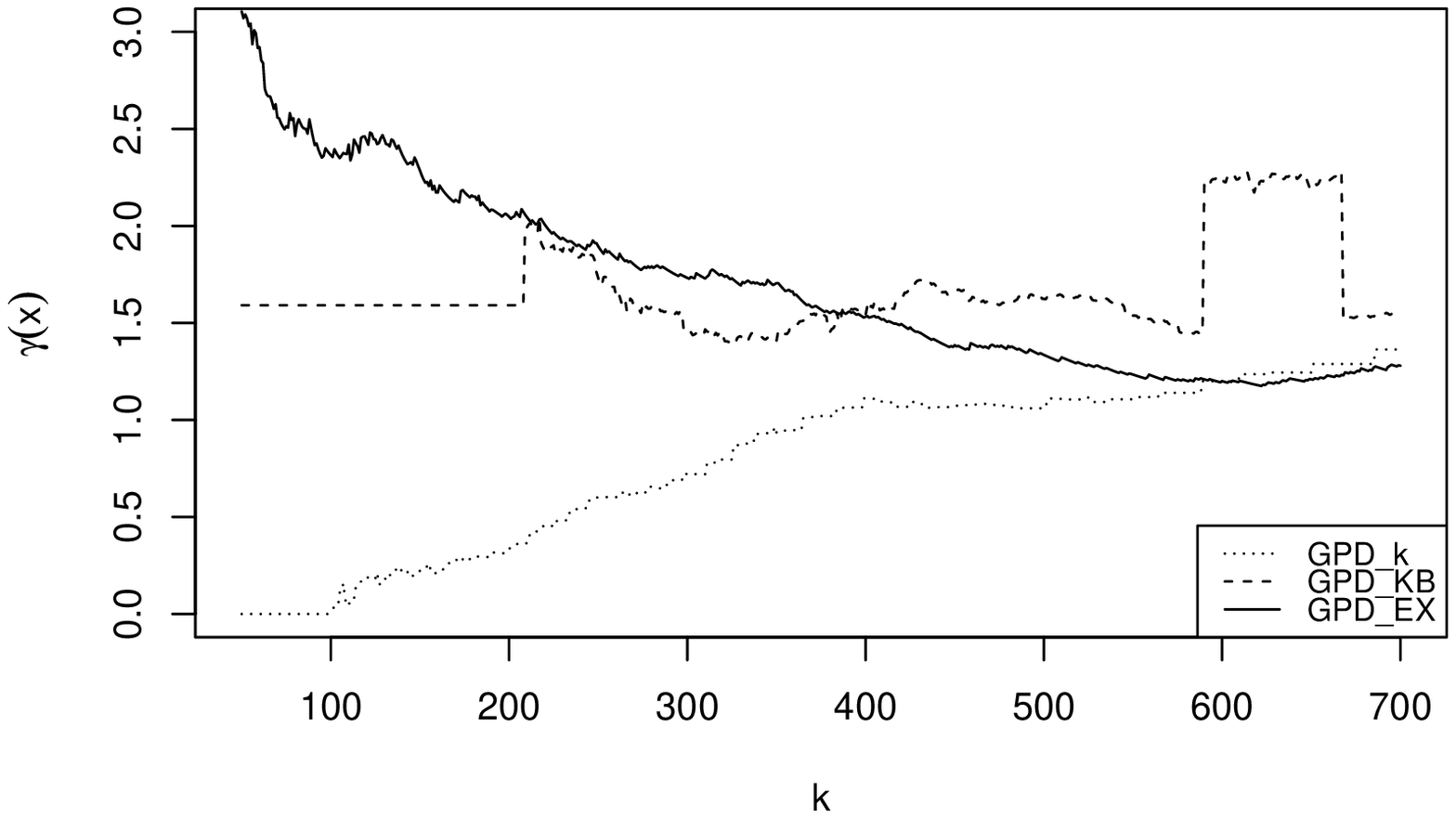}}\\
	\caption{Estimates of $\gamma(x):$ Top left, $x=0.10;$ top right,  $x=0.40;$ bottom left, $x=0.70;$ and bottom right, $x=0.90.$}
	\label{gam}
\end{figure}

We illustrate the estimation of the tail index of the distribution of time until claims payment conditional on claim amount at four levels. These $x$-levels were chosen to cover the scaled covariate space, 0 to 1. The results of the estimates of $\gamma(x)$ at various $x$ and $k$ values are presented in Figure \ref{gam}. The top panel shows the values of $\gamma(x)$ where claim amounts are smaller. From the results of our simulation study, we are inclined to choose the value of $\gamma(x)$ using the expectile threshold. The left panel has stability for all the estimators between $200\le k\le 400,$ and hence, $\gamma(x)$ can be chosen in this region to be approximately 0.018 and 0.098 respectively at $x=0.10$ and $x=0.4.$

In addition, the lower panel indicates longer waiting times, and hence, the constant threshold provides the best estimators according to our simulation studies. Therefore, using this guideline, we can estimate the value of $\gamma(x)$ between the range of $k\in [400,  600],$ where the estimators especially the one based on constant threshold seems relatively constant. The estimate of $\gamma(x)$ is approximately 0.4 and 1.0 at $x=0.70$ and $x=0.90$ respectively. With these estimates of  $\gamma(x)$ together with that of the scale and the thresholds, questions such as how long a customer will wait for a specific amount and the exceedance probabilities over a specific time, given an amount involved, can be obtained.

\section{Conclusion}\label{P4}

In this paper, we reviewed threshold selection in the presence of covariate information in Extreme Value Theory and introduced a covariate dependent threshold based on expectiles. The expectile based threshold was compared with the existing ones in the literature, constant and the quantile regression thresholds, for estimating the tail index of the Generalised Pareto distribution via a simulation study. The results show that no threshold selection method is universally the best. However, for the exponential growth data considered, we find that the expectile threshold outperforms the others when the response variable has smaller to medium values. Also, for larger values of the response variable, the constant threshold is generally the best method. We illustrated the application of these thresholds in estimating the tail index of an insurance claims data. We remark that other functional forms for the tail index as well as extreme quantile estimation are the subject of future studies.
\bibliographystyle{apalike}

\begin{thebibliography}{}
	
	\bibitem[Balkema and de~Haan, 1974]{Balkema1974}
	Balkema, A.~A. and de~Haan, L. (1974).
	\newblock {Residual life time at great age}.
	\newblock {\em Annals of Probability}, 2(5):792--804.
	
	\bibitem[Beirlant and Goegebeur, 2004]{Beirlant2004}
	Beirlant, J. and Goegebeur, Y. (2004).
	\newblock {Local polynomial maximum likelihood estimation for Pareto-type
		distributions}.
	\newblock {\em Journal of Multivariate Analysis}, 89(1):97--118.
	
	\bibitem[Beirlant et~al., 2004]{Beirlant2005}
	Beirlant, J., Goegebeur, Y., Segers, J., and Teugels, J.~L. (2004).
	\newblock {\em {Statistics of Extremes: Theory and Applications}}.
	\newblock Wiley, England.
	
	\bibitem[Burr, 1942]{Burr1942}
	Burr, I.~W. (1942).
	\newblock {Cumulative frequency functions}.
	\newblock {\em The Annals of Mathematical Statistics}, 13(2):215--232.
	
	\bibitem[Coles, 2001]{Coles2001}
	Coles, S. (2001).
	\newblock {\em {An introduction to statistical modelling of extreme values}}.
	\newblock Springer, London.
	
	\bibitem[Daouia et~al., 2018]{Daouia2018}
	Daouia, A., Girard, S., and Stupfler, G. (2018).
	\newblock Estimation of tail risk based on extreme expectiles.
	\newblock {\em Journal of the Royal Statistical Society: Series B (Statistical
		Methodology)}, 80(2):263--292.
	
	\bibitem[Davison and Smith, 1990]{Davison1990}
	Davison, A.~C. and Smith, R.~L. (1990).
	\newblock {Models for exceedances over high thresholds}.
	\newblock {\em Journal of the Royal Statistical Society: Series B (Statistical
		Methodology)}, 52(3):393--442.
	
	\bibitem[Harlow, 2002]{Harlow2002}
	Harlow, D.~G. (2002).
	\newblock {Applications of the Fr{\'{e}}chet distribution function}.
	\newblock {\em International Journal of Materials and Product Technology},
	17(576):482--495.
	
	\bibitem[Koenker and Bassett, 1978]{Koenker1978}
	Koenker, R. and Bassett, G. (1978).
	\newblock {Regression quantiles}.
	\newblock {\em Econometrica}, 46(1):33--50.
	
	\bibitem[Kuan et~al., 2009]{Kuan2009}
	Kuan, C.-M., Yeh, J.-H., and Hsu, Y.-C. (2009).
	\newblock Assessing value at risk with care, the conditional autoregressive
	expectile models.
	\newblock {\em Journal of Econometrics}, 150(2):261--270.
	
	\bibitem[Ndao et~al., 2014]{Ndao2014}
	Ndao, P., Diop, A., and Dupuy, J.~F. (2014).
	\newblock {Nonparametric estimation of the conditional tail index and extreme
		quantiles under random censoring}.
	\newblock {\em Computational Statistics and Data Analysis}, 79:63--79.
	
	\bibitem[Newey and Powell, 1987]{Newey1987}
	Newey, W.~K. and Powell, J.~L. (1987).
	\newblock Asymmetric least squares estimation and testing.
	\newblock {\em Econometrica}, 55(4):819--847.
	
	\bibitem[Northrop and Jonathan, 2011]{Northrop2011}
	Northrop, P.~J. and Jonathan, P. (2011).
	\newblock {Threshold modelling of spatially dependent non-stationary extremes
		with application to hurricane-induced wave heights}.
	\newblock {\em Environmetrics}, 22(7):799--809.
	
	\bibitem[{Pickands III}, 1975]{Pickands1975}
	{Pickands III}, J. (1975).
	\newblock {Statistical Inference Using Extreme Order Statistics}.
	\newblock {\em The Annals of Statistics}, 3(1):119--131.
	
	\bibitem[Scarrott and Macdonald, 2012]{Scarrott2012}
	Scarrott, C. and Macdonald, A. (2012).
	\newblock {A review of extreme value threshold estimation and uncertainty
		quantification}.
	\newblock {\em REVSTAT}, 10(1):33--60.
	
	\bibitem[Smith, 1989]{Smith1989}
	Smith, R.~L. (1989).
	\newblock {Extreme value analysis of environmental time series: an application
		to trend detection in ground-level ozone}.
	\newblock {\em Statistical Science}, 4(4):367--377.
	
	\bibitem[Tancredi et~al., 2006]{Tancredi2006}
	Tancredi, A., Anderson, C., and O'Hagan, A. (2006).
	\newblock {Accounting for threshold uncertainty in extreme value estimation}.
	\newblock {\em Extremes}, 9:87--106.
	
	\bibitem[Thompson et~al., 2009]{Thompson2009}
	Thompson, P., Cai, Y., Reeve, D., and Stander, J. (2009).
	\newblock {Automated threshold selection methods for extreme wave analysis}.
	\newblock {\em Coastal Engineering}, 56(10):1013--1021.
	
	\bibitem[Wang and Tsai, 2009]{Wang2009}
	Wang, H. and Tsai, C.-L. (2009).
	\newblock {Tail index regression}.
	\newblock {\em Journal of the American Statistical Association},
	104(487):1233--1240.
	
\end{thebibliography}

\appendix
\section*{Appendices}
\addcontentsline{toc}{section}{Appendices}
\renewcommand{\thesubsection}{\Alph{subsection}}

\subsection{Pareto Distribution}	

\begin{figure}[htpb!]
	\centering
	
	\subfloat[]{%
		\includegraphics[height=6cm,width=.33\textwidth]{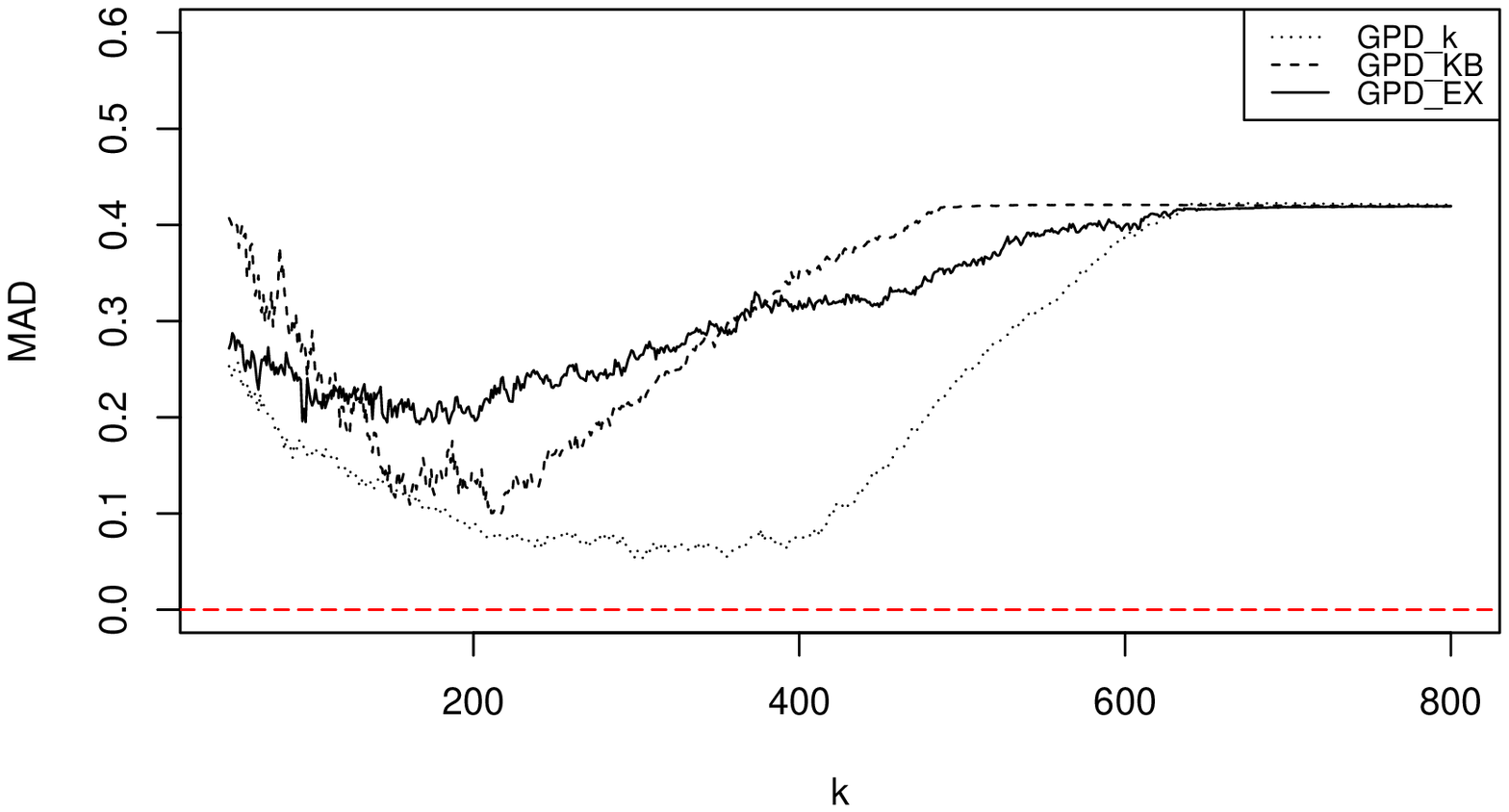}}\hfill
	\subfloat[]{%
		\includegraphics[height=6cm,width=.33\textwidth]{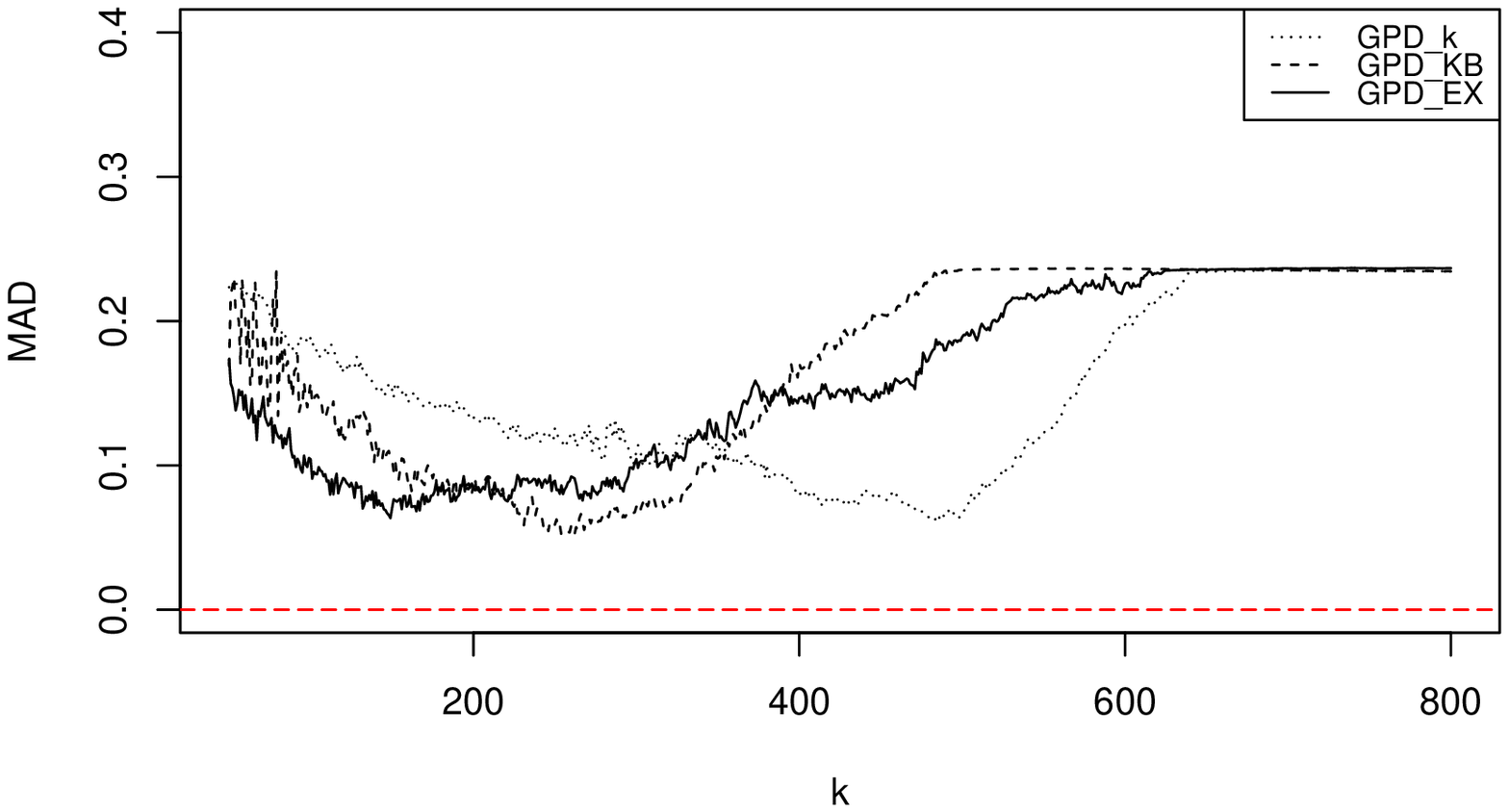}}\hfill
	\subfloat[ ]{%
		\includegraphics[height=6cm,width=.33\textwidth]{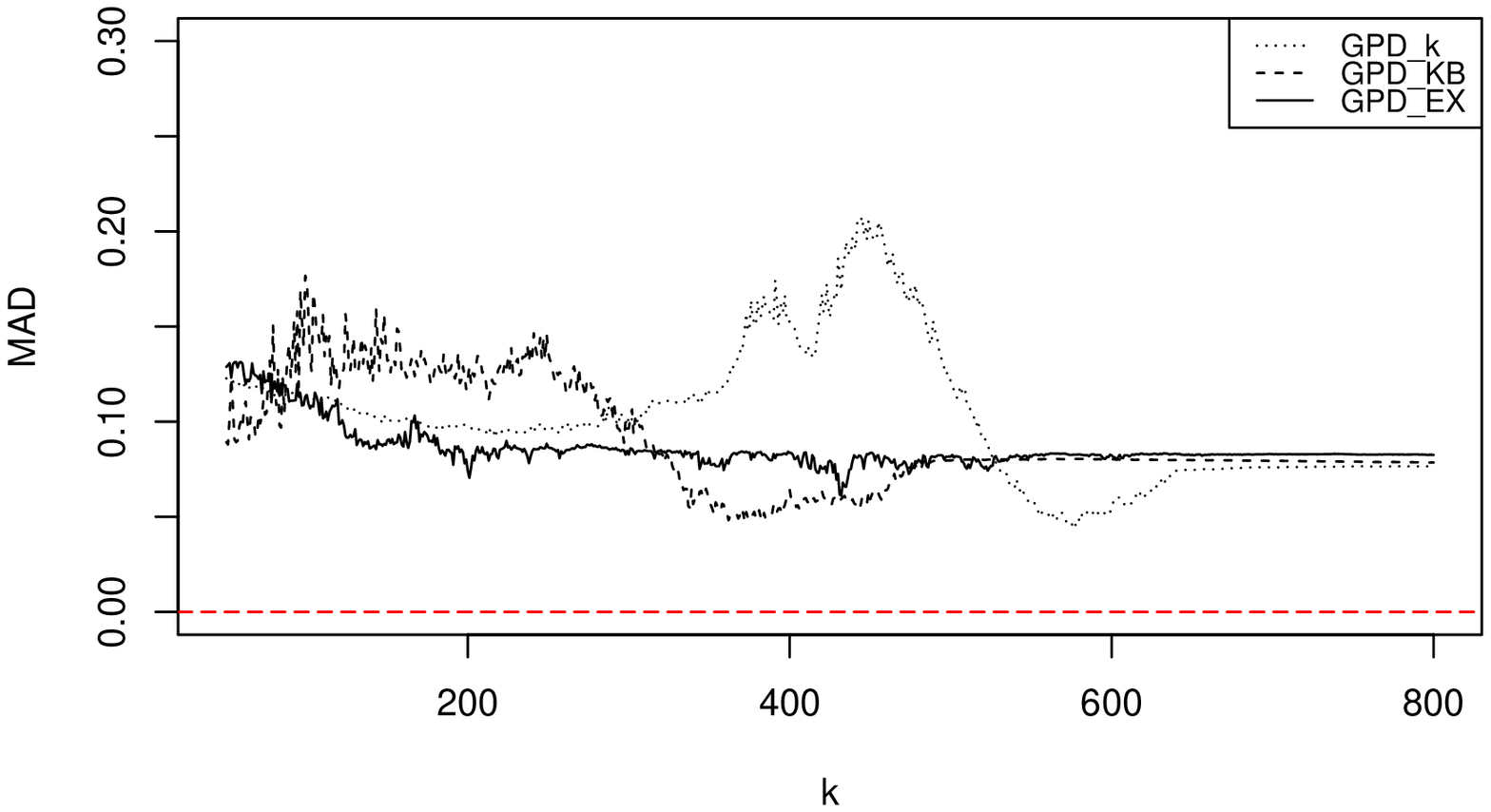}}\\
	\subfloat[]{%
		\includegraphics[height=6cm,width=.33\textwidth]{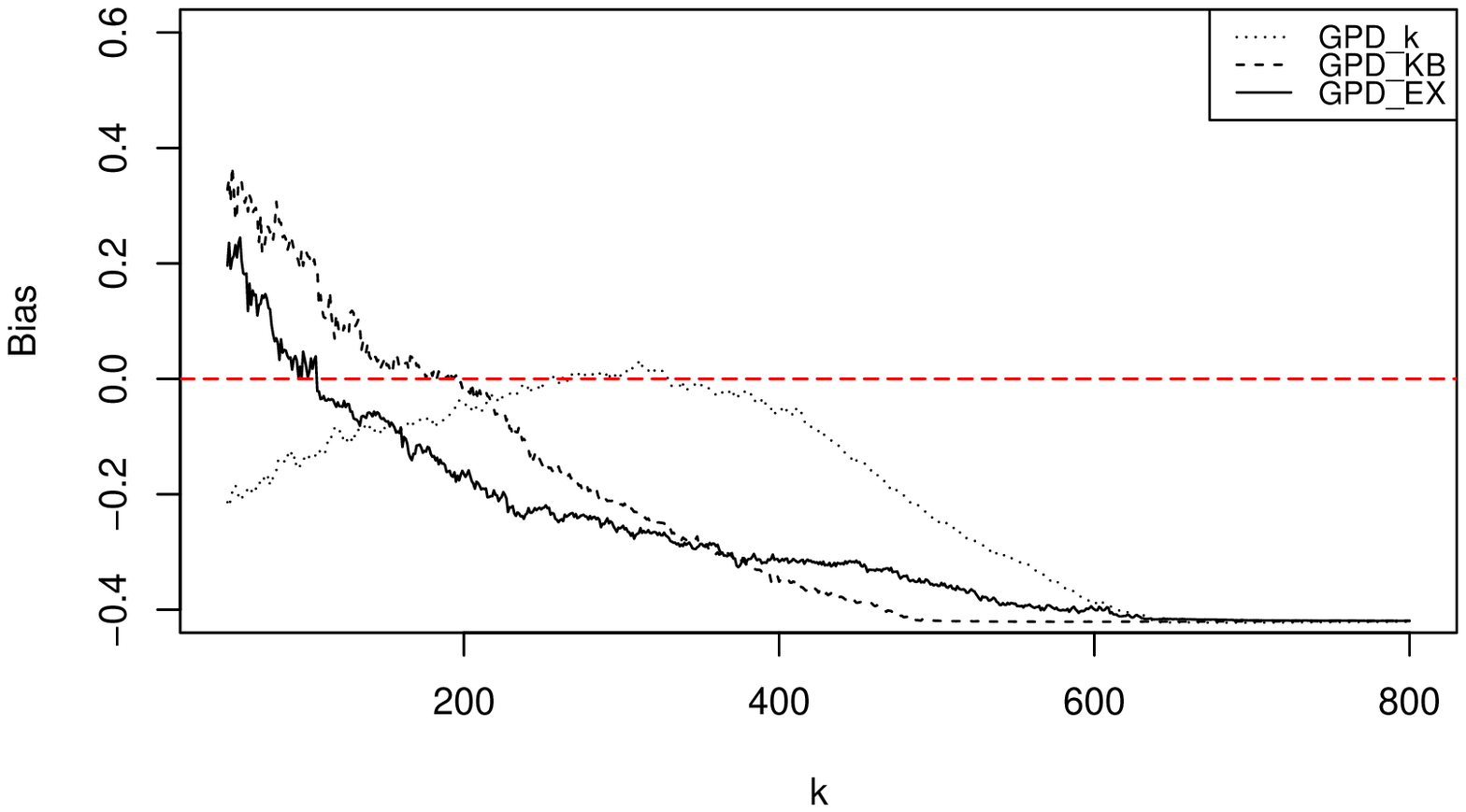}}\hfill
	\subfloat[]{%
		\includegraphics[height=6cm,width=.33\textwidth]{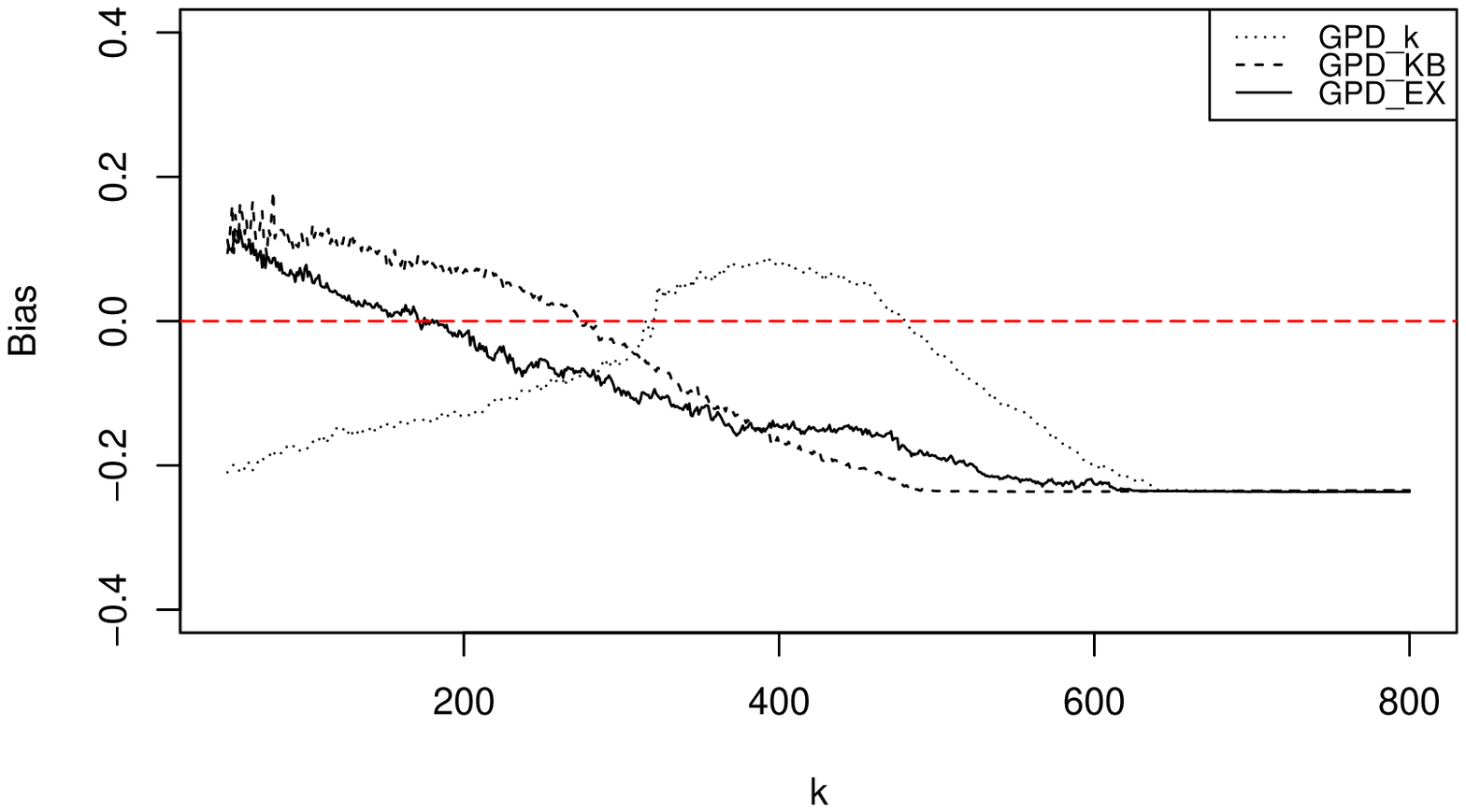}}\hfill
	\subfloat[ ]{%
		\includegraphics[height=6cm,width=.33\textwidth]{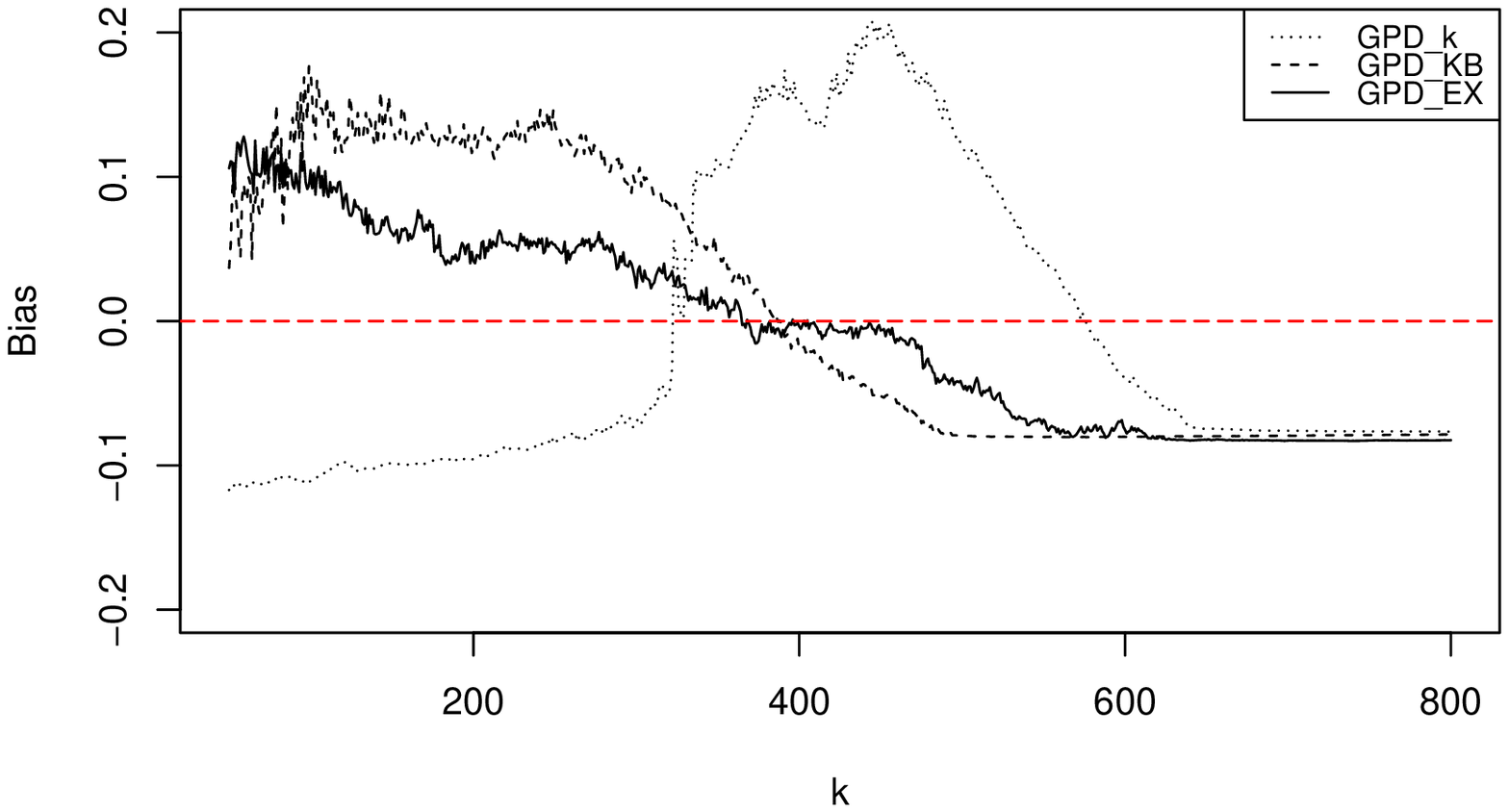}}\\
	\caption{Burr Distribution with $n=1000:$ Left Column: $\gamma(x)=0.50;$ Middle Column: $\gamma(x)=0.30;$ and Right Column: $\gamma(x)=0.13$}
	\label{Par1}
\end{figure}

\begin{figure}[htpb!]
	\centering
	
	\subfloat[]{%
		\includegraphics[height=6cm,width=.33\textwidth]{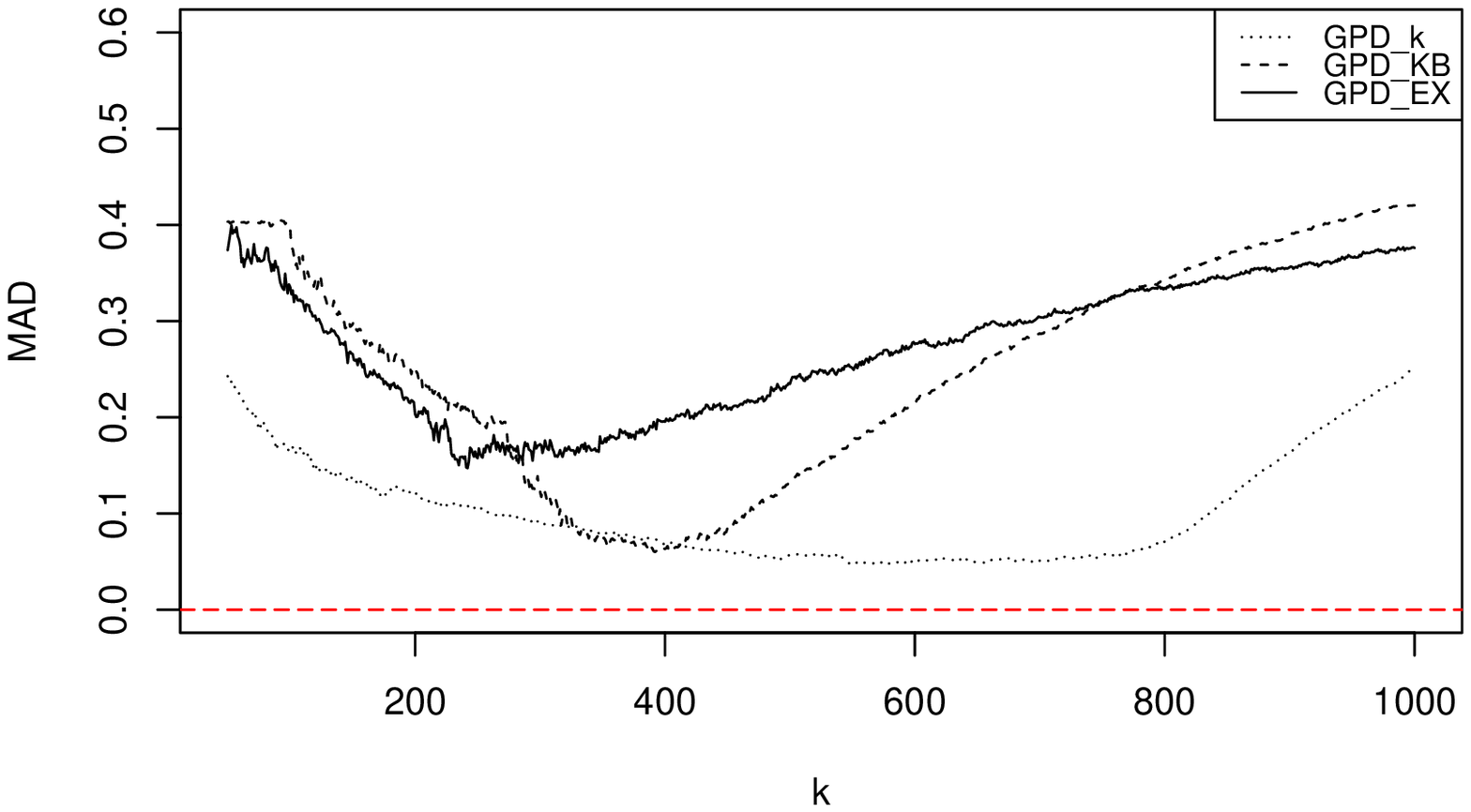}}\hfill
	\subfloat[]{%
		\includegraphics[height=6cm,width=.33\textwidth]{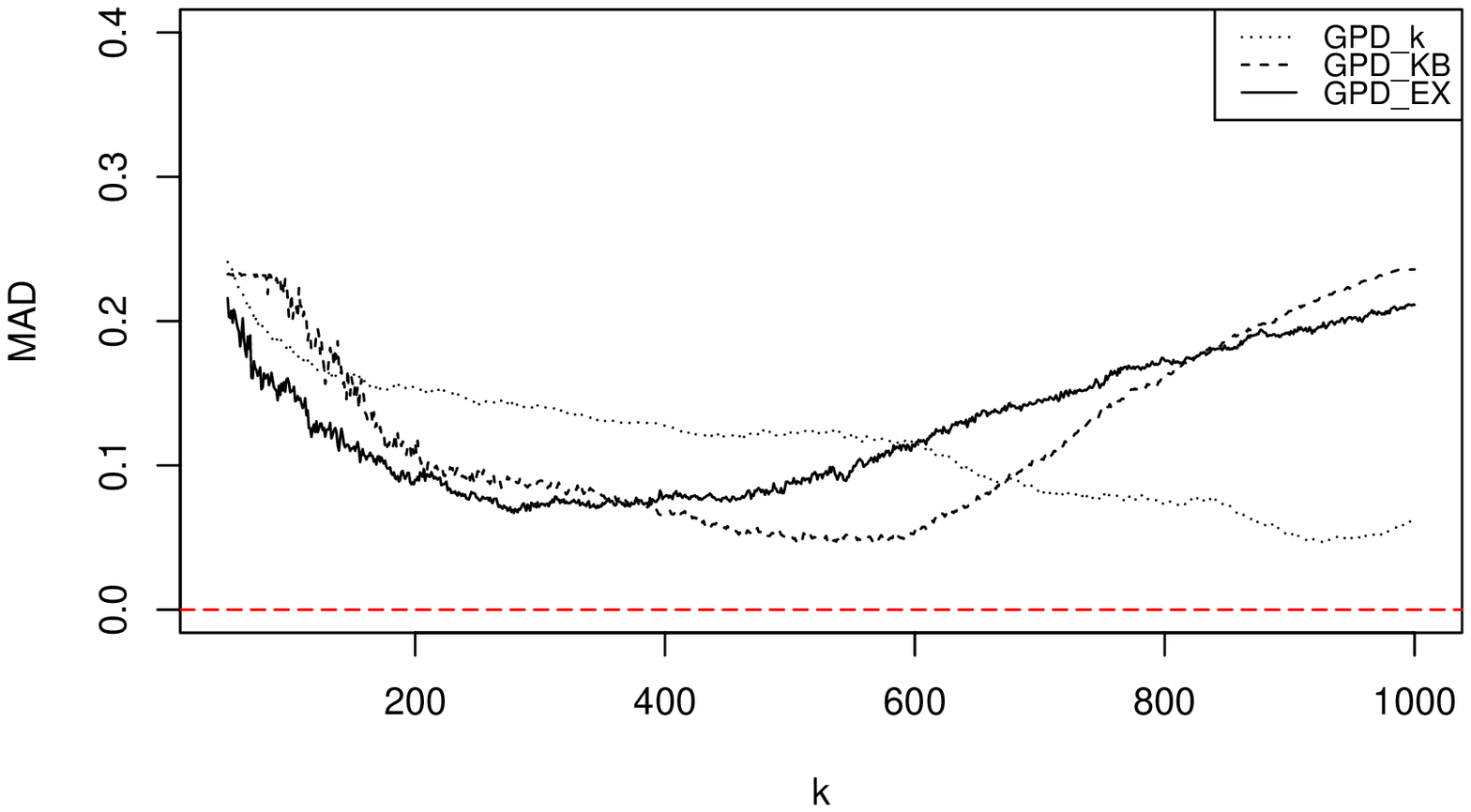}}\hfill
	\subfloat[ ]{%
		\includegraphics[height=6cm,width=.33\textwidth]{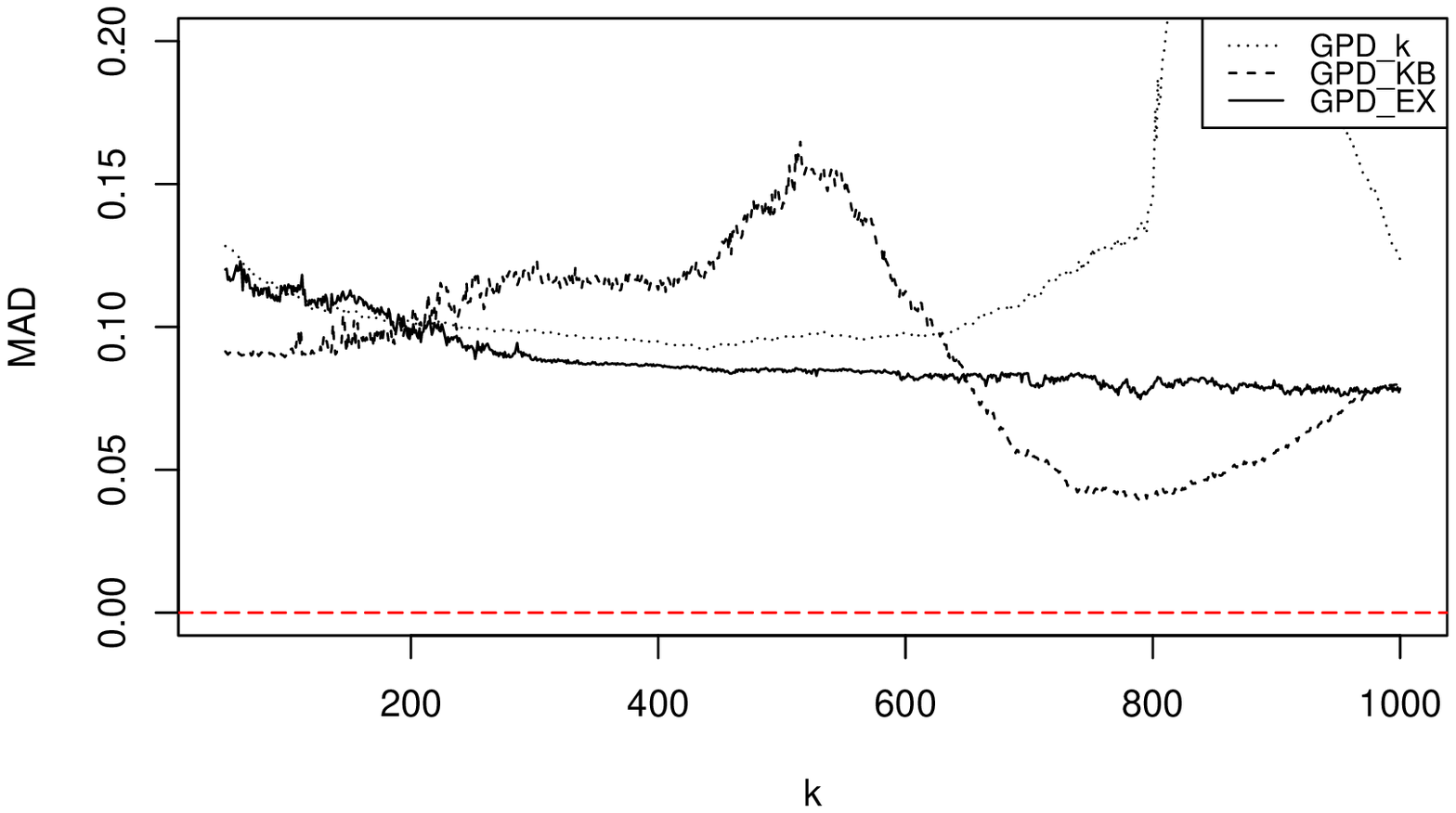}}\\
	\subfloat[]{%
		\includegraphics[height=6cm,width=.33\textwidth]{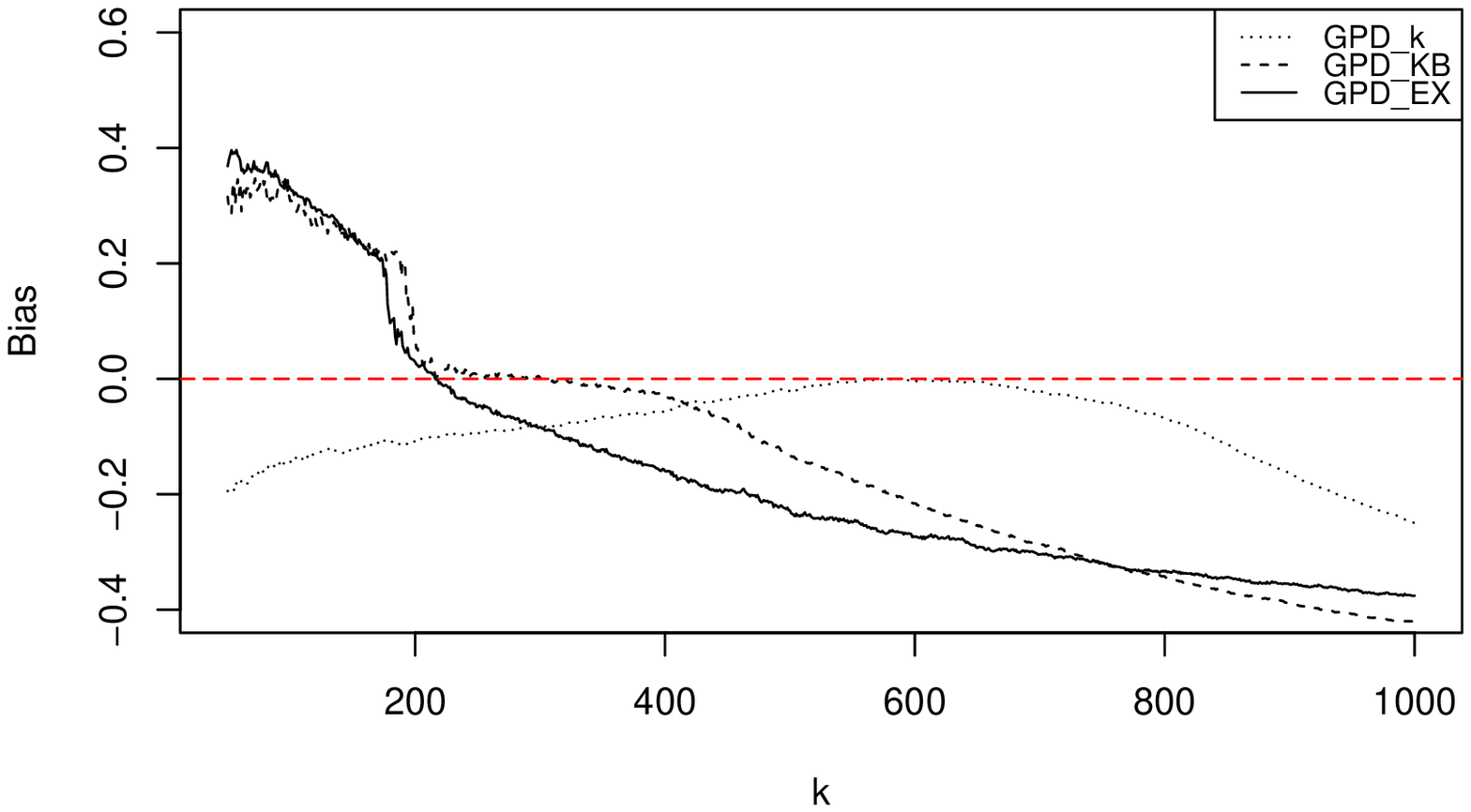}}\hfill
	\subfloat[]{%
		\includegraphics[height=6cm,width=.33\textwidth]{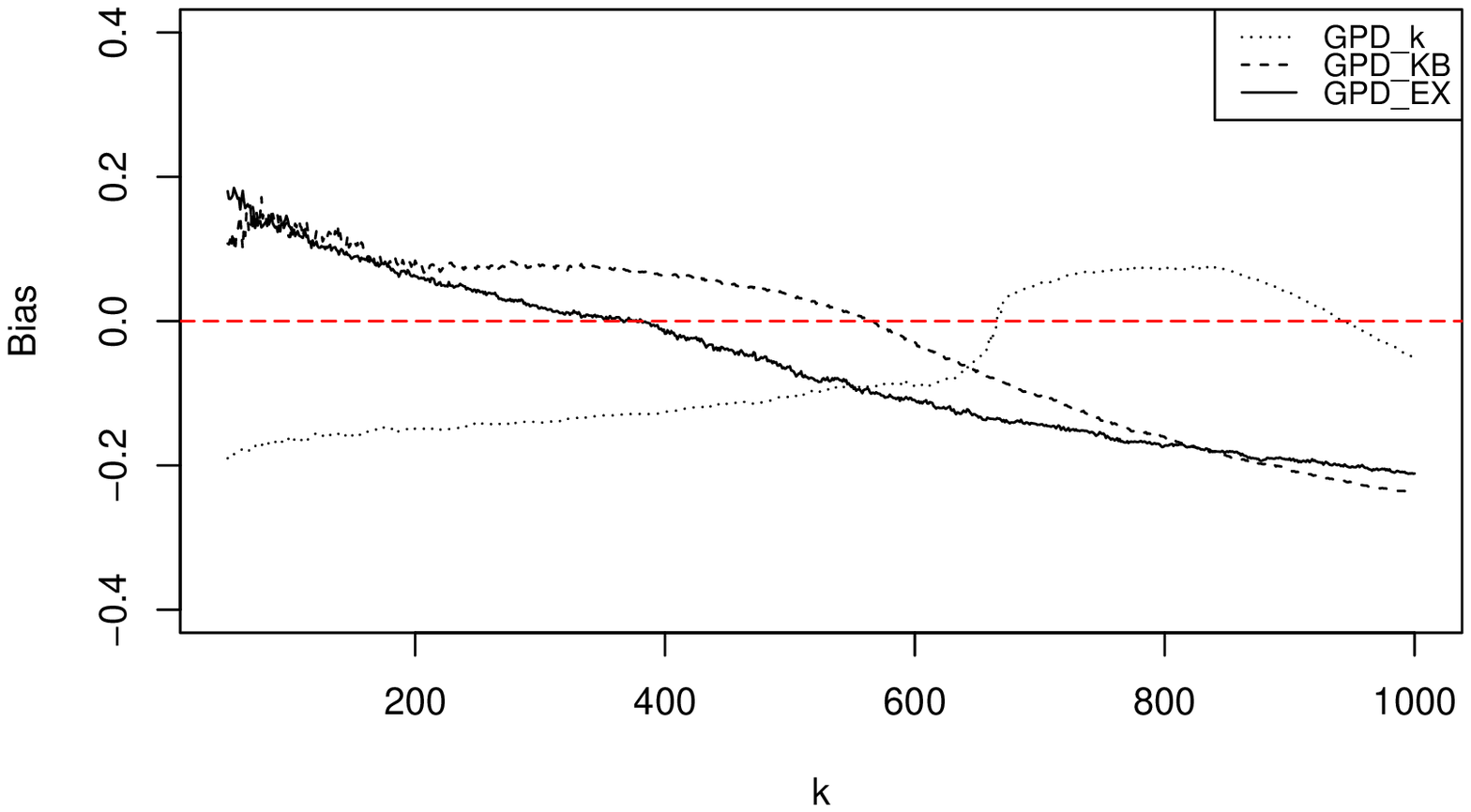}}\hfill
	\subfloat[ ]{%
		\includegraphics[height=6cm,width=.33\textwidth]{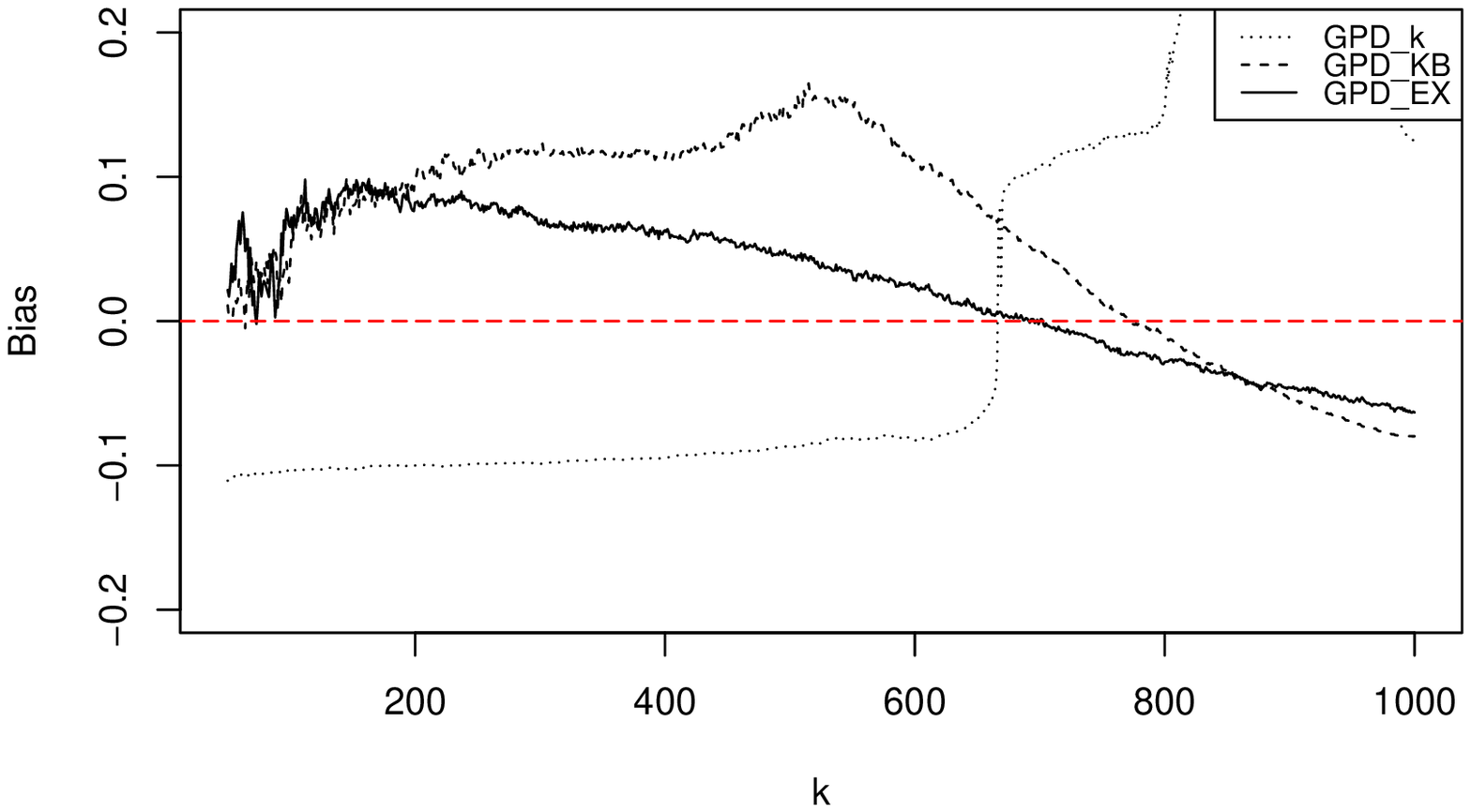}}\\
	\caption{Burr Distribution with $n=2000:$ Left Column: $\gamma(x)=0.50;$ Middle Column: $\gamma(x)=0.30;$ and Right Column: $\gamma(x)=0.13$}
	\label{Par2}
\end{figure}

\begin{figure}[htpb!]
	\centering
	
	\subfloat[]{%
		\includegraphics[height=6cm,width=.33\textwidth]{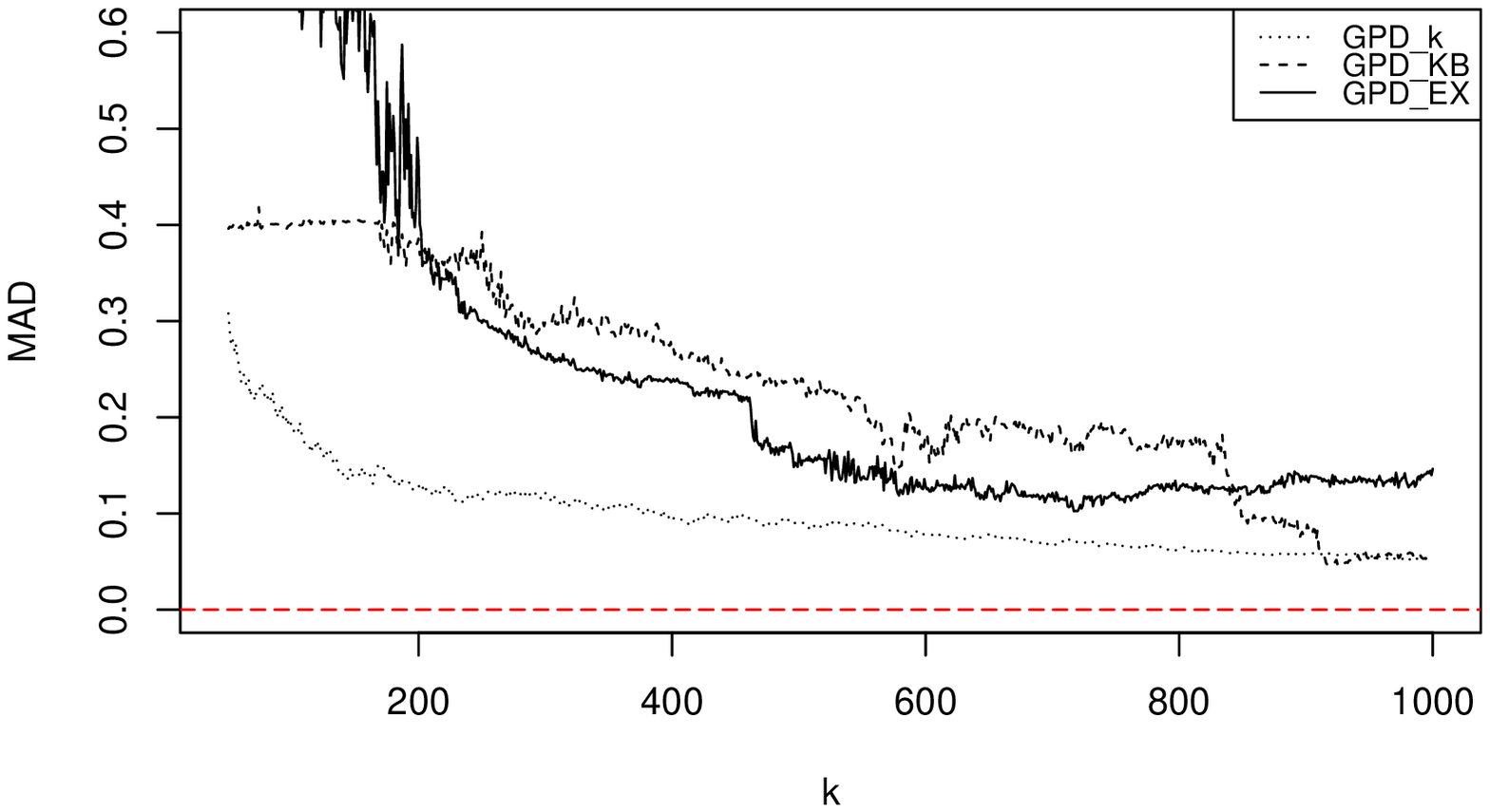}}\hfill
	\subfloat[]{%
		\includegraphics[height=6cm,width=.33\textwidth]{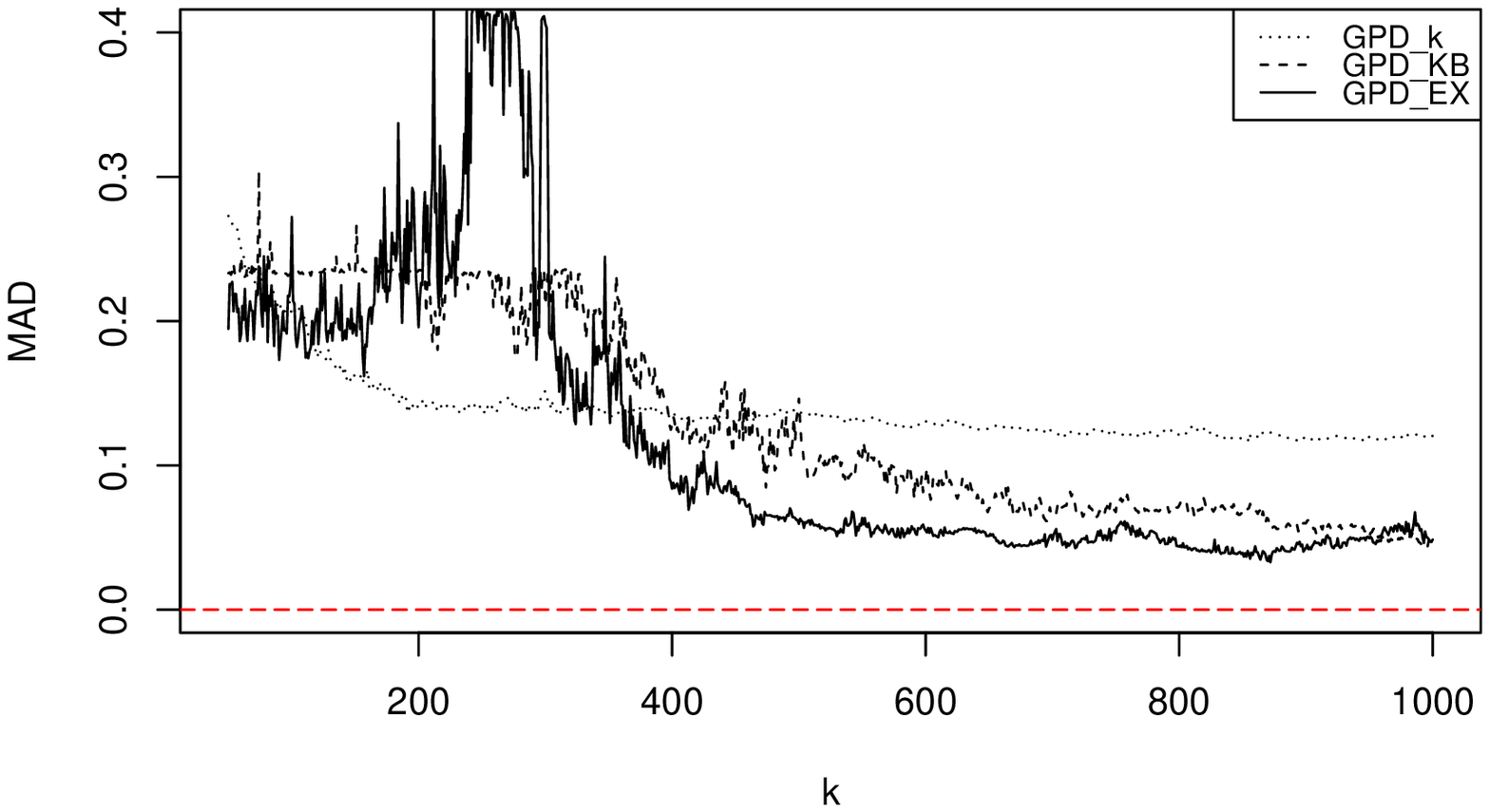}}\hfill
	\subfloat[ ]{%
		\includegraphics[height=6cm,width=.33\textwidth]{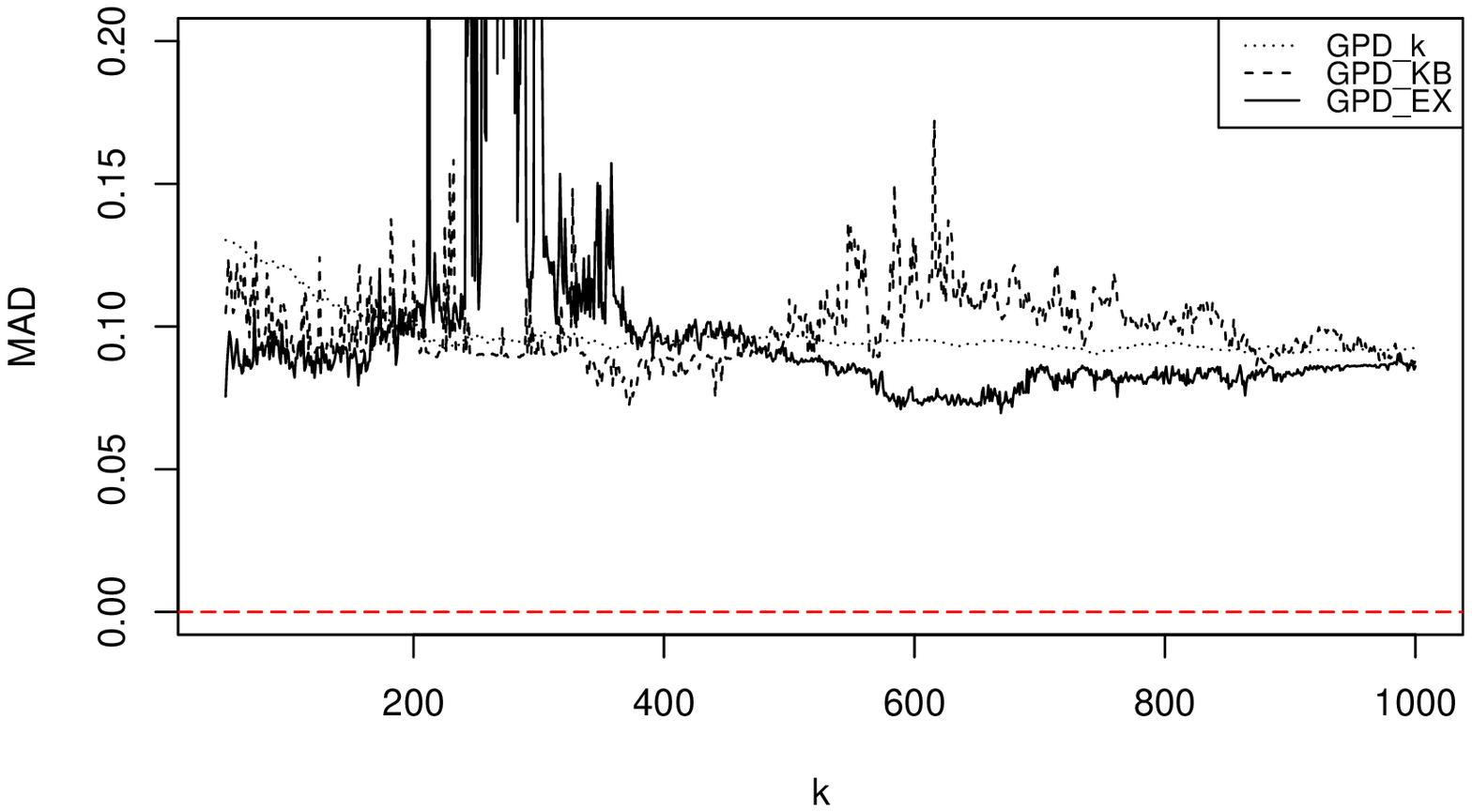}}\\
	\subfloat[]{%
		\includegraphics[height=6cm,width=.33\textwidth]{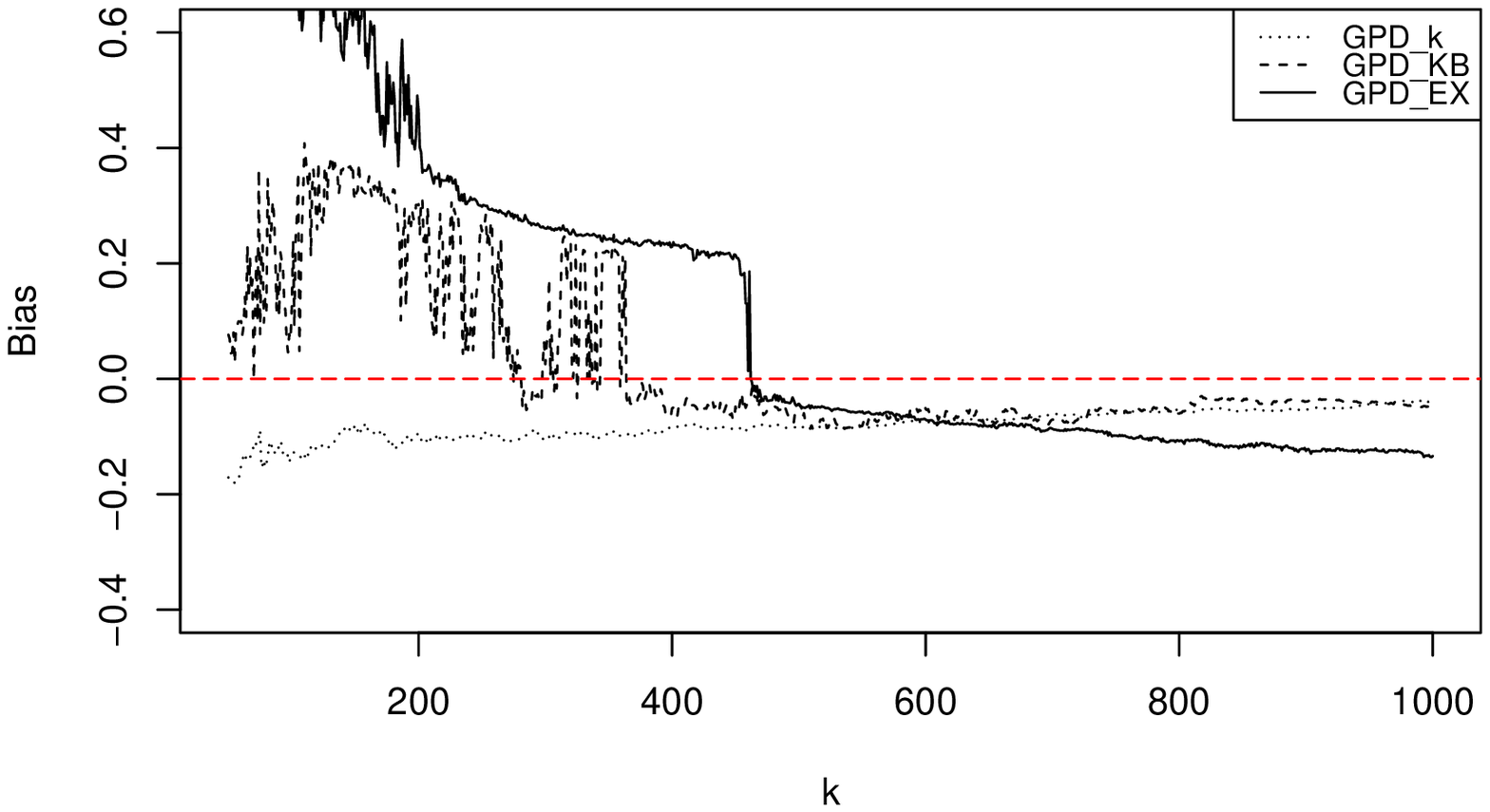}}\hfill
	\subfloat[]{%
		\includegraphics[height=6cm,width=.33\textwidth]{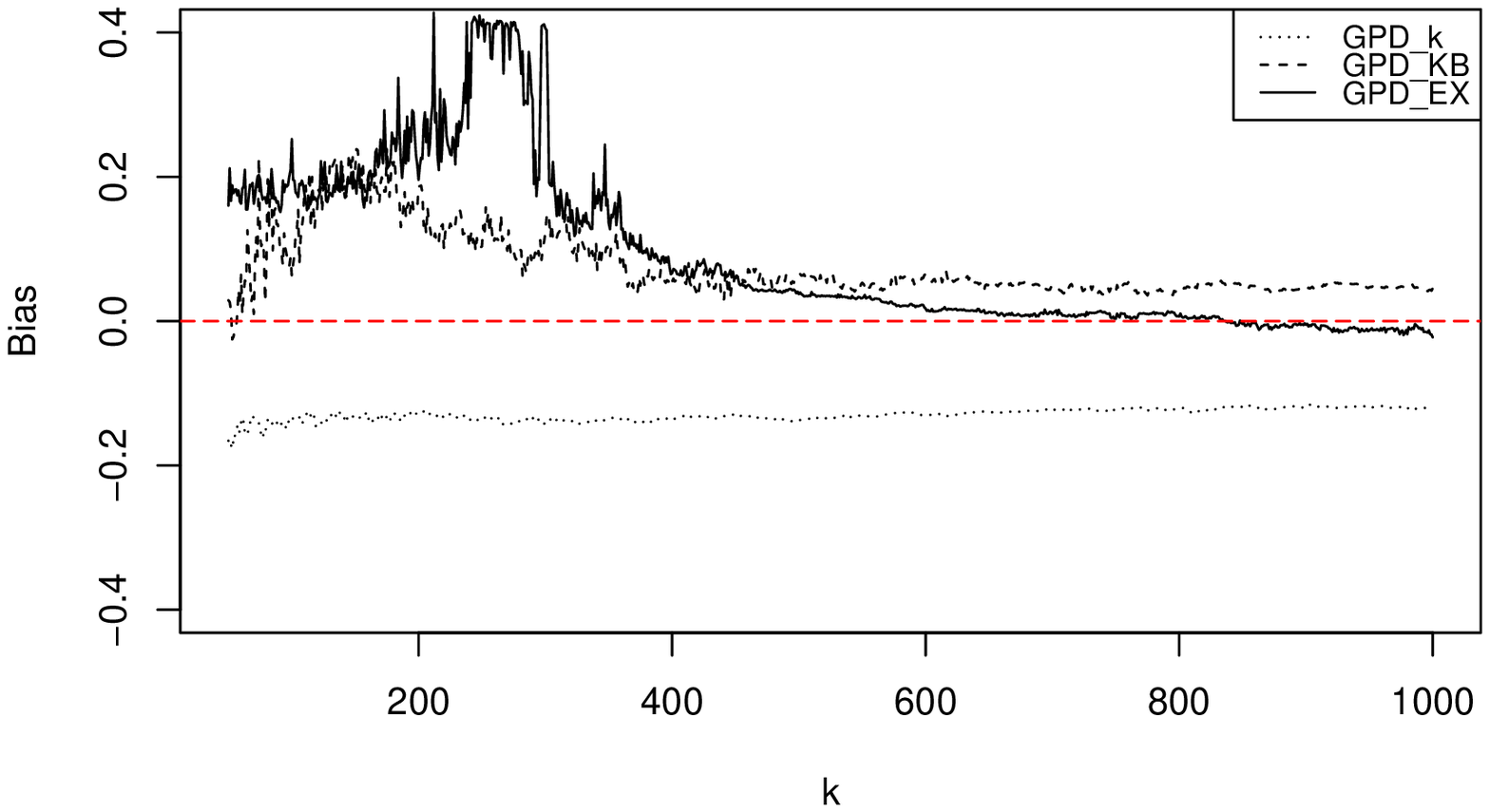}}\hfill
	\subfloat[ ]{%
		\includegraphics[height=6cm,width=.33\textwidth]{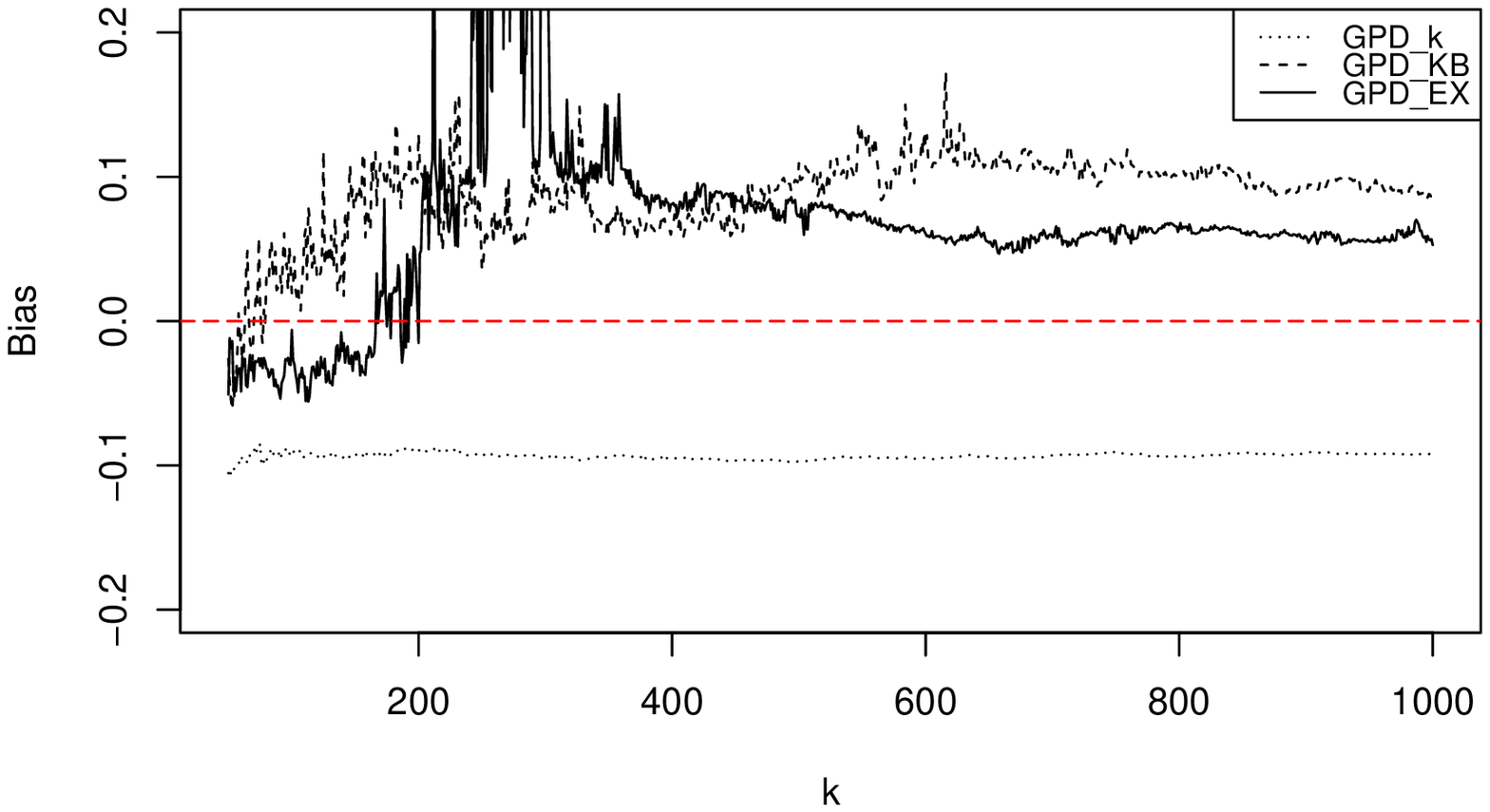}}\\
	\caption{Burr Distribution with $n=5000:$ Left Column: $\gamma(x)=0.50;$ Middle Column: $\gamma(x)=0.30;$ and Right Column: $\gamma(x)=0.13$}
	\label{Par3}
\end{figure}

\newpage
\subsection{Fr\'{e}chet Distribution}	
\begin{figure}[htpb!]
	\centering
	
	\subfloat[]{%
		\includegraphics[height=6cm,width=.33\textwidth]{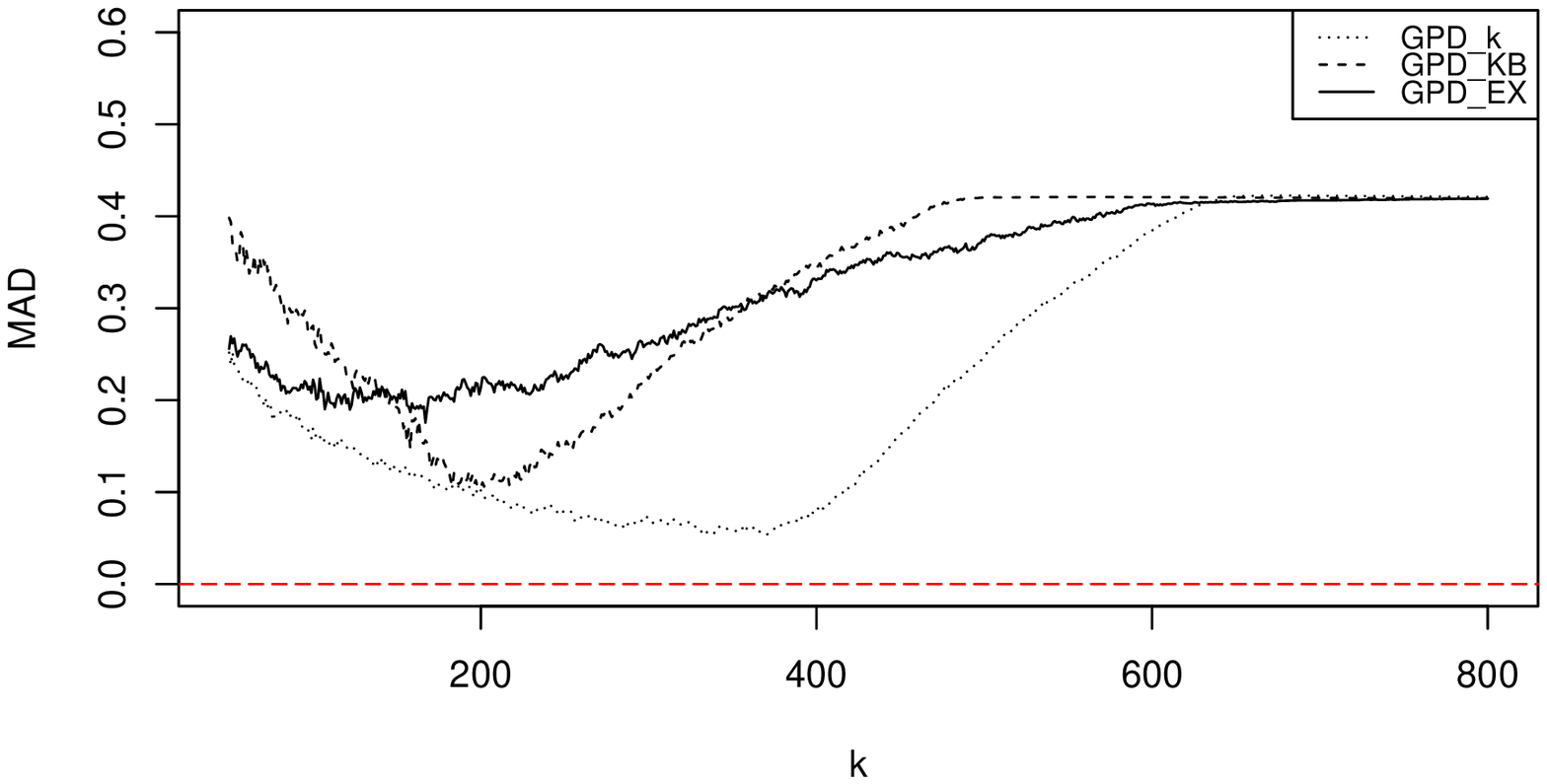}}\hfill
	\subfloat[]{%
		\includegraphics[height=6cm,width=.33\textwidth]{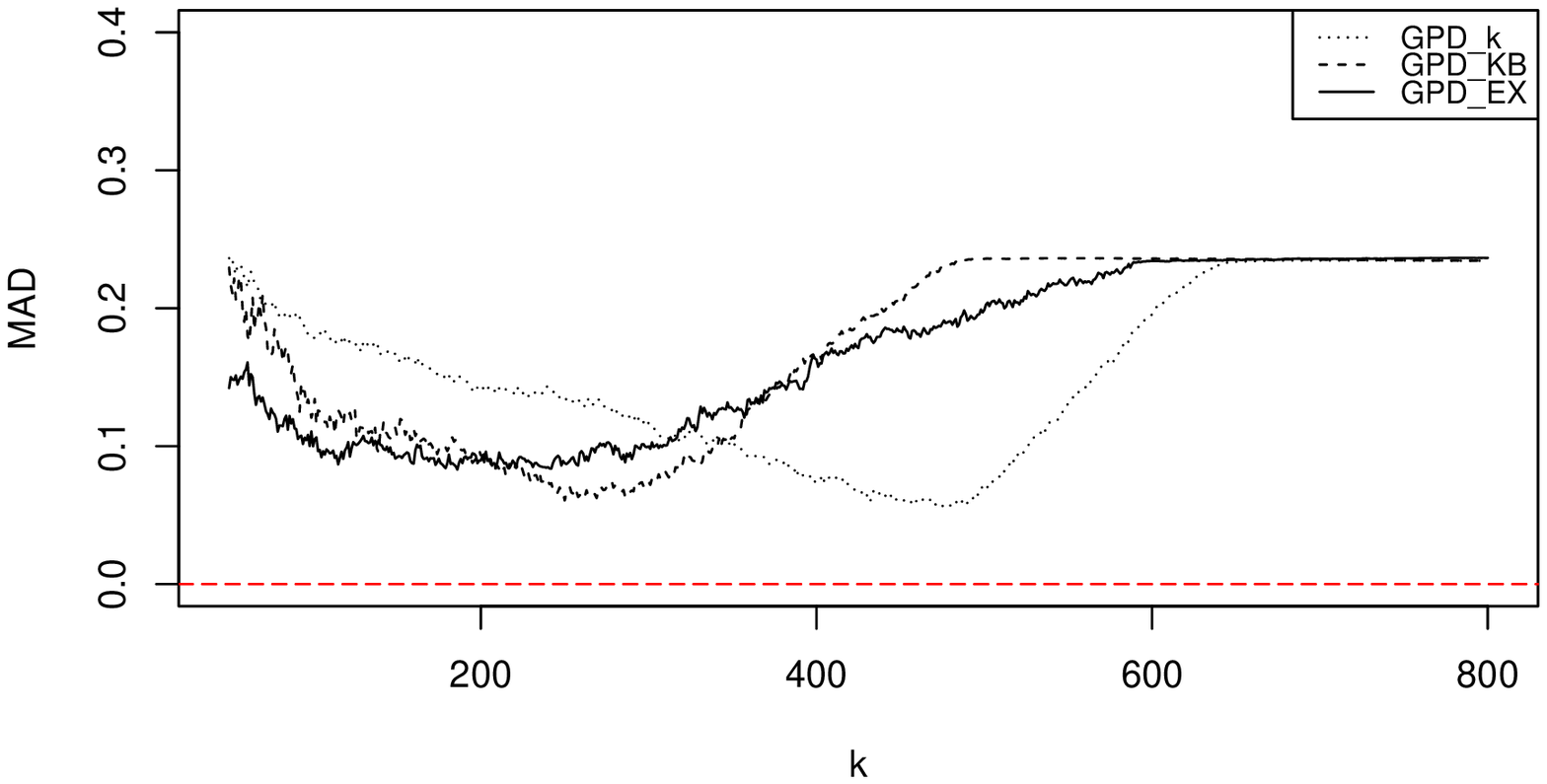}}\hfill
	\subfloat[ ]{%
		\includegraphics[height=6cm,width=.33\textwidth]{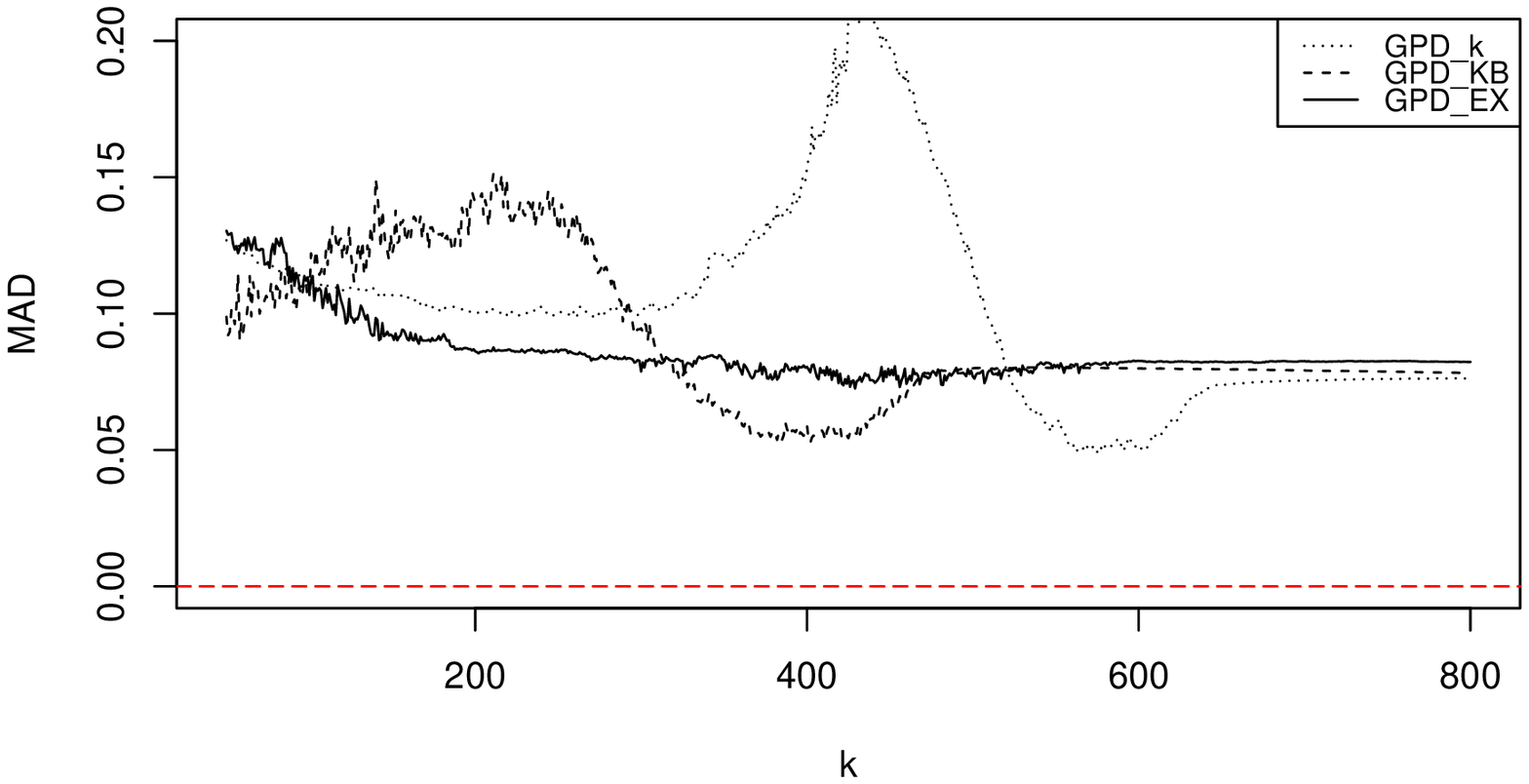}}\\
	\subfloat[]{%
		\includegraphics[height=6cm,width=.33\textwidth]{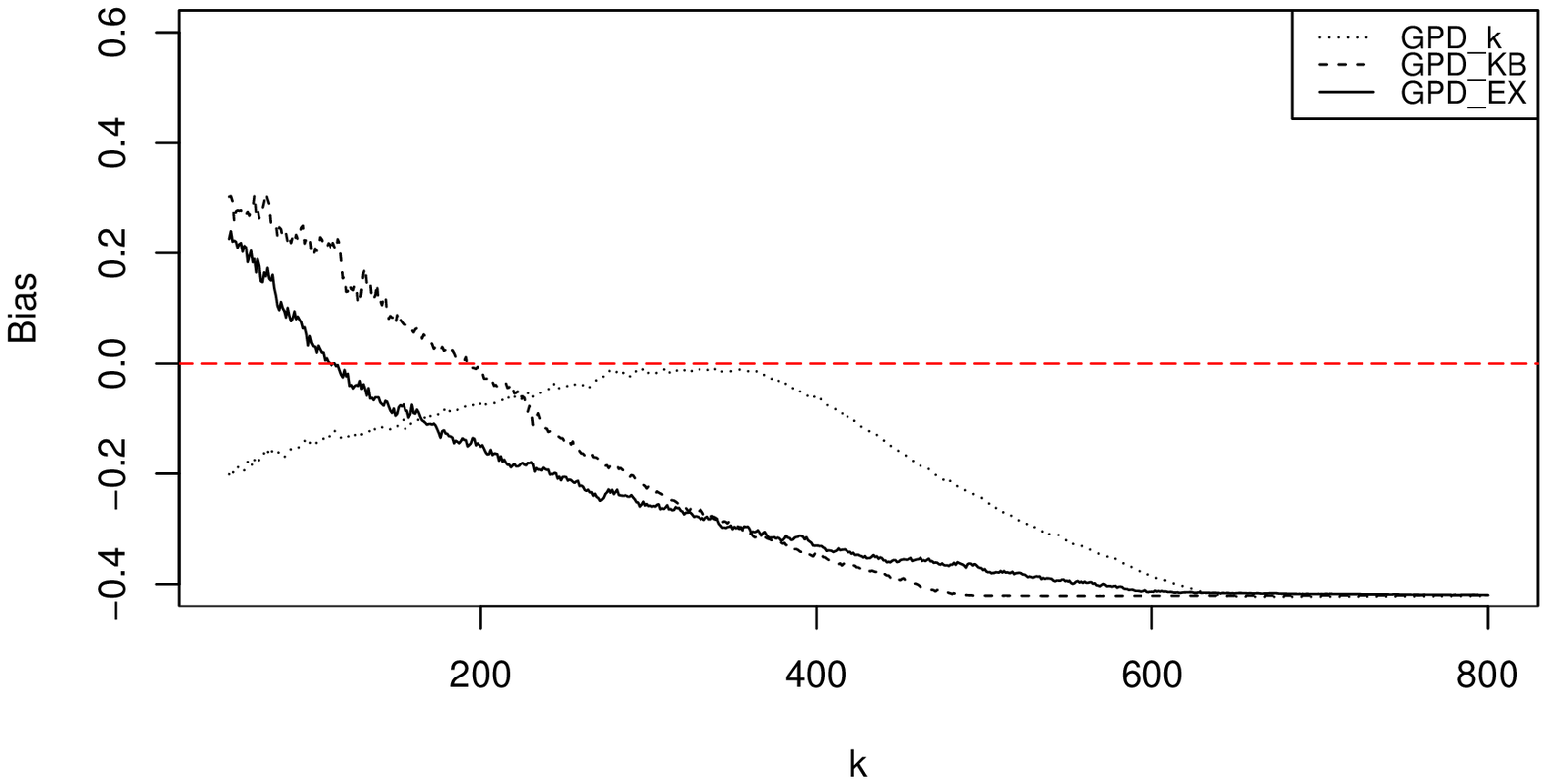}}\hfill
	\subfloat[]{%
		\includegraphics[height=6cm,width=.33\textwidth]{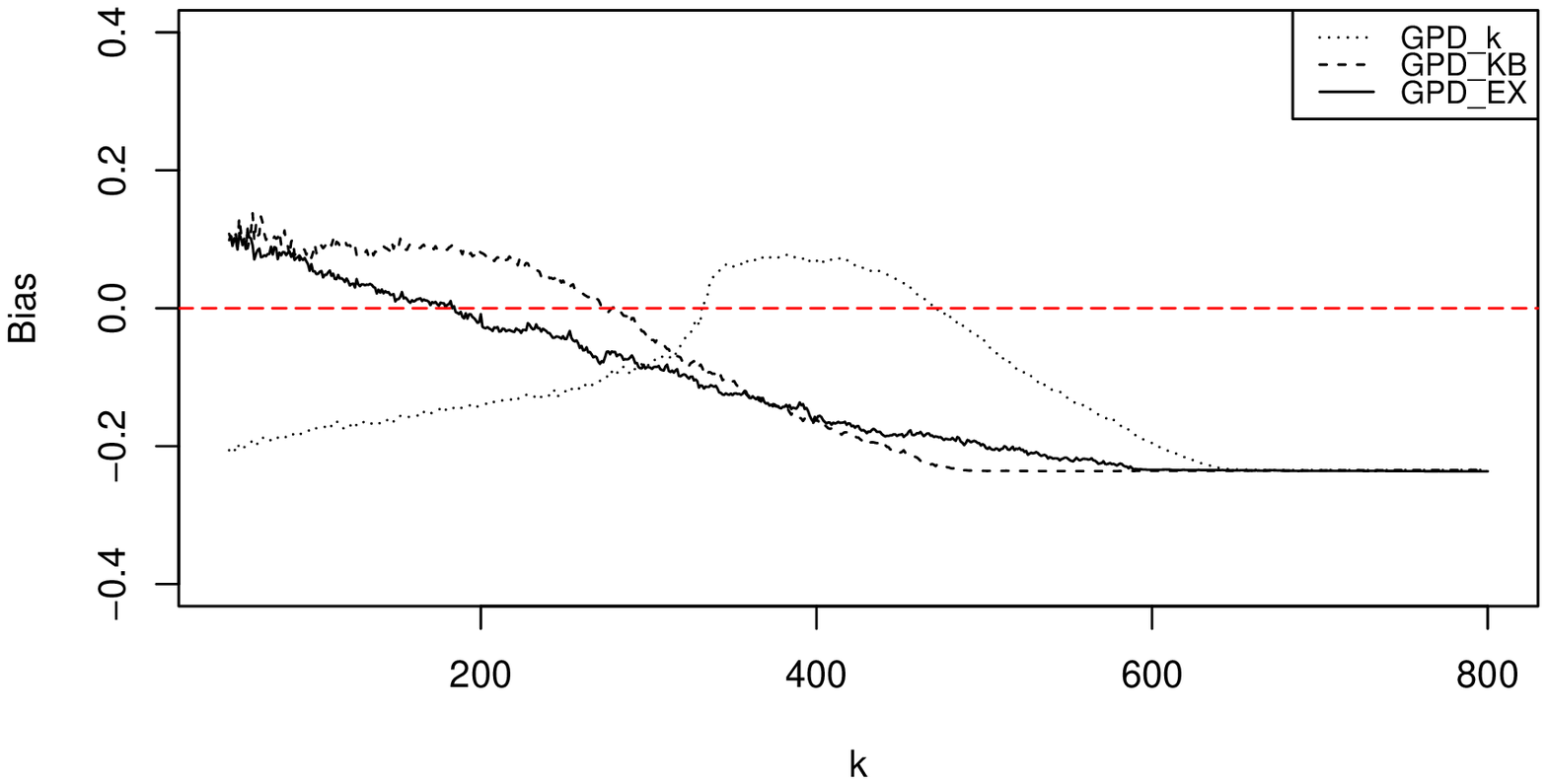}}\hfill
	\subfloat[ ]{%
		\includegraphics[height=6cm,width=.33\textwidth]{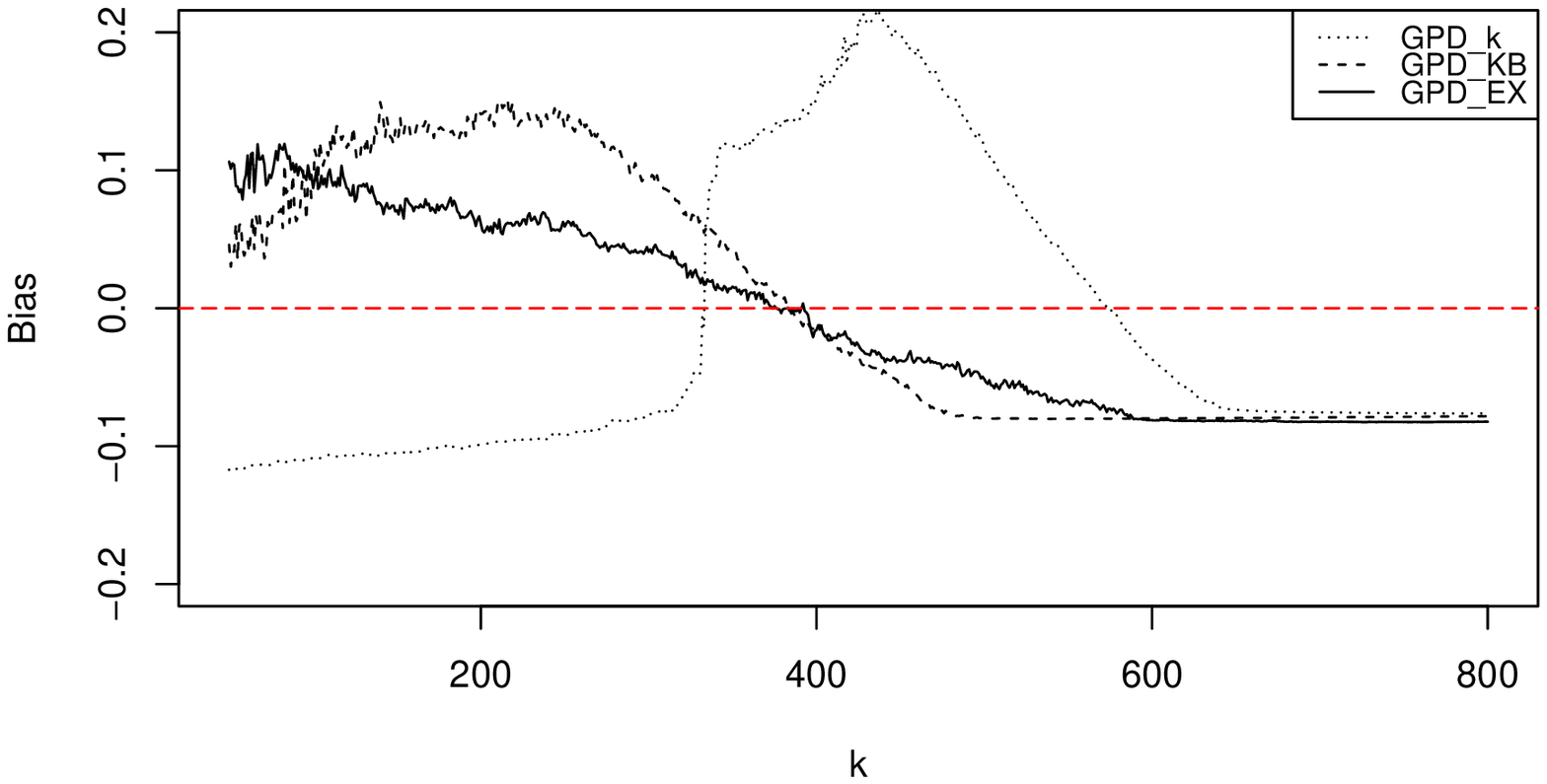}}\\
	\caption{Fr\'{e}chet Distribution with $n=1000:$ Left Column: $\gamma(x)=0.50;$ Middle Column: $\gamma(x)=0.30;$ and Right Column: $\gamma(x)=0.13$}
	\label{Fre1}
\end{figure}

\begin{figure}[htpb!]
	\centering
	
	\subfloat[]{%
		\includegraphics[height=6cm,width=.33\textwidth]{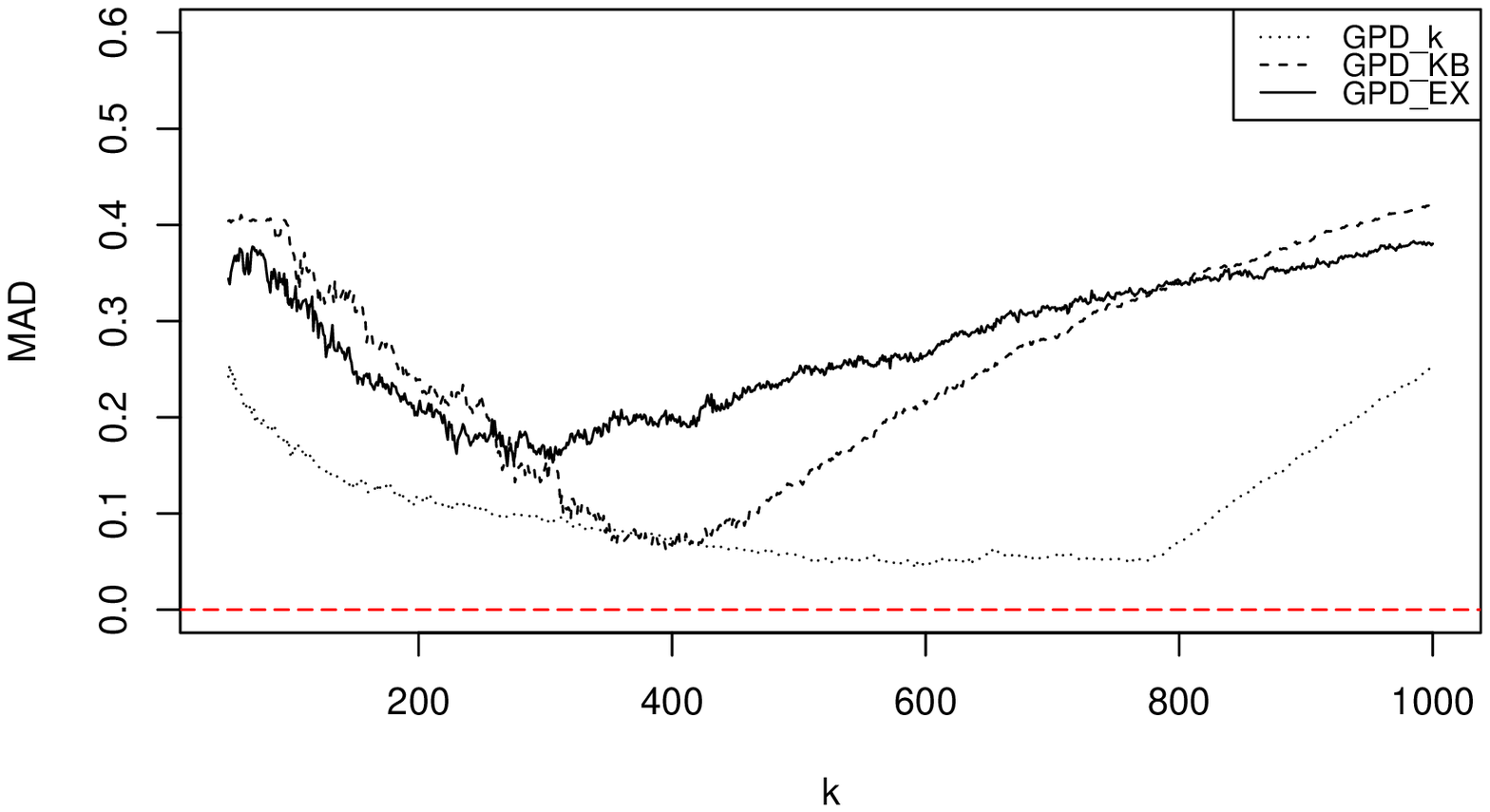}}\hfill
	\subfloat[]{%
		\includegraphics[height=6cm,width=.33\textwidth]{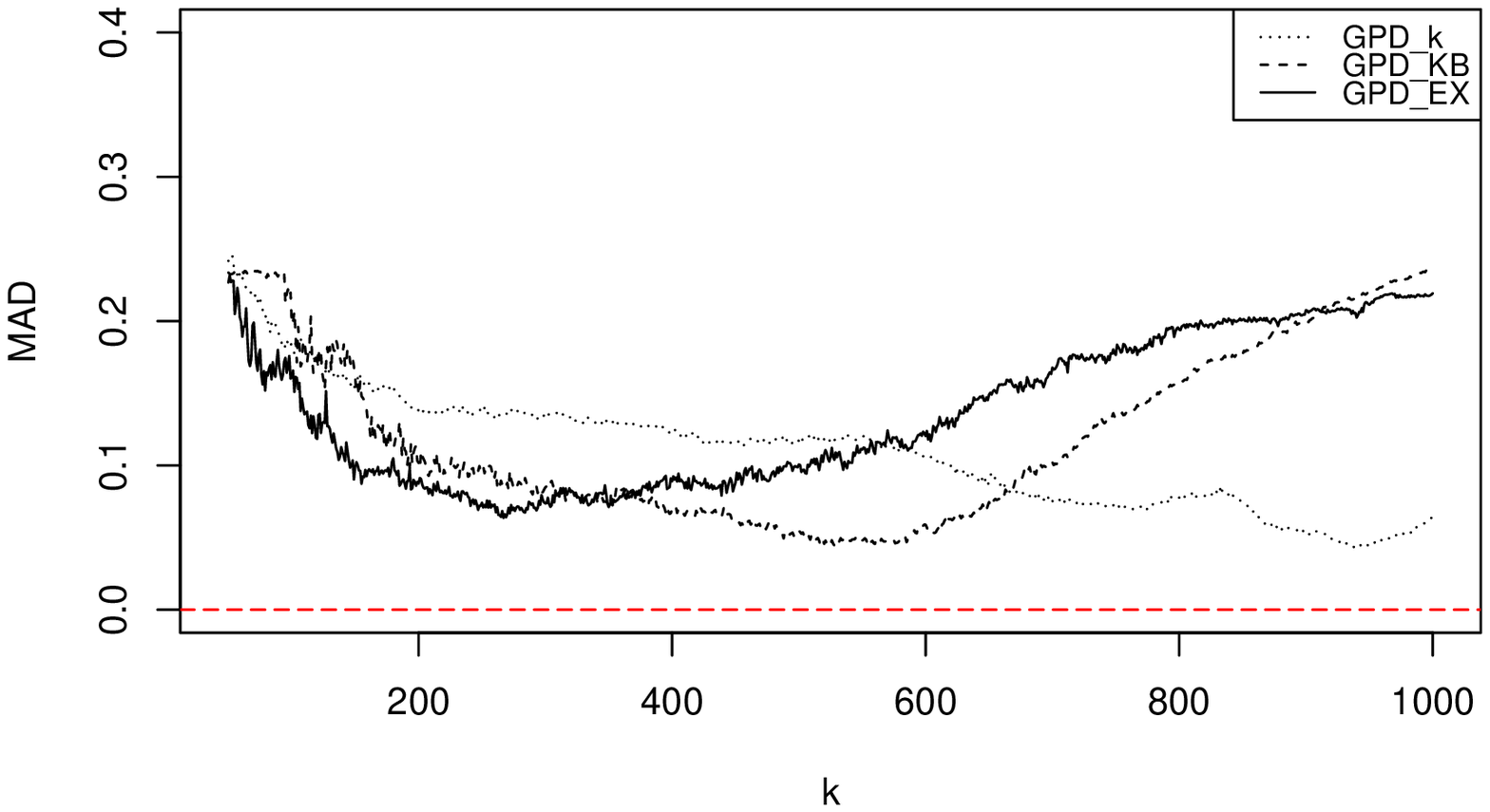}}\hfill
	\subfloat[ ]{%
		\includegraphics[height=6cm,width=.33\textwidth]{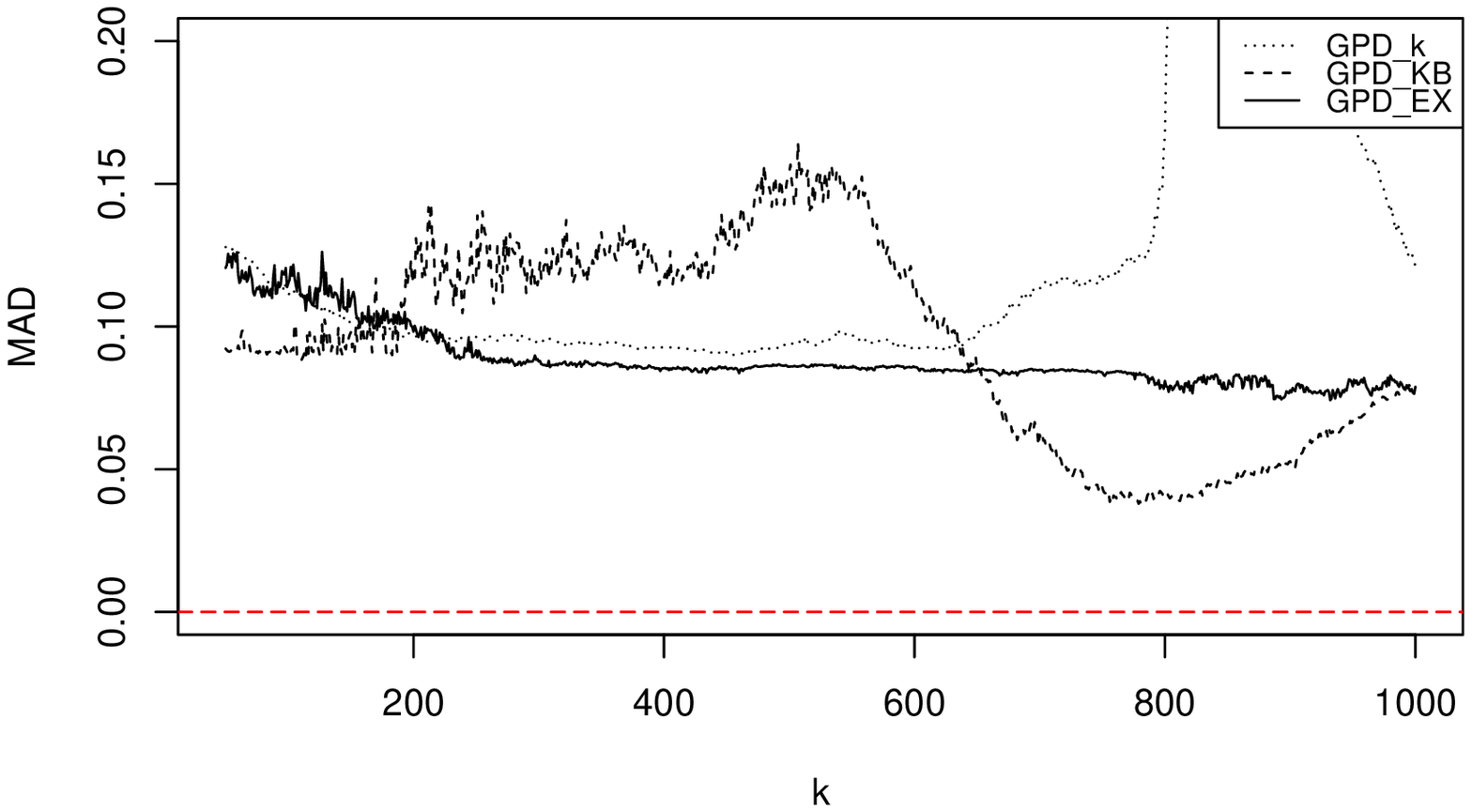}}\\
	\subfloat[]{%
		\includegraphics[height=6cm,width=.33\textwidth]{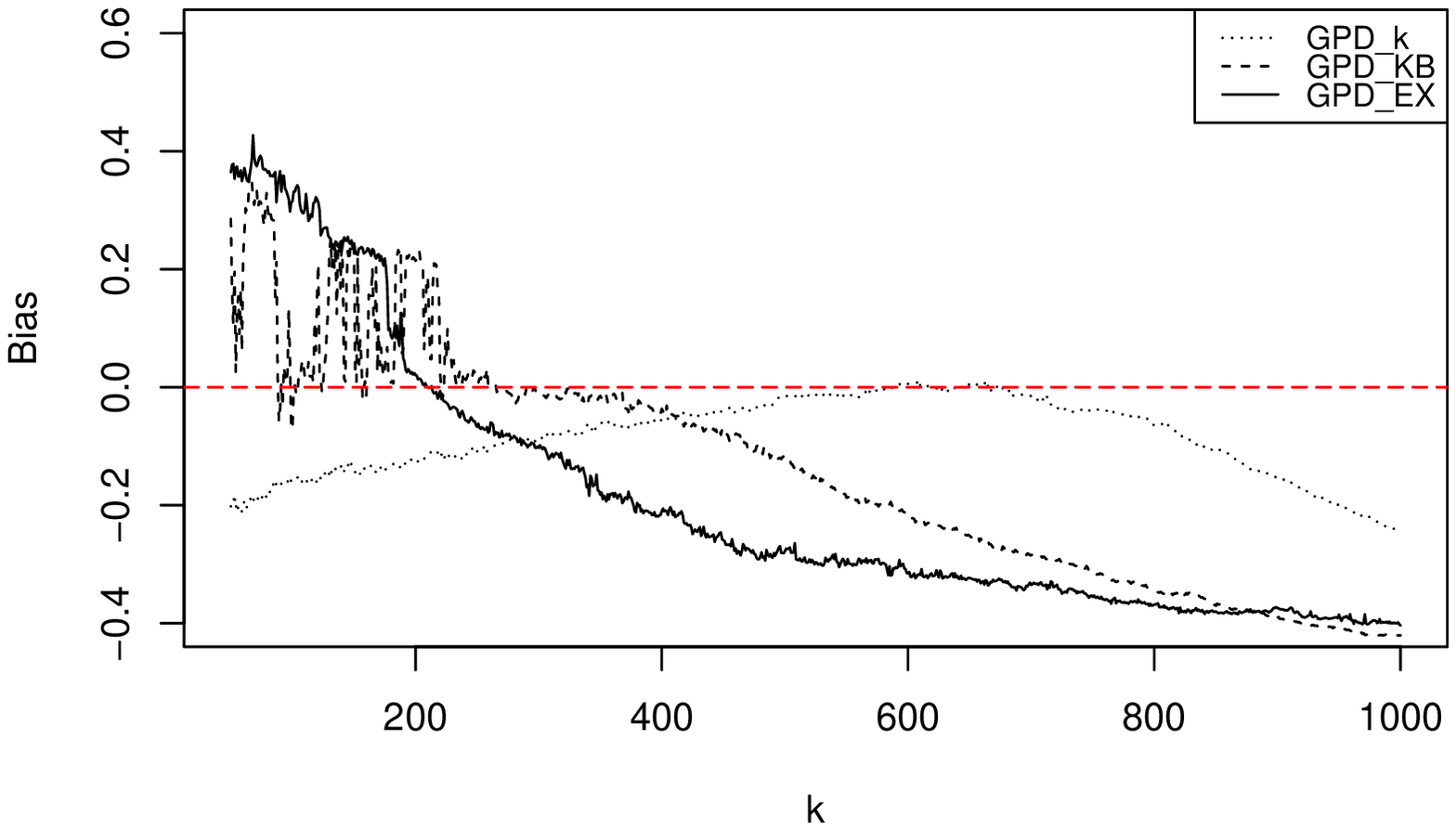}}\hfill
	\subfloat[]{%
		\includegraphics[height=6cm,width=.33\textwidth]{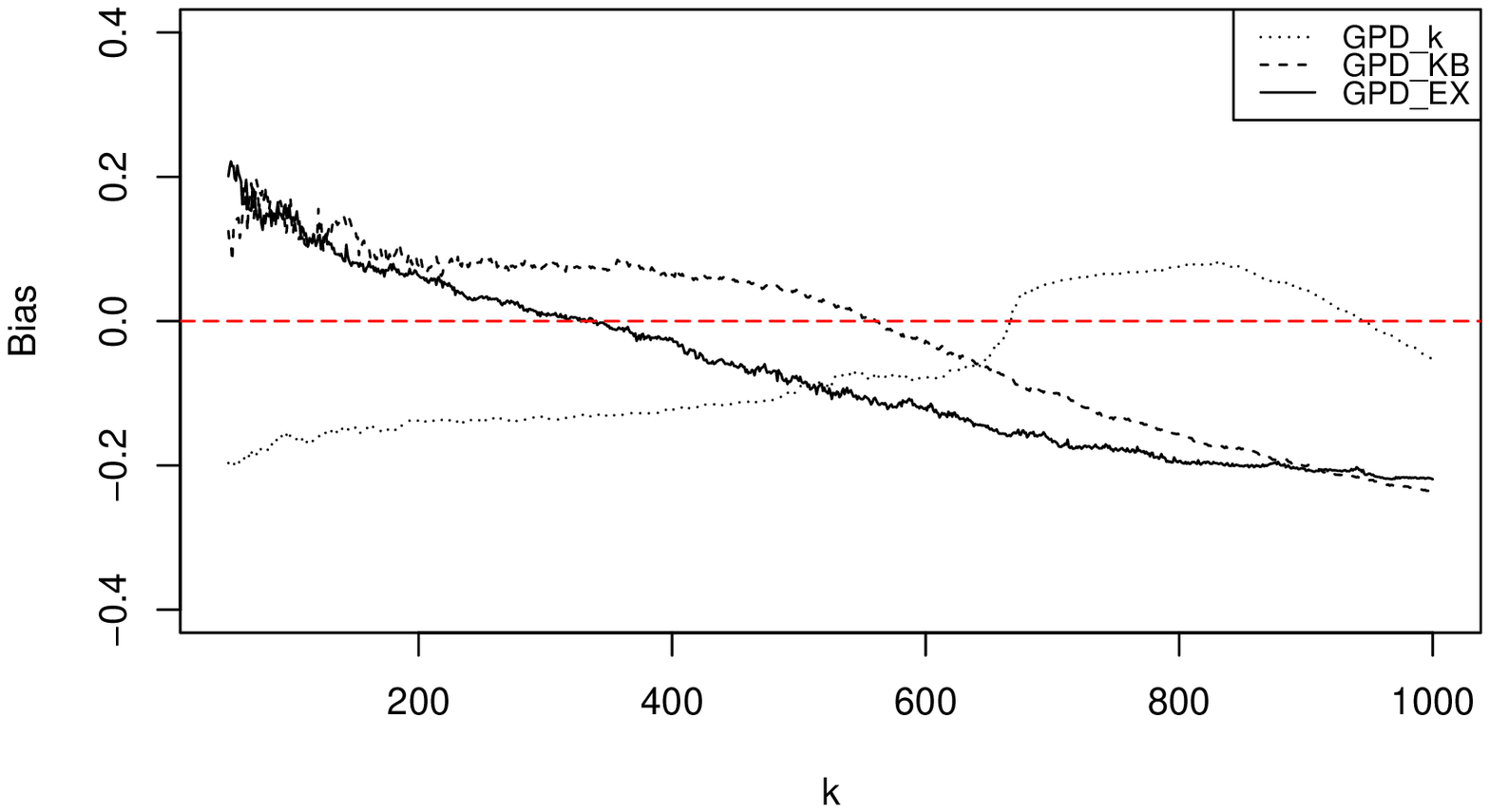}}\hfill
	\subfloat[ ]{%
		\includegraphics[height=6cm,width=.33\textwidth]{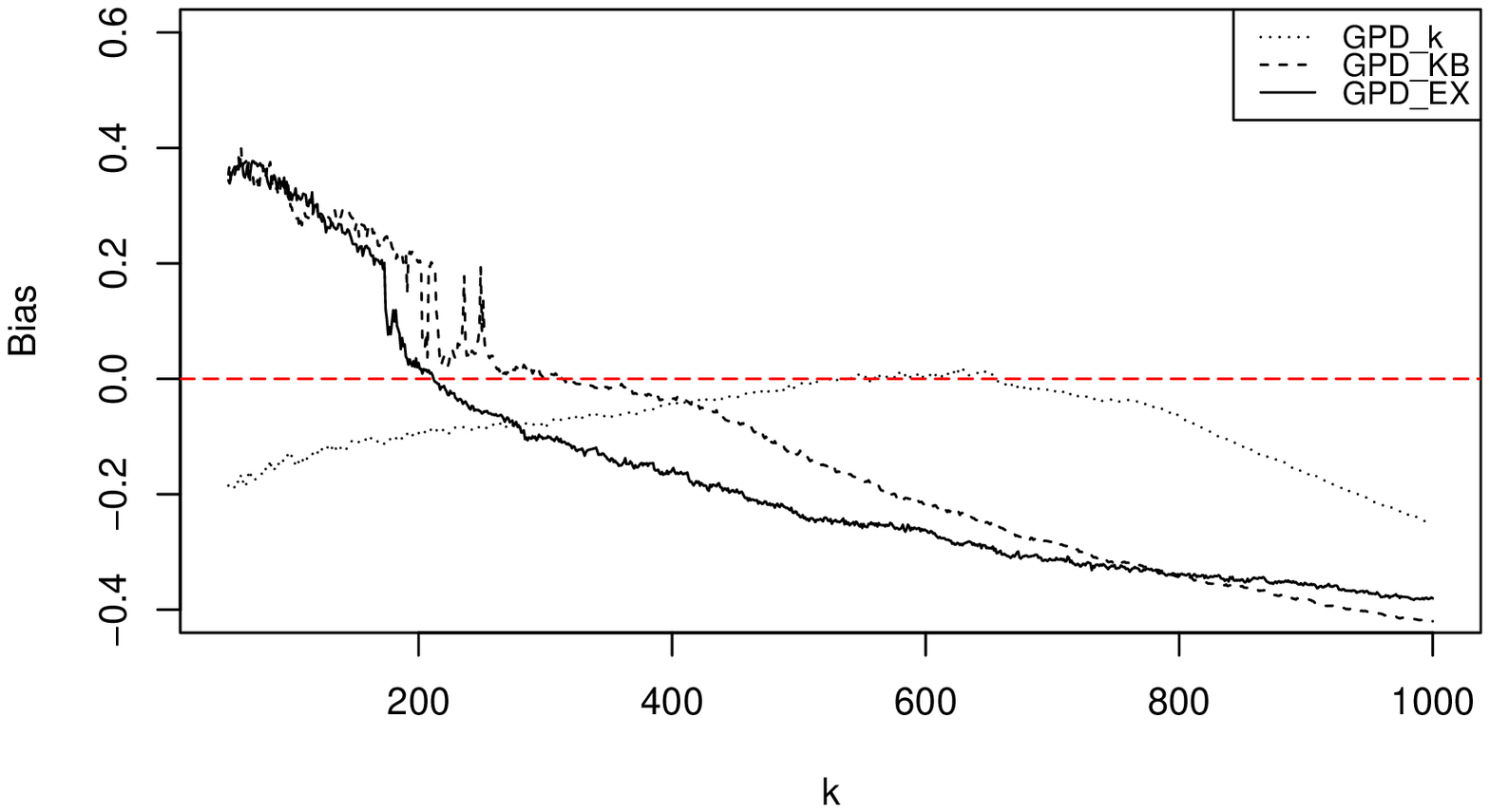}}\\
	\caption{Fr\'{e}chet Distribution with $n=2000:$ Left Column: $\gamma(x)=0.50;$ Middle Column: $\gamma(x)=0.30;$ and Right Column: $\gamma(x)=0.13$}
	\label{Fre2}
\end{figure}

\begin{figure}[htpb!]
	\centering
	
	\subfloat[]{%
		\includegraphics[height=6cm,width=.33\textwidth]{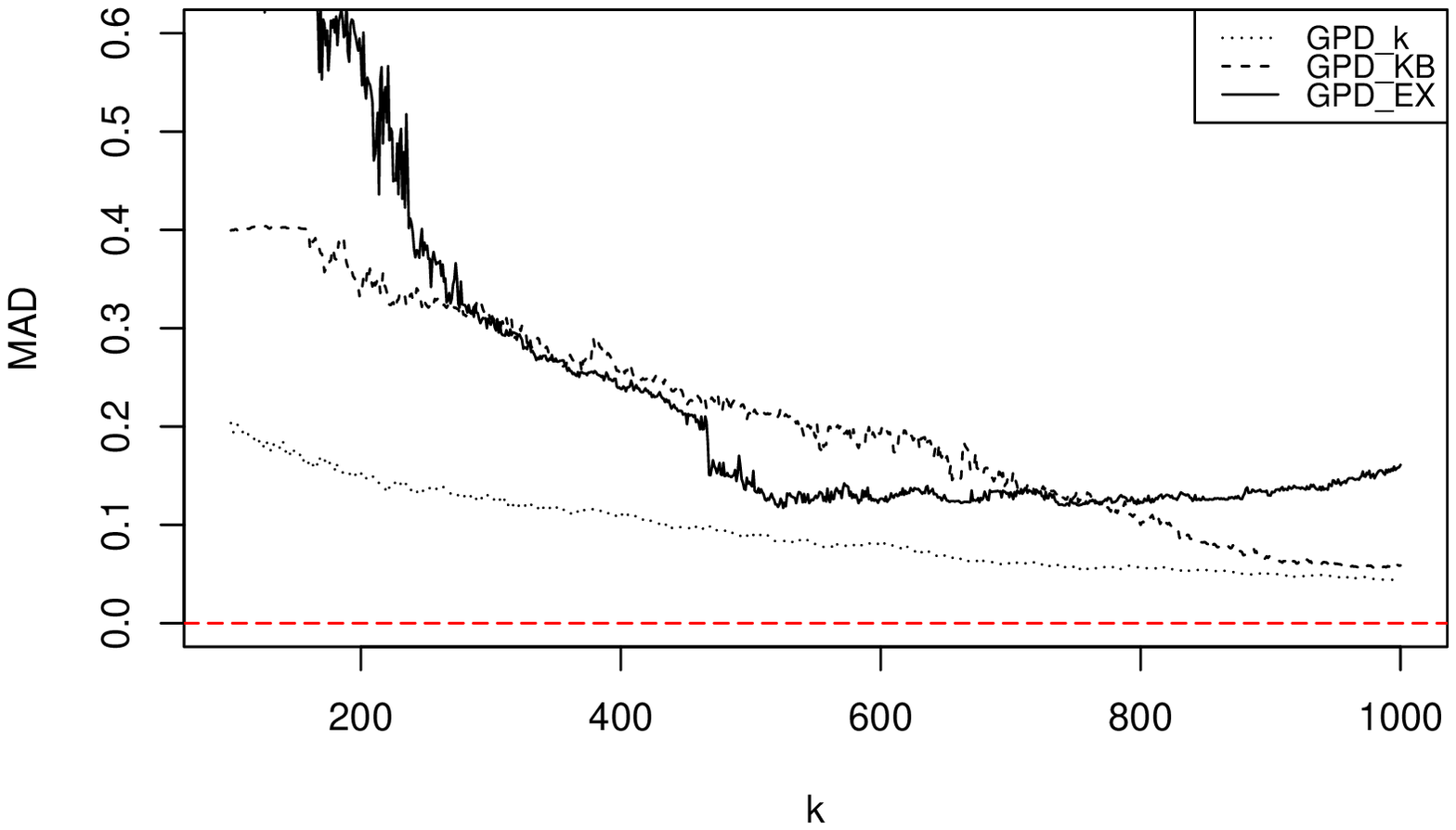}}\hfill
	\subfloat[]{%
		\includegraphics[height=6cm,width=.33\textwidth]{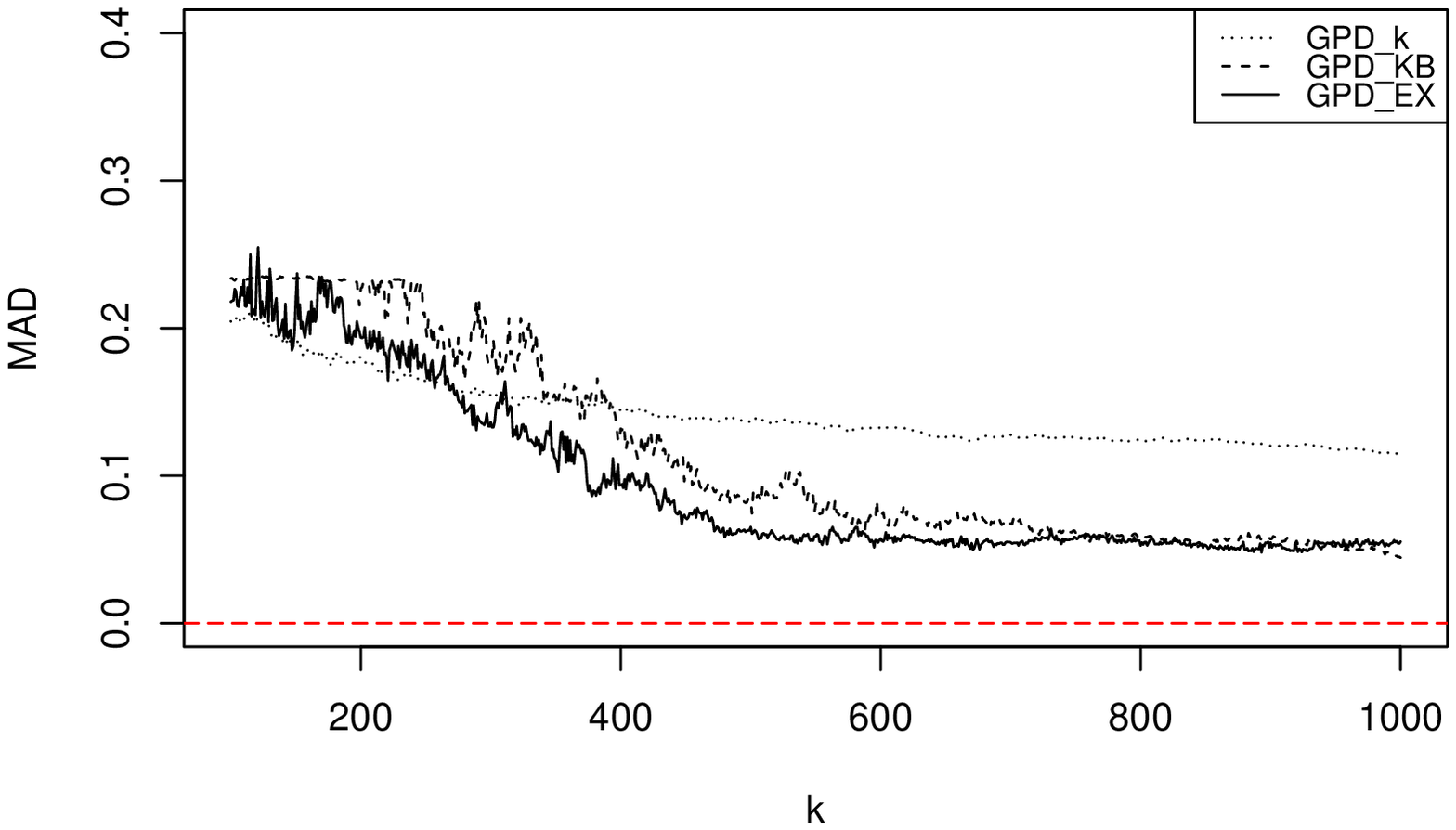}}\hfill
	\subfloat[ ]{%
		\includegraphics[height=6cm,width=.33\textwidth]{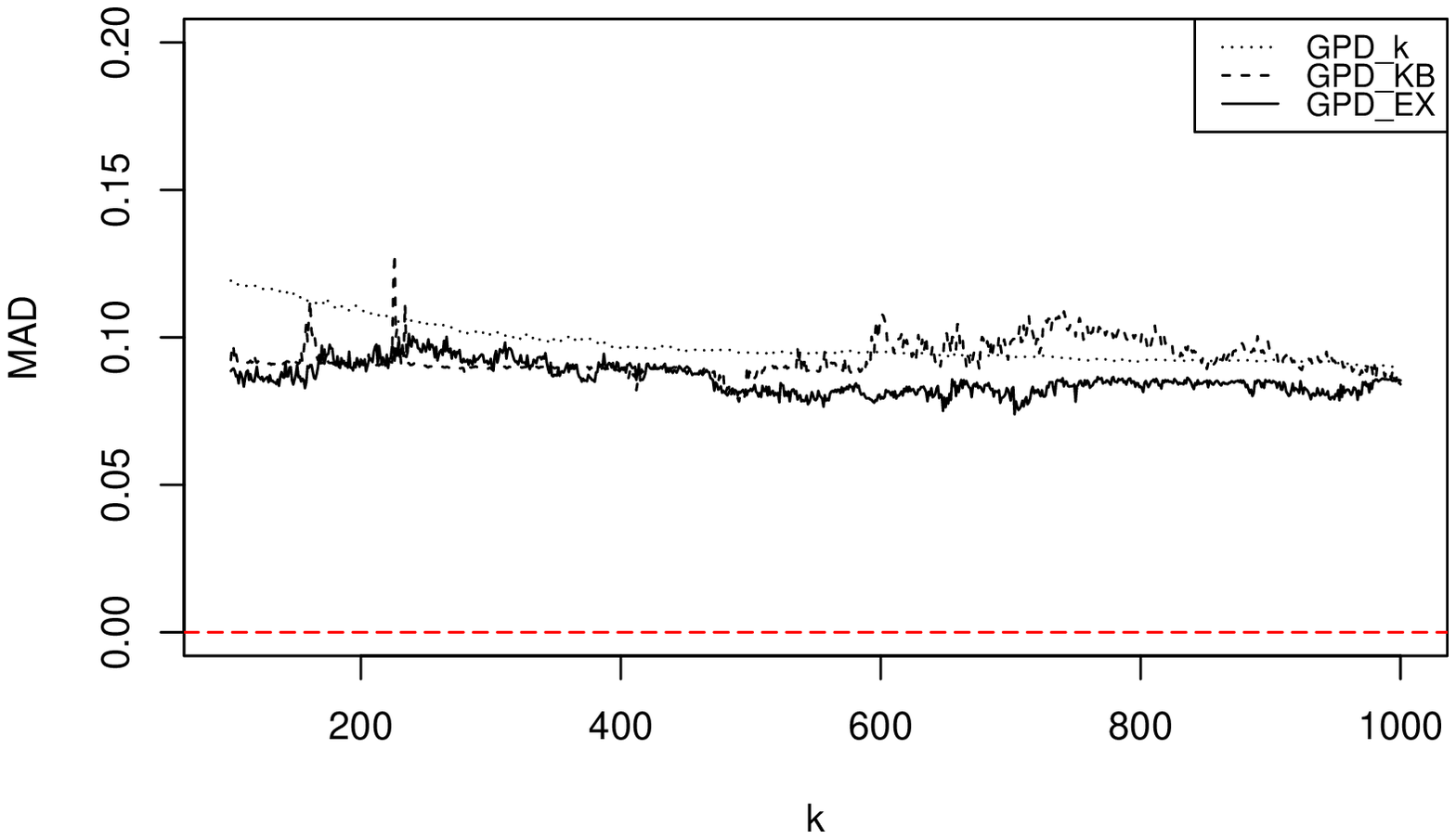}}\\
	\subfloat[]{%
		\includegraphics[height=6cm,width=.33\textwidth]{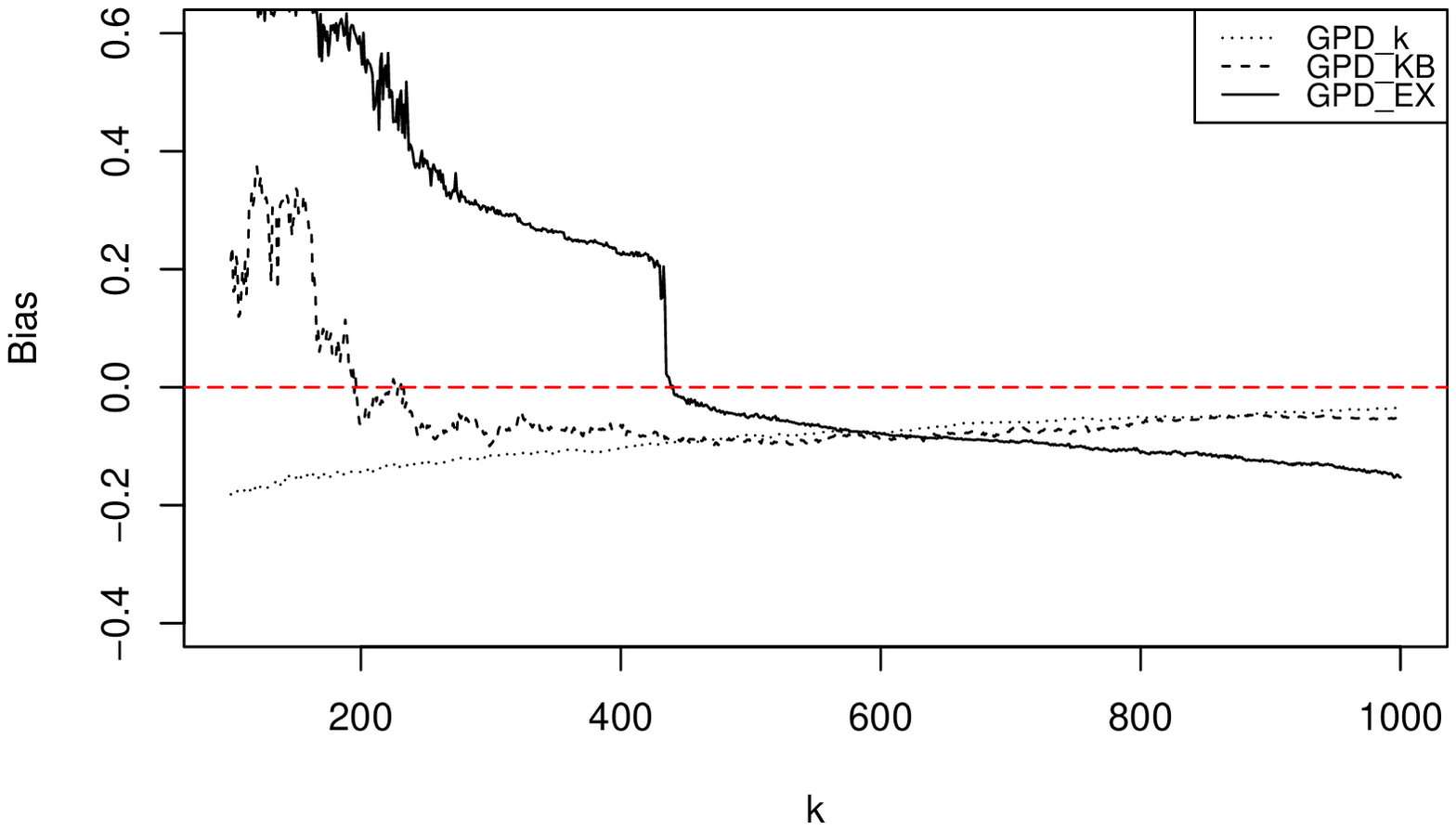}}\hfill
	\subfloat[]{%
		\includegraphics[height=6cm,width=.33\textwidth]{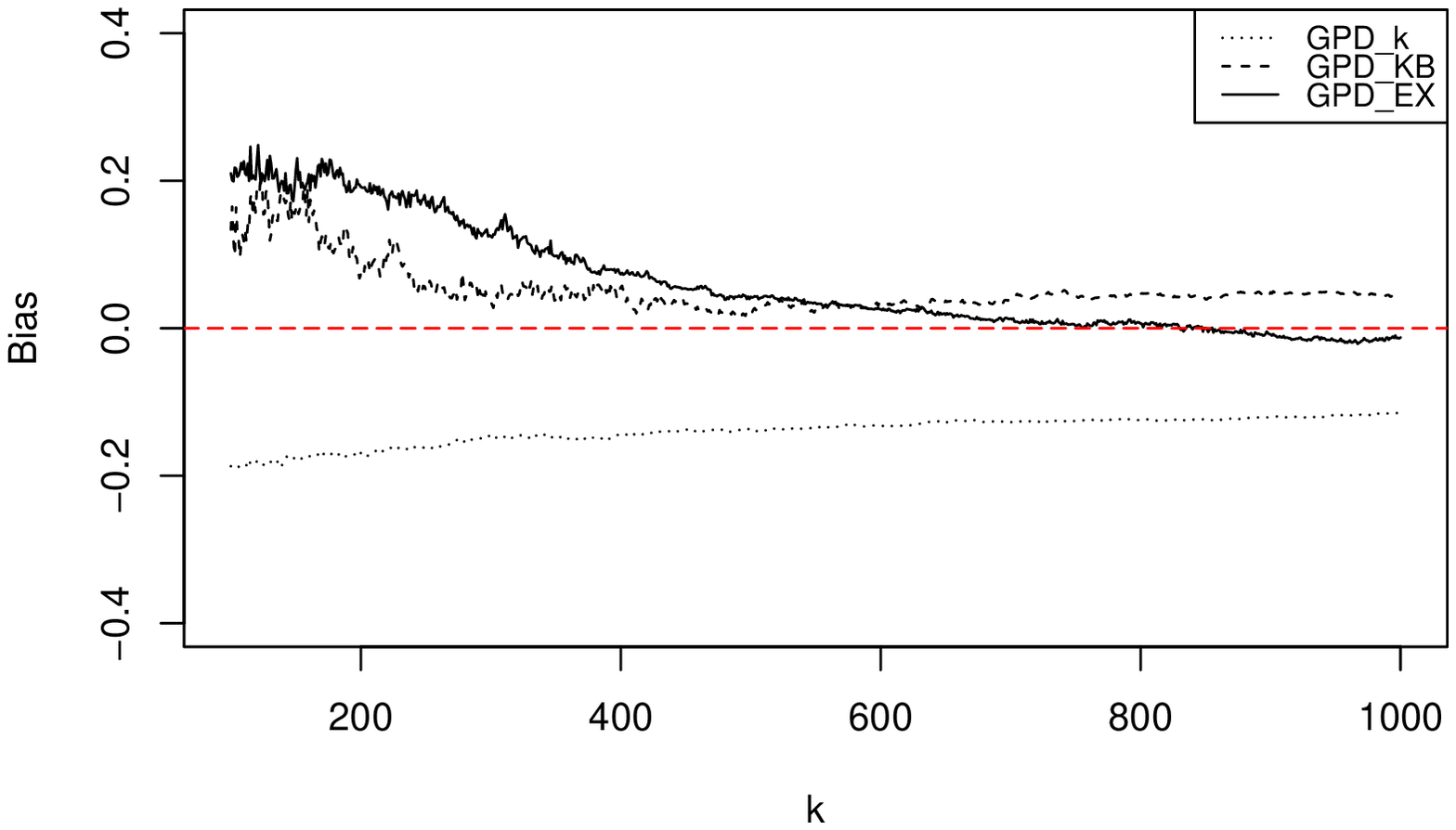}}\hfill
	\subfloat[ ]{%
		\includegraphics[height=6cm,width=.33\textwidth]{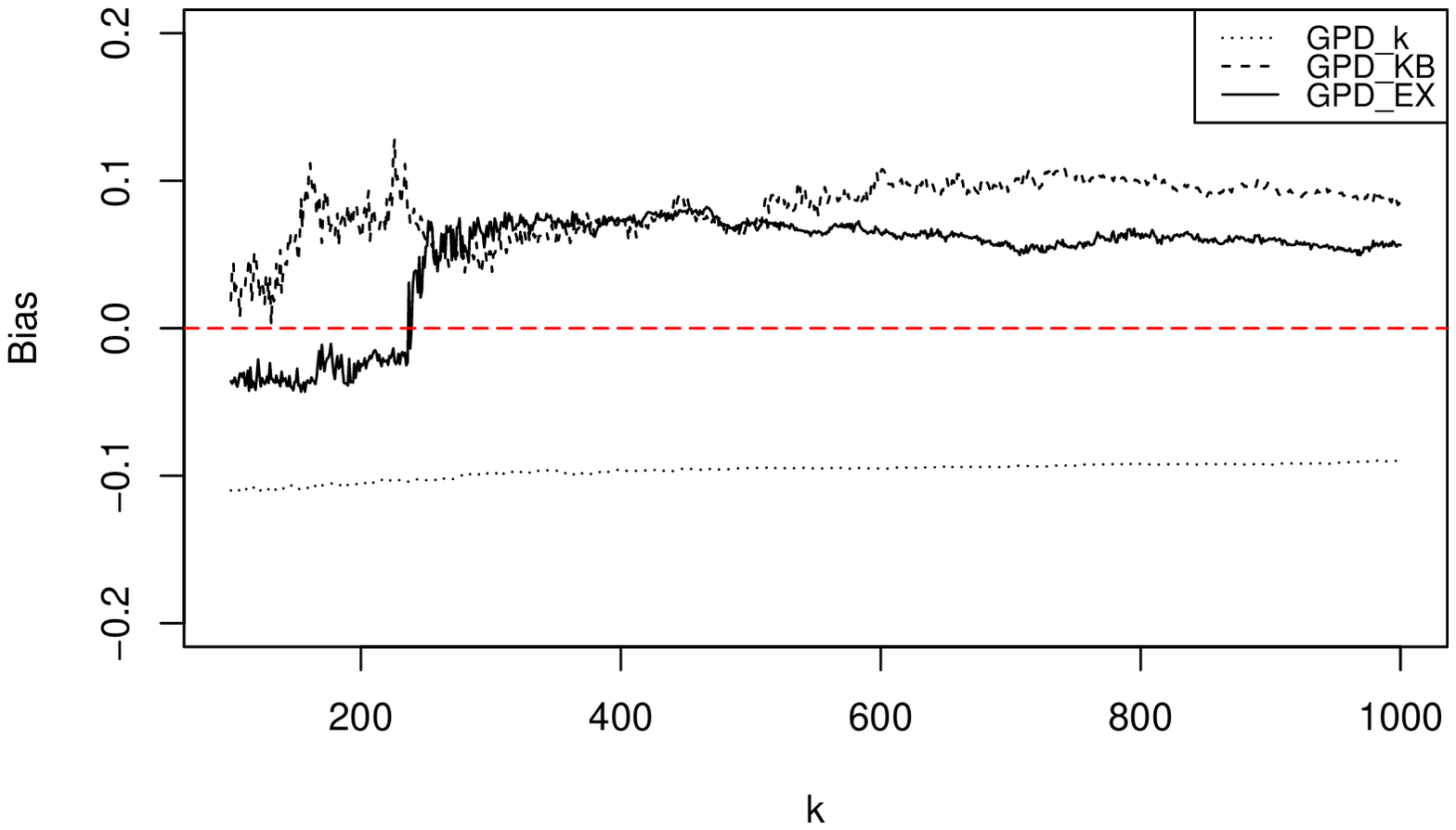}}\\
	\caption{Fr\'{e}chet Distribution with $n=5000:$ Left Column: $\gamma(x)=0.50;$ Middle Column: $\gamma(x)=0.30;$ and Right Column: $\gamma(x)=0.13$}
	\label{Fre3}
\end{figure}

\end{document}